\include{macpap2}

\documentclass[12pt]{ociamthesis}

\usepackage{amssymb,epsfig}
\usepackage{youngtab}
\usepackage[colorlinks=true,linkcolor=black,citecolor=black,plainpages=false,pdfpagelabels]{hyperref} 
\usepackage{dsfont}
\usepackage{amsmath}
\usepackage{amssymb}
\usepackage{amsthm}
\usepackage{graphicx}
\usepackage{mathrsfs}
\usepackage{units}

\newcommand{\comment}[1]{}

\newcommand{\abs}[1]{\ensuremath{|#1|}}

\newcommand{\norm}[2]{\ensuremath{|\!|#1|\!|_{#2}}}

\newcommand{\normBig}[2]{\ensuremath{\Big|\!\Big|#1\Big|\!\Big|_{#2}}}
\newcommand{\Norm}[2]{\ensuremath{\left|\!\left|#1\right|\!\right|_{#2}}}

\newcommand{\tr}{\textnormal{tr}}
\newcommand{\trace}[1]{\ensuremath{\tr (#1)}}
\newcommand{\Trace}[1]{\ensuremath{\tr \left( #1 \right)}}
\newcommand{\tracebig}[1]{\ensuremath{\tr \big( #1 \big)}}
\newcommand{\ptr}[1]{\textnormal{tr}_{\textnormal{\tiny #1}}}
\newcommand{\ptrace}[2]{\ensuremath{\ptr{#1} (#2)}}
\newcommand{\Ptrace}[2]{\ensuremath{\ptr{#1} \left(#2\right)}}

\newcommand{\supp}[1]{\textnormal{supp}\, \{ #1 \}}

\newcommand{\idx}[2]{{#1}_{\textnormal{\tiny #2}}}

\newcommand{\ket}[1]{| #1 \rangle}

\newcommand{\bra}[1]{\langle #1 |}

\newcommand{\braket}[2]{\langle #1 | #2 \rangle}

\newcommand{\bracket}[3]{\langle #1 | #2 | #3 \rangle}

\newcommand{\proj}[2]{| #1 \rangle\!\langle #2 |}
\newcommand{\proji}[3]{| #1 \rangle\!\langle #2 |_{\textnormal{\tiny #3}}}

\renewcommand{\d}[1]{\ensuremath{\textnormal{d}#1}}

\newcommand{\eps}{\varepsilon}

\newcommand{\hi}[1]{\ensuremath{\mathcal{H}_{\textnormal{\tiny #1}}}}

\newcommand{\hAB}{\hi{AB}}

\newcommand{\id}{\ensuremath{\mathds{1}}}
\newcommand{\idi}[1]{\ensuremath{\mathds{1}_{\textnormal{\tiny #1}}}}
\newcommand{\idA}{\idi{A}}

\newcommand{\opid}{\ensuremath{\mathcal{I}}}
\newcommand{\opidi}[1]{\ensuremath{\mathcal{I}_{\textnormal{\tiny #1}}}}

\newcommand{\linops}[1]{\ensuremath{\mathcal{L}(#1)}}
\newcommand{\hermops}[1]{\ensuremath{\mathcal{L}^\dagger(#1)}}
\newcommand{\posops}[1]{\ensuremath{\mathcal{P}(#1)}}

\newcommand{\normstates}[1]{\ensuremath{\mathcal{S}_{=}(#1)}}
\newcommand{\subnormstates}[1]{\ensuremath{\mathcal{S}_{\leq}(#1)}}

\newcommand{\cA}{\mathcal{A}}
\newcommand{\cB}{\mathcal{B}}

\newcommand{\cD}{\mathcal{D}}
\newcommand{\cE}{\mathcal{E}}
\newcommand{\cF}{\mathcal{F}}
\newcommand{\cG}{\mathcal{G}}
\newcommand{\cH}{\mathcal{H}}

\newcommand{\cN}{\mathcal{N}}

\newcommand{\cP}{\mathcal{P}}

\newcommand{\cT}{\mathcal{T}}

\newcommand{\cX}{\mathcal{X}}
\newcommand{\cY}{\mathcal{Y}}

\newcommand{\rhoA}{\ensuremath{\idx{\rho}{A}}}
\newcommand{\rhoB}{\ensuremath{\idx{\rho}{B}}}

\newcommand{\rhoAB}{\ensuremath{\idx{\rho}{AB}}}

\newcommand{\rhoAR}{\ensuremath{\idx{\rho}{AR}}}

\newcommand{\rhoR}{\ensuremath{\idx{\rho}{R}}}

\newcommand{\sigmaB}{\ensuremath{\idx{\sigma}{B}}}

\newcommand{\sigmaE}{\ensuremath{\idx{\sigma}{E}}}

\newcommand{\zetaR}{\ensuremath{\idx{\zeta}{R}}}

\newcommand{\cTAE}{\ensuremath{\idx{\cT}{A$\rightarrow$E}}}
\newcommand{\HA}{\ensuremath{\idx{\cH}{A}}}
\newcommand{\HB}{\ensuremath{\idx{\cH}{B}}}
\newcommand{\HE}{\ensuremath{\idx{\cH}{E}}}

\newcommand{\hh}[4]{\ensuremath{H_{#1}^{#2}(\textnormal{#3})_{#4}}}
\newcommand{\chh}[5]{\ensuremath{H_{#1}^{#2}(\textnormal{#3}|\textnormal{#4})_{#5}}}

\newcommand{\hmin}[2]{\hh{\textnormal{min}}{}{#1}{#2}} 

\newcommand{\chmin}[3]{\chh{\textnormal{min}}{}{#1}{#2}{#3}}
\newcommand{\chmineps}[3]{\chh{\textnormal{min}}{\eps}{#1}{#2}{#3}}
\newcommand{\chmineeps}[4]{\chh{\textnormal{min}}{#1}{#2}{#3}{#4}}

\newcommand{\chtwo}[3]{\chh{\textnormal{2}}{}{#1}{#2}{#3}}

\newcommand{\epsball}[1]{\ensuremath{\mathcal{B}^\eps(#1)}}

\theoremstyle{plain}

\theoremstyle{definition}
\newtheorem{definition}{Definition}

\include{babel}

\begin{document}




\begin{titlepage}
\begin{center}
\vspace*{5cm}
\noindent {\LARGE \textsc{MASTER THESIS:}} \\
\vspace*{0.8cm}
\noindent {\Huge \textbf{Decoupling Theorems}} \\
\vspace*{5cm}
\noindent \LARGE Oleg Szehr\\
\vspace*{0.8cm}
\noindent \large Supervisors:\\
\vspace*{0.2cm}
\noindent \Large Marco Tomamichel, Fr\'ed\'eric Dupuis\\
and Renato Renner\\

\vspace*{4cm}
\noindent \large Institut f\"ur Theoretische Physik\\
\vspace*{0.2cm}
\noindent \large ETH Z\"{u}rich, January 2011 \\
\vspace*{0.5cm}
\end{center}
\end{titlepage}
\sloppy

\titlepage

\baselineskip=18pt plus1pt
\setcounter{secnumdepth}{3}
\setcounter{tocdepth}{3}

\begin{acknowledgements}
I thank Marco Tomamichel, Fr\'{e}d\'{e}ric Dupuis and Renato Renner for the supervision of this project.
\end{acknowledgements}

\begin{abstract}
Decoupling theorems have proven useful in various applications in the area of quantum information theory. This thesis builds upon preceding work by Fr\'{e}d\'{e}ric Dupuis \cite{Fred:PHD}, where a general decoupling theorem is obtained and its implications for quantum coding theory are studied. At first we generalize this theorem to the case where the average is taken over an approximate unitary 2-design. The second part of this thesis tackles the question whether or not it is possible to decorrelate CQ-states with classical operations. We obtain results similar to the pivotal Leftover Hash Lemma. Finally we analyze the decoupling power of permutation operators in a fully quantum context and show a general procedure that yields decoupling theorems with permutations.
\end{abstract}

\begin{romanpages}
\tableofcontents
\end{romanpages}

\chapter{Introduction}
\section{Overview}
Quantum mechanical correlations are at the core of Quantum Information Theory. Correlated systems behave in a predictable way, such that the probability distributions obtained from measuring some observable on the one system partially determine possible measurement outcomes of the other. Knowledge about the physical state of one system implies knowledge about the state of any system correlated to it and correlations can be thought of as the carrier of quantum mechanical information.
It is a basic issue in information theory to examine the correlations between different systems, since it includes the question of how much information one physical system contains over the other. The fact that two quantum systems are (almost) uncorrelated has significant consequences not only for information processing tasks but also for the physical behavior of these systems.
In particular, this fact can be used to show that other systems are strongly correlated, which bonds the behavior of the one system to the other. Thus, a theorem that states conditions under which different systems are close to being uncorrelated provides substantial insight to the information theory of those and other systems. In \cite{Fred:PHD} a very general decoupling theorem is derived (in the sequel called the Decoupling Theorem) and its impact on the theory of quantum coding is studied. Roughly speaking, the theorem applies in a situation where a joint system $AR$ with possible correlations between the subsystems is given and a unitary evolution followed by an arbitrary physical process take place on the $A$ part of the whole system. It provides a bound on how far a typical resulting state is from a completely uncorrelated state. Denoting the state of the system $AR$ with $\rhoAR$, the unitary evolution with $\idx{U}{A}$, and the arbitrary physical process following the unitary evolution with $\cT$, the theorem states that 
$$\int_{\mathbb{U}(A)} {\left\| \mathcal{T}((\idx{U}{A} \otimes \idi{R}) \ \rhoAR \  (\idx{U}{A}^\dagger \otimes \idi{R})) - \idx{\omega}{E} \otimes \idx{\rho}{R} \right\|}\: \d{U} \leq
2^{-\frac{1}{2} \chmin{A'}{E}{\omega} - \frac{1}{2}\chmin{A}{R}{\rho} }.$$
The integration goes over the whole group of unitary matrices and is with respect to the normalized Haar measure. A quantum state with tensor product structure as $\idx{\omega}{E} \otimes \idx{\rho}{R}$ in Quantum Mechanics represents a joint system whose subsystem are not correlated. The Decoupling Theorem bounds the average distance of the state $\mathcal{T}((\idx{U}{A} \otimes \idi{R}) \ \rhoAR \  (\idx{U}{A}^\dagger \otimes \idi{R}))$ from a tensor product state. The right hand side of the above inequality is given in terms of the conditional $\idx{H}{min}$-entropy, which is a prevalent measure of uncertainty. The term $\chmin{A}{R}{\rho}$ for instance quantifies the uncertainty an observer with access to the $R$ subsystem of $AR$ has about the state of $A$.\\
Though stated in the context of coding theory, this theorem has far reaching implications in various areas of theoretical physics.
Among other things, decoupling arguments and corollaries of the Decoupling Theorem deepened the insight into thermodynamics \cite{ThermoNeg} and  Black Hole Information theory \cite{BLACK}. Nevertheless, it turns out that for physical systems taking the average over all unitary matrices as it is done in the statement of the theorem is too strong an assumption. In a real system the internal dynamics are governed by the laws of nature and typically these very laws restrict the possible unitary evolutions of the system. One cannot expect that such dynamics produce arbitrary unitaries evenly distributed according to the Haar measure. Instead it was recently shown \cite{RQC} that in a many qubit system with random local two-particle interactions the possible evolutions of the system constitute a unitary $\varepsilon$-almost 2-design.\\
This thesis discusses several decoupling results. The first chapter introduces the notation used and gives a brief overview of the quantum mechanical background. In the second chapter an alternative proof of the Decoupling Theorem resulting in a slightly tighter bound (as compared to \cite{Fred:PHD}) is shown. The following two chapters are devoted to a generalization of the Decoupling Theorem to the case when the average is taken over an $\varepsilon$-almost 2-design instead of the whole unitary group, opening applications of this theorem in various physical situations. Chapters five, six and seven analyze the decoupling behavior of permutations. Potential applications lie in the areas of quantum cryptography and coding theory. The results of the chapters five and six are related to the General Leftover Hash Lemma, which is of significance for quantum cryptography \cite{Renner:PHD}. Chapter seven aims at providing theorems for classical coding theory.

\section{Notation and a glance at quantum mechanics}
Throughout this thesis we will abide by the notational rules introduced in this section. One of the core objects of the mathematical description of some physical system $A$ according to quantum mechanics is a complex Hilbert space $\idx{\cH}{A}$, which we always will assume to have finite dimension, $\idx{d}{A}$. The space of linear operators on a Hilbert space $\cH$ will be denoted by $\linops{\cH}$, the subspace of hermitian operators by $\hermops{\cH}$ and the set of positive-semidefinite operators is given by $\posops{\cH}$. The set of normalized positive operators is given by $\normstates{\cH} := \{\rho\in\posops{\cH}\:\mid\:\tr\rho=1\}$ and the set of sub normalized positive operators is $\subnormstates{\cH} := \{\rho\in\posops{\cH}\:\mid\:\tr\rho\leq1\}$. For those sets one has the following trivial inclusions: 
$\normstates{\cH}\subset\subnormstates{\cH}\subset\posops{\cH}\subset\hermops{\cH}\subset\linops{\cH}$. More generally the vector space of homomorphisms of some Hilbert space $\idx{\cH}{A}$ to $\idx{\cH}{B}$ will be denoted by $\textnormal{Hom}(\idx{\cH}{A},\idx{\cH}{B})$. For $\varphi\in\cH$, the corresponding elements of the spaces $\textnormal{Hom}(\mathbb{C},\cH)$ and $\textnormal{Hom}(\cH,\mathbb{C})$ are denoted by $\ket{\varphi}$ and $\bra{\varphi}$ respectively. Thus for example $\proj{\varphi}{\varphi}\in\posops{\cH}$ is a projector. Since the spaces $\idx{\cH}{A}$ and $\textnormal{Hom}(\mathbb{C},\cH)$ are isomorphic, we sometimes will treat $\ket{\varphi}$ as if it was an element of $\idx{\cH}{A}$. In this cases we implicitly mean the unique corresponding element in $\idx{\cH}{A}$.\\
If the quantum state of the system $A$ is known with certainty, according to quantum mechanics it is represented by an element
\begin{align}
[\idx{\varphi}{A}]:=\{e^{i\alpha}\idx{\varphi}{A}\:|\:\idx{\varphi}{A}\in\idx{\cH}{A}; \Norm{\idx{\varphi}{A}}{}=1;\alpha\in[0,2\pi]\}\nonumber
\end{align}
of the projective Hilbert space belonging to $\idx{\cH}{A}$, where an index letter following some mathematical object denotes to which physical system it belongs.
More generally the quantum states of the system $A$ are in one to one correspondence with the elements $\idx{\rho}{A}$ of $\normstates{\idx{\cH}{A}}$, even if the state of the system is not fully known to its observer. Since in quantum information theory usually only partial knowledge about some system $A$ is given, the core object of study will be the density operator $\rhoA$, and we call $\rhoA$ just the state of the system. Sometimes one considers joint systems $AB$, which, due to the postulates of quantum mechanics, are represented by a tensor product space $\idx{\cH}{A}\otimes\idx{\cH}{B}=:\idx{\cH}{AB}$. Then the corresponding density operator will have a double index, too: $\rhoAB$. However, when it is clear which systems are represented by the density operators we might drop the indices to simplify the notation. We will denote with $\idi{A}$ the identity operator on $\HA$ and with $\idx{\pi}{A}:= \frac{\idi{A}}{\idx{d}{A}}$ the completely mixed state on $A$. Moreover $\idx{\Phi}{AB}$ is the completely entangled state on $AB$ i.e. $\idx{\Phi}{AB}:=\idx{\ket{\Phi}\bra{\Phi}}{AB}$ , where $\idx{\ket{\Phi}}{AB}:=\frac{1}{\sqrt{\idx{d}{A}}}\sum_i\idx{\ket{i}}{A}\otimes\idx{\ket{i}}{B}$ and $\idx{d}{A}=d_B$ and the $\idx{\ket{i}}{A}$, $\idx{\ket{i}}{B}$ form an orthonormal basis for $\HA$ and for $\HB$, which is isomorphic to $\HA$.\\
Linear maps from $\linops{\idx{\cH}{A}}$ to $\linops{\idx{\cH}{B}}$ will be denoted by the calligraphic letters $\idx{\cT}{A$\rightarrow$B}$, $\idx{\cE}{A$\rightarrow$B}$, $\idx{\cN}{A$\rightarrow$B}$,... 
Quantum operations are in one to one correspondence to the trace preserving and completely positive maps (TPCPM) $\idx{\cT}{A$\rightarrow$B}$ which map density operators to density operators. We sometimes will call a TPCP map also a \textit{quantum channel}, if we want to emphasize the use of the TPCPM under consideration for information processing. The TPCPM we will encounter most often is the partial trace (over the system $B$), which is given by a map $\idx{\cE}{AB$\rightarrow$A}$, defined to be the adjoint mapping $\mathcal{T}^{\dagger}$ of $\mathcal{T}(\idx{\xi}{A}) = \idx{\xi}{A} \otimes\idi{B}\  ;\  \xi\in\hermops{\cH}$ with respect to the Schmidt scalar product 
$\langle A,B\rangle \ := \ \tr(A^{\dagger}B)$. That means $\tr(\mathcal{T}(\xi)\zeta) = \tr( \xi \mathcal{T}^{\dagger}(\zeta))$. It is used to determine the expectation values for measurements on a subsystem $A$ if the state of some bipartite system $AB$ is given. Because of its crucial importance the partial trace has a special notation: 

$$\idx{\cE}{AB$\rightarrow$A}(\idx{\zeta}{AB})\ =:\ \Ptrace{B}{\idx{\zeta}{AB}},$$
where $\idx{\zeta}{AB}$ is any operator in $\hermops{\idx{\cH}{AB}}$. If we have a bipartite state $\idx{\xi}{AB}$ and we would like to consider a subsystem only, we will denote by $\idx{\xi}{A}$ the partial trace of $\idx{\xi}{AB}$ over the system $B$, $\idx{\xi}{A}=\ptr{B}{\idx{\xi}{AB}}$.\\
We will call $\idx{\omega}{A'E}:=(\cTAE\otimes\idx{\opid}{A'})(\idx{\Phi}{AA'})$ for any $\cTAE\in \textnormal{Hom}(\linops{\HA},\linops{\HE})$ the Choi-Jamiolkowski representation \cite{CHOI,JAM} of $\cTAE$, where $\idx{\cH}{A'}$ is a copy of $\HA$ and $\idx{\opid}{A'}\in\textnormal{Hom}(\linops{\idx{\cH}{A'}},\linops{\idx{\cH}{A'}})$ denotes the operator identity. (Note that writing $\idx{\omega}{A'E}$ we slightly abuse notation, since, strictly speaking, $\idx{\omega}{A'E}$ is by definition an operator in $\linops{\idx{\cH}{E}\otimes\idx{\cH}{A'}}$ and not in $\linops{\idx{\cH}{A'}\otimes\idx{\cH}{E}}$ as the notation indicates. This will be the only exception.) To keep the notation as simple as possible we have the convention that if a map $\cTAE$ acts on a bipartite state with subsystem $A$, we mean that implicitly the identity is applied on the other subsystem. Thus we for example write $\cTAE(\idx{\Phi}{AA'}):=(\cTAE\otimes\idx{\opid}{A'})(\idx{\Phi}{AA'})$ for the Choi-Jamiolkowski representation. \\
Any TPCPM can be viewed as a unitary operation on some larger system. More precisely speaking we have the following lemma \cite{Renner:Skript}:\\
\\
\textbf{Lemma 1:} (\textnormal{Stinespring dilation})
\textit{Let $\cT$ be a TPCPM from $\linops{\HA}$ to $\linops{\HB}$. Then there exists some isometry $U\in\textnormal{Hom}(\HA,\HB\otimes\idx{\cH}{R})$ for some Hilbert space $\idx{\cH}{R}$, such that
$$\cT\ :\ \rhoA \mapsto \Ptrace{R}{U\rhoA U^\dagger}.$$}
\\
Thus, $\cT$ can be viewed as a concatenation of two TPCP maps: First conjugating $\rhoA$ with $U$ and afterwards taking the partial trace over the system $R$. We will write shortly $\cT = \ptr{R}\circ U\cdot$ for this operation.\\
To quantify whether a quantum state is preserved by some quantum operation (or more generally by some mapping $\cT$) or not, we have to introduce distance measures on the set of density operators. 
For any operator in $\xi\in\linops{\cH}$ we denote by $\left\|\xi\right\|_1$ the Schatten 1-norm, by $\left\|\xi\right\|_2$ the Schatten 2-norm, by $\left\|\xi\right\|_F$ the Frobenius-norm and by $\left\|\xi\right\|_\infty$ the $\infty$-norm of $\xi$ which are defined to be
\begin{align}&\left\|\xi\right\|_1 := \tr\abs{\xi} := \trace{\sqrt{\xi^{\dagger}\xi}}\\
&\left\|\xi\right\|_2 := \sqrt{\trace{\xi^{\dagger}\xi}}\\
&\left\|\xi\right\|_F := \sqrt{\sum_{i,j}{\abs{\xi_{ij}}^2}}\\
&\left\|\xi\right\|_\infty := \sqrt{\lambda_{\max}(\xi^{\dagger}\xi)}\label{expleig}
\end{align}
respectively.
Particularly that means that if $\xi$ is in $\hermops{\cH},\ \left\|\xi\right\|_1$ is just equal to the sum of the absolute values of the eigenvalues of $\xi$. For our purposes it is sufficient to think of the Schatten 2-norm as the induced norm by the Schmidt scalar product. In contrast to the other norms above, the Frobenius-norm is an ``entrywise'' norm: $\xi_{ij}$ are the entries of the matrix corresponding to the operator $\xi$ in some basis. (We work in finite dimensions only.) It is a priori not clear that the norm defined in the above way is well defined but a short calculation reveals that 
$\Norm{\xi}{F}=\Norm{\xi}{2}$ which also proofs that it actually can be seen as an operator norm (\cite{Bhatia}). Finally the $\infty$-norm is given by the square root of the biggest eigenvalue $\lambda_{\max}$ of $\xi^{\dagger}\xi$  for any $\xi\in\linops{\cH}$. We will frequently have to link the above norms in terms of inequalities. This is achieved using the following standard result (\cite{Watrous}, \textit{Lecture 3}):\\
\\
\textbf{Lemma 2:} (\textnormal{Norm Inequalities})
\textit{For any $A,B,C\in\linops{\cH}$, the following inequalities hold
\begin{align}
&\Norm{ABC}{\infty}\leq\Norm{A}{\infty}\Norm{B}{\infty}\Norm{C}{\infty},\nonumber\\
&\Norm{ABC}{1}\leq\Norm{A}{\infty}\Norm{B}{1}\Norm{C}{\infty},\nonumber\\
&\Norm{ABC}{2}\leq\Norm{A}{\infty}\Norm{B}{2}\Norm{C}{\infty}.\nonumber
\end{align}}
\\

The Schatten 1-norm induces a metric on $\subnormstates{\cH}$, which we call the $trace\ distance$ and is defined (in this thesis) to be:
$$D(\rho,\sigma):= \left\|\rho-\sigma\right\|_1$$
In the special case that $\rho$ and $\sigma$ are elements of $\normstates{\cH}$, this gives a good distance measure in the sense that any measurement performed on states that are close in trace distance gives rise to probability distributions $p$, $q$ that are close in the sense that the the maximum difference of the probabilities that some event $S$ occurs with respect to the different probability distributions $\max_S(\sum_{x\in S}p_x - \sum_{x\in S}q_x)$ is small, too \cite{NielsenChuang}. That means that those density operators cannot be well distinguished by any measurement. For positive operators with trace not equal to one this distance is not convenient. One therefore introduces a \textit{generalized trace distance}, $\bar{D}(\rho,\sigma)$, which gives a good distance measure for sub-normalized states \cite{TCR09}:
$$\bar{D}(\rho,\sigma):= \left\|\rho-\sigma\right\|_1+\abs{\tr\rho-\tr\sigma},$$
for any $\rho$, $\sigma$ $\in\posops{\cH}$.
In the case of normalized states it reduces to the usual trace distance defined above.\\
Another commonly used measure of distance is the \textit{fidelity}:
$$F(\rho,\sigma) := \left\|\sqrt{\rho}\sqrt{\sigma}\right\|_1,$$
for any density operators $\rho$, $\sigma$. The fidelity is not a metric on the states of some quantum system, but there are several types of metrics derived from it. In this thesis we will deal with the following one:
$$P(\rho,\sigma):= \sqrt{1-F(\rho,\sigma)^2}$$
Again this type of distance measure is not convenient for quantum information theory if the density operators are not normalized. In \cite{TCR09}
the \textit{purified distance} $\bar{P}(\rho,\sigma)$ is introduced, which generalizes $P(\rho,\sigma)$:
$$\bar{P}(\rho,\sigma):= \sqrt{1-\bar{F}(\rho,\sigma)^2},$$
where $\bar{F}(\rho,\sigma)$ is the \textit{generalized fidelity}:
$$\bar{F}(\rho,\sigma):= F(\rho,\sigma)+\sqrt{(1-\tr\rho)(1-\tr\sigma)},$$
for $\rho$, $\sigma$ in $\subnormstates{\cH}$. Note that, if $\rho$ or $\sigma$ is in $\normstates{\cH}$, the generalized fidelity reduces to the usual one.\\
Both distance measures $D(\rho,\sigma)$ and $P(\rho,\sigma)$ introduced above are essentially equivalent, due to the following fundamental \textit{Fuchs- van der Graaf inequalities} \cite{NielsenChuang}.\\
\\
\textbf{Lemma 3:} (\textnormal{Fuchs- van der Graaf inequalities})
\textit{Let $\rho$, $\sigma$ be in $\normstates{\cH}$, then
$$
\frac{1}{2}{\left\|\rho-\sigma\right\|}_1\leq P(\rho,\sigma)\leq\sqrt{{\left\|\rho-\sigma\right\|}_1}.
$$
}Using the generalized versions of trace distance and the fidelity the authors derive in \cite{TCR09} the analogous relations for the generalized quantities:\\
\\
\textbf{Lemma 4:} (\textnormal{Generalized Fuchs- van der Graaf inequalities})
\textit{Let $\rho$, $\sigma$ be in $\subnormstates{\cH}$, then
$$
\frac{1}{2}\bar{D}(\rho,\sigma)\leq \bar{P}(\rho,\sigma)\leq\sqrt{\bar{D}(\rho,\sigma)}.
$$
}
\\
To quantify the uncertainty of our knowledge about some quantum state we use entropy measures. Various such measures are treated in the literature, for us the most important will be the quantum two-entropy and the min-entropy:
\begin{definition}
Let $\idx{\rho}{A}$ be in $\posops{\idx{\cH}{A}}$. Then its min-entropy is defined to be
$$\hmin{A}{\rho}\::=\:-\log\min\{\lambda\in\mathbb{R}\mid\idx{\rho}{A}\leq\lambda\idi{A}\}.$$
\end{definition}
This is just the negative logarithm of the largest eigenvalue of $\idx{\rho}{A}$. We also require a conditional version of the min-entropy. Given a bipartite quantum system, conditional entropies in general aim to quantify the uncertainty, which we have about one of the subsystems if the state of the other subsystem is known.
\begin{definition}
Let $\rhoAB\in\subnormstates{\hAB}$, then the min-entropy of $A$ conditioned on $B$ of $\rhoAB$ is defined as \cite{Renner:PHD,TCR09}
$$\chmin{A}{B}{\rho}\::=\: \max_{\idx{\sigma}{B}\in\normstates{\HB}}\sup\{\lambda\in\mathbb{R}\mid\rhoAB\leq2^{-\lambda}\idA\otimes\sigmaB\}.$$
\end{definition}
Finally, we define the quantum conditional 2-entropy, which will occur in the statement of the decoupling theorem, but is only an auxiliary quantity from the point of view of information theory:
\begin{definition}
Let $\rhoAB\in\subnormstates{\hAB}$, then the 2-entropy of $A$ conditioned on $B$ of $\rhoAB$ is defined as
$$\chtwo{A}{B}{\rho}\::=\: -\log\min_{\sigmaB\in\normstates{\HB}}\frac{1}{\Trace{\idx{\rho}{AB}}}\tr\Big(((\idi{A}\otimes\idx{\sigma}{B})^{-1/2}\rhoAB)^2\Big)$$
\end{definition}
The minimum is attained for a $\sigmaB$ with $\supp{\sigmaB}\supset\supp{\rhoB}$, where $\supp{\cdot}$ denotes the support of some operator.
Between the conditional min-entropy and the conditional two-entropy we have the following important relation, which links the auxiliary quantity $H_2$ to the physical quantity $H_{\textnormal{min}}$:\\
\textbf{Lemma 5:}
\textit{Let $\rhoAB\in\subnormstates{\hAB}$, then
$$\chmin{A}{B}{\rho}\:\leq\:\chtwo{A}{B}{\rho}$$}

This is just remark (5.3.2) in \cite{Renner:PHD}. The proof of the statement can be found, there.

\section{Induced neighborhoods and the smoothed conditional min-entropy}

One can define a smoothed version of $H_\textnormal{min}$ in the following way: Instead of evaluating $H_\textnormal{min}$ at $\rho$ directly one maximizes the min-entropy over a set of states that are $\varepsilon$-close to $\rho$. Obviously the crucial question is which distance measure should be used to determine $\varepsilon$-closeness?
In \cite{TCR09} the authors show that the following type of $\varepsilon$-neighborhoods is particularly useful:
\begin{definition}[\cite{TCR09}]
Let $\varepsilon\geq0$ and $\rho\in\subnormstates{\cH}$ with $\sqrt{\tr\rho}>\varepsilon$, then 
$$\epsball{\cH;\rho}\::=\:\{\sigma\in\subnormstates{\cH}\mid \bar{P}(\sigma,\rho)\leq\varepsilon\}.$$
\end{definition}
This $\varepsilon$-ball can be used to define a smoothed version of the min-entropy:
\begin{definition}[\cite{Renner:PHD,TCR09}]
Let $\varepsilon\geq0$ and $\rhoAB\in\subnormstates{\idx{\cH}{AB}}$, then the \textit{$\varepsilon$-smooth\ min-entropy} of $A$ conditioned on $B$ of $\rhoAB$ is defined as
$$\chmineps{A}{B}{\rho}\:\:=\:\max_{\tilde{\rho}\in\epsball{H;\rho}}\chmin{A}{B}{\tilde{\rho}}.$$
\end{definition}
Many important properties of this entropy, including the proof of the fact that it actually is a continuous function of $\rho$, can be found in \cite{TCR09}.
\chapter{Decoupling via the Schatten 2-norm}

One of the main results of this thesis is a generalization of the decoupling theorem of \cite{Fred:PHD} and \cite{Semthesis} to the case when unitary almost 2-designs are considered only. The proof of this formula will be a generalized proof of the decoupling theorem, with major parts resembling the original proof in \cite{Fred:PHD}. In this chapter we present a shorter proof of the original decoupling theorem with integration over the unitary group $\mathbb{U}(A)$. The methodology developed here will be relevant when proving the theorem's generalization in the next chapter. First we prove a lemma which provides an easy way of performing the required integration. This lemma will rely on a consideration of the Schatten 2-norm instead of the Schatten 1-norm and on working with the state $\idx{\xi}{A$\tilde{\textnormal{A}}$}:=\idx{\Phi}{A$\tilde{\textnormal{A}}$}-\idx{\pi}{A}\otimes\idx{\pi}{$\tilde{\textnormal{A}}$}$.
As a result we will obtain a slightly better bound than is given by the original decoupling theorem.

\section{Lemma: Decoupling with the Schatten 2-norm}

Many of the difficulties one has to overcome during the derivation of the decoupling theorem as in \cite{Fred:PHD} arise from the fact that untill now we cannot integrate the square-root function. The idea therefore is to consider an integrand which does not contain any roots, instead. For this reason we will work with the Schatten 2-norm. In return the derived statement will be an equality, such that the converse of the theorem is valid automatically.\\
\\
\textbf{Lemma:} (Decoupling Lemma)\textit{
Let $\rhoAR\in\hermops{\idx{\cH}{A}\otimes\idx{\cH}{R}}$  and let $\cTAE\in\textnormal{Hom}(\linops{\HA},\linops{\HE})$ be a linear map with Choi-Jamiolkowski representation $\idx{\omega}{A'E}\in\hermops{\idx{\cH}{E}\otimes\idx{\cH}{A'}}$, then
\begin{align}\int_{\mathbb{U}(A)} &{\left\| \mathcal{T}(\idx{U}{A} \otimes \idi{R} \ \rhoAR \  \idx{U}{A}^\dagger \otimes \idi{R}) - \idx{\omega}{E} \otimes \idx{\rho}{R} \right\|}_2^2 \d{U}\nonumber\\
&=\frac{\idx{d}{A}^2}{\idx{d}{A}^2-1}\:{\left\|\idx{\rho}{AR}-\idx{\pi}{A}\otimes\rhoR\right\|}_2^2\: {\left\|\idx{\omega}{A'E}-\idx{\pi}{A'}\otimes\idx{\omega}{E}\right\|}_2^2\nonumber
\end{align}
where the integration goes over all unitaries and with respect to the probability Haar measure $\d U$.
}\\
\\
For the proof it will be convenient to reformulate the argument of the integral in a more symmetric way. We introduce the map $\idx{\cE}{$\tilde{\textnormal{A}}\rightarrow$R}$, which we define to be the unique Choi-Jamiolkowski preimage of the state $\rhoAR$ i.e. $\idx{\cE}{$\tilde{\textnormal{A}}\rightarrow$R}(\idx{\Phi}{A$\tilde{\textnormal{A}}$})=\rhoAR$, where $\tilde{A}$ is just a copy of $A$. Note that $\cE$ is not trace-preserving in general. Because $\cE$ acts only on the $\tilde{A}$ subsystem and the identity is applied to the $A$ part, taking the partial trace over $A$ commutes with applying the map $\cE$ to $\idx{\Phi}{A$\tilde{\textnormal{A}}$}$. We thus have
\begin{align}
\rhoR&=\ptrace{A}{\cE(\idx{\Phi}{A$\tilde{\textnormal{A}}$})}\\
&=\cE(\idx{\pi}{$\tilde{\textnormal{A}}$}).\label{trvrnew}
\end{align}
An analogous relation is also valid for $\cTAE$ and we can write for any unitary $\idx{U}{A}$:
\begin{align}
&\mathcal{T}((\idx{U}{A} \otimes \idi{R}) \ \rhoAR \  (\idx{U}{A}^\dagger \otimes \idi{R})) - \idx{\omega}{E} \otimes \idx{\rho}{R} \label{advxi}\nonumber\\
&= \mathcal{T}((\idx{U}{A} \otimes \idi{R}) \cE(\idx{\Phi}{A$\tilde{\textnormal{A}}$})(\idx{U}{A}^\dagger \otimes \idi{R})) - \cT(\idx{\pi}{A})\otimes \cE(\idx{\pi}{$\tilde{\textnormal{A}}$})\\\label{comm1new}
&=(\mathcal{T}\otimes\cE)((\idx{U}{A} \otimes \idi{$\tilde{A}$}) \idx{\Phi}{A$\tilde{\textnormal{A}}$}(\idx{U}{A}^\dagger \otimes \idi{$\tilde{A}$})) - (\cT\otimes\cE)(\idx{\pi}{A}\otimes\idx{\pi}{$\tilde{\textnormal{A}}$})\\
&=(\cT\otimes\cE)((\idx{U}{A} \otimes \idi{$\tilde{A}$}) (\idx{\Phi}{A$\tilde{\textnormal{A}}$} - \idx{\pi}{A}\otimes\idx{\pi}{$\tilde{\textnormal{A}}$})(\idx{U}{A}^\dagger \otimes \idi{$\tilde{A}$}))\\
&=(\cT\otimes\cE)((\idx{U}{A} \otimes \idi{$\tilde{A}$}) (\idx{\xi}{A$\tilde{\textnormal{A}}$})(\idx{U}{A}^\dagger \otimes \idi{$\tilde{A}$}))\label{intxinew}
\end{align}
In equation \eqref{comm1new}, we used the fact that the unitary applied only acts on the $A$ subsystem in contrast to $\cE$ which acts on $R$ only and therefore the operations of conjugation with $\idx{U}{A}$ and applying $\cE$ commute. In the last equation \eqref{intxinew} we introduce the \textit{decoupling state} $\idx{\xi}{A$\tilde{\textnormal{A}}$}:=\idx{\Phi}{A$\tilde{\textnormal{A}}$}-\idx{\pi}{A}\otimes\idx{\pi}{$\tilde{\textnormal{A}}$}$ for notational convenience. We will later see an interesting property of this state.\\
This way of writing the integrand of the usual decoupling theorem yields a shorter proof of the theorem, since the mixed terms in \cite{Fred:PHD} do not have to be considered anymore. Moreover the proof gets ``symmetric'' in the treatment of $\rhoAR$ and $\idx{\omega}{A'E}$ which will be of crucial relevance, at the moment when we will have to apply the definition of the almost 2-design and find upper bounds in the case of the generalized theorem.
Note that by \eqref{trvrnew} we have in addition that
\begin{align}
\cE(\idx{\xi}{A$\tilde{\textnormal{A}}$})=\rhoAR-\idx{\pi}{A}\otimes\rhoR\ \wedge\ \cT(\idx{\xi}{A$\tilde{\textnormal{A}}$})=\idx{\omega}{$\tilde{\textnormal{A}}$E}-\idx{\pi}{$\tilde{\textnormal{A}}$}\otimes\idx{\omega}{E}.
\end{align}
Thus the stated lemma can be rewritten equivalently in terms of the decoupling state. We note that \begin{align}\frac{\idx{d}{A}^2}{\idx{d}{A}^2-1}=\frac{1}{{\left\|\idx{\xi}{A$\tilde{\textnormal{A}}$}\right\|}_2^2}\end{align} and obtain:\\
\\
\textbf{Lemma:} (Decoupling Lemma)\textit{
Let $\idx{\xi}{A$\tilde{\textnormal{A}}$}=\idx{\Phi}{A$\tilde{\textnormal{A}}$}-\idx{\pi}{$\tilde{\textnormal{A}}$}\otimes\idx{\pi}{$\tilde{\textnormal{A}}$}$  and let $\cTAE\in\textnormal{Hom}(\linops{\HA},\linops{\HE})$ and $\idx{\cE}{$\tilde{\textnormal{A}}\rightarrow$R}\in\textnormal{Hom}(\linops{\idx{\cH}{$\tilde{\textnormal{A}}$}},\linops{\idx{\cH}{R}})$ be linear maps then
\begin{align}\int_{\mathbb{U}(A)} {\left\|(\cT\otimes\cE)(\idx{U}{A} \otimes \idi{$\tilde{A}$}\  \idx{\xi}{A$\tilde{\textnormal{A}}$}\   \idx{U}{A}^\dagger \otimes \idi{$\tilde{A}$}) \right\|}_2^2 \d{U}
=\frac{{\left\|\cE(\idx{\xi}{A$\tilde{\textnormal{A}}$})\right\|}_2^2\: {\left\|\cT(\idx{\xi}{A$\tilde{\textnormal{A}}$})\right\|}_2^2}{{\left\|\idx{\xi}{A$\tilde{\textnormal{A}}$}\right\|}_2^2}\nonumber
\end{align}
where the integration goes over all unitaries and with respect to the probability Haar measure $\d U$.
}\\
\\
This formulation is especially convenient for the proof. We have that
\begin{align}
&\int_{\mathbb{U}(A)}{{\left\|(\cT\otimes\cE)(\idx{U}{A} \otimes \idi{$\tilde{A}$}\  \idx{\xi}{A$\tilde{\textnormal{A}}$}\   \idx{U}{A}^\dagger \otimes \idi{$\tilde{A}$}) \right\|}_2^2}\d{U}\nonumber\\
&=\int_{\mathbb{U}(A)}{\Trace{(\cT\otimes\cE)(\idx{U}{A} \otimes \idi{$\tilde{A}$}\  \idx{\xi}{A$\tilde{\textnormal{A}}$}\   \idx{U}{A}^\dagger \otimes \idi{$\tilde{A}$})^2}}\d{U}\\
&=\int_{\mathbb{U}(A)}{\Trace{(\cT\otimes\cE)^{\otimes2}\left((\idx{U}{A} \otimes \idi{$\tilde{A}$})^{\otimes2}\  (\idx{\xi}{A$\tilde{\textnormal{A}}$})^{\otimes2}\   (\idx{U}{A}^\dagger \otimes \idi{$\tilde{A}$})^{\otimes2}\right)\:\idx{\cF}{ER}}}\d{U}\label{thswp}\\
&=\int_{\mathbb{U}(A)}{\Trace{\left((\idx{U}{A} \otimes \idi{$\tilde{A}$})^{\otimes2}\  (\idx{\xi}{A$\tilde{\textnormal{A}}$})^{\otimes2}\   (\idx{U}{A}^\dagger \otimes \idi{$\tilde{A}$})^{\otimes2}\right)\:(\cT^\dagger)^{\otimes2}[\idx{\cF}{E}]\otimes(\cE^\dagger)^{\otimes2}[\idx{\cF}{R}]}}\d{U}.\label{dfndgr}
\end{align}
We introduced two further copies $A'$ and $\tilde{A}'$ of $A$ when using the swap trick (see Appendix C) in equation \eqref{thswp}, i.e. $(\idx{\xi}{A$\tilde{\textnormal{A}}$})^{\otimes2}\:=\:\idx{\xi}{A$\tilde{\textnormal{A}}$}\otimes\idx{\xi}{A'$\tilde{\textnormal{A'}}$}$.
In equation \eqref{dfndgr} we used the definition of the adjoint of the mapping $(\tilde{\mathcal{T}}\otimes\tilde{\cE})^{\otimes2}$ with respect to the Schmidt scalar product. Note that this map has product structure and since we apply it on a product state the two different parts $(\tilde{\mathcal{T}})^{\otimes2}$ and $(\tilde{\cE})^{\otimes2}$ can be applied separately.\\
At this point it is not difficult to perform the integration on $(\idx{\xi}{A$\tilde{\textnormal{A}}$})^{\otimes2}$ directly but it is known from \cite{Fred:PHD,Merging} that
\begin{align}
\int{[(\idx{U}{A})^{\dagger\otimes2}(\tilde{\mathcal{T}}^{\dagger})^{\otimes2}(\idx{\mathcal{F}}{E})(\idx{U}{A})^{\otimes2}]}\d{U}
= \alpha \idi{AA'} + \beta\idx{\cF}{A}\label{intgrlpr},
\end{align}
with the coefficients $\alpha$ and $\beta$ satisfying
\begin{align}
\alpha &=\trace{\idx{\omega}{E}^2}\left(\frac{\idx{d}{A}^2-\frac{\idx{d}{A}\:\Trace{\idx{\omega}{A'E}^2}}{\Trace{\idx{\omega}{E}^2}}}{\idx{d}{A}^2-1}\right)\\
\beta &=\trace{\idx{\omega}{A'E}^2}\left(\frac{\idx{d}{A}^2-\frac{\idx{d}{A}\:\Trace{\idx{\omega}{E}^2}}{\Trace{\idx{\omega}{A'E}^2}}}{\idx{d}{A}^2-1}\right).
\end{align}
Thus it is even less work to perform the integration over $(\tilde{\mathcal{T}}^{\dagger})^{\otimes2}(\idx{\cF}{E})$. We get that
\begin{align}
&\int_{\mathbb{U}(A)}{{\left\|(\cT\otimes\cE)(\idx{U}{A} \otimes \idi{$\tilde{A}$}\  \idx{\xi}{A$\tilde{\textnormal{A}}$}\   \idx{U}{A}^\dagger \otimes \idi{$\tilde{A}$}) \right\|}_2^2}\d{U}\nonumber\\
&=\int_{\mathbb{U}(A)}{\Trace{(\idx{\xi}{A$\tilde{\textnormal{A}}$})^{\otimes2}\  
(\idx{U}{A}^\dagger )^{\otimes2}(\cT^\dagger)^{\otimes2}[\idx{\cF}{E}](\idx{U}{A})^{\otimes2}\otimes(\cE^\dagger)^{\otimes2}[\idx{\cF}{R}]}}\d{U}\\
&=\Trace{(\idx{\xi}{A$\tilde{\textnormal{A}}$})^{\otimes2}\{ \alpha \idi{AA'} + \beta\idx{\cF}{A}\}\otimes(\cE^\dagger)^{\otimes2}[\idx{\cF}{R}]}\\
&=\alpha\:\Trace{(\idx{\xi}{A$\tilde{\textnormal{A}}$})^{\otimes2}\ \idi{AA'}\otimes(\cE^\dagger)^{\otimes2}[\idx{\cF}{R}]}+\beta\:\Trace{(\idx{\xi}{A$\tilde{\textnormal{A}}$})^{\otimes2}\ \idx{\cF}{A}\otimes(\cE^\dagger)^{\otimes2}[\idx{\cF}{R}]}\\
&=\beta\:\Trace{(\idx{\xi}{A$\tilde{\textnormal{A}}$})^{\otimes2}\ \idx{\cF}{A}\otimes(\cE^\dagger)^{\otimes2}[\idx{\cF}{R}]}.
\end{align}
In the last step we used that tracing out one of the subsystems $A$, $\tilde{A}$ of $(\idx{\xi}{A$\tilde{\textnormal{A}}$})$ gives the zero state. Using the definition of the adjoint of $\cE$ we find
\begin{align}
\beta\:\Trace{(\idx{\xi}{A$\tilde{\textnormal{A}}$})^{\otimes2}\ \idx{\cF}{A}\otimes(\cE^\dagger)^{\otimes2}[\idx{\cF}{R}]}
&=\beta\:\Trace{\cE(\idx{\xi}{A$\tilde{\textnormal{A}}$})^{\otimes2}\ \idx{\cF}{AR}}\\
&=\beta\:\Trace{\cE(\idx{\xi}{A$\tilde{\textnormal{A}}$})^2}\\
&=\beta\:\Norm{\cE(\idx{\xi}{A$\tilde{\textnormal{A}}$})}{2}^2.\label{withbet}
\end{align}
Rewriting $\beta$ we find that
\begin{align}
\beta &=\trace{\idx{\omega}{A'E}^2}\left(\frac{\idx{d}{A}^2-\frac{\idx{d}{A}\:\Trace{\idx{\omega}{E}^2}}{\Trace{\idx{\omega}{A'E}^2}}}{\idx{d}{A}^2-1}\right)\label{shbet1}\\
&=\frac{\idx{d}{A}^2}{\idx{d}{A}^2-1}\:\left(\Trace{\idx{\omega}{A'E}^2}-\frac{1}{\idx{d}{A}}\:\Trace{\idx{\omega}{E}^2}\right)\\
&=\frac{\idx{d}{A}^2}{\idx{d}{A}^2-1}\:\left(\Trace{\idx{\omega}{A'E}^2}-2\frac{1}{\idx{d}{A}}\:\Trace{\idx{\omega}{E}^2}+\frac{1}{\idx{d}{A}}\:\Trace{\idx{\omega}{E}^2}\right)\\
&=\frac{\idx{d}{A}^2}{\idx{d}{A}^2-1}\:\left(\Trace{\idx{\omega}{A'E}^2}-2\frac{1}{\idx{d}{A}}\:\Trace{\idi{A'}\otimes\idx{\omega}{E}\:\idx{\omega}{A'E}}+\Trace{\idx{\pi}{A}^2\otimes\idx{\omega}{E}^2}\right)\\
&=\frac{\idx{d}{A}^2}{\idx{d}{A}^2-1}\:\Trace{(\idx{\omega}{A'E}-\idx{\pi}{A'}\otimes\idx{\omega}{E})^2}\\
&=\frac{\idx{d}{A}^2}{\idx{d}{A}^2-1}\:\Trace{\cT(\idx{\xi}{A$\tilde{\textnormal{A}}$})^2}\\
&=\frac{\idx{d}{A}^2}{\idx{d}{A}^2-1}\:\Norm{\cT(\idx{\xi}{A$\tilde{\textnormal{A}}$})}{2}^2.\label{shbet2}
\end{align}
We conclude plugging this into equation \eqref{withbet} that
\begin{align}
&\int_{\mathbb{U}(A)}{{\left\|(\cT\otimes\cE)(\idx{U}{A} \otimes \idi{$\tilde{A}$}\  \idx{\xi}{A$\tilde{\textnormal{A}}$}\   \idx{U}{A}^\dagger \otimes \idi{$\tilde{A}$}) \right\|}_2^2}\d{U}\nonumber\\
&=\frac{\idx{d}{A}^2}{\idx{d}{A}^2-1}\:\Norm{\cT(\idx{\xi}{A$\tilde{\textnormal{A}}$})}{2}^2\:\Norm{\cE(\idx{\xi}{A$\tilde{\textnormal{A}}$})}{2}^2,
\end{align}
which proofs the decoupling lemma.

\section{An alternative proof of the decoupling theorem}
As an application of the last section's lemma, we shortly rederive the decoupling theorem of \cite{Fred:PHD} in the formulation which is given in \cite{Semthesis}.\\
\\
\textbf{Theorem:} (Decoupling theorem)\textit{
Let $\rhoAR\in\subnormstates{\idx{\cH}{A}\otimes\idx{\cH}{R}}$ be a sub normalized density operator and let $\cTAE$ be a completely positive linear map going from $\subnormstates{\idx{\cH}{A}\otimes\idx{\cH}{R}}$ to $\posops{\idx{\cH}{E} \otimes \idx{\cH}{R}}$ with Choi-Jamiolkowski representation $\idx{\omega}{A'E}\in\subnormstates{\idx{\cH}{E}\otimes\idx{\cH}{A'}}$, then
$$\int_{\mathbb{U}(A)} {\left\| \mathcal{T}((\idx{U}{A} \otimes \idi{R}) \ \rhoAR \  (\idx{U}{A}^\dagger \otimes \idi{R})) - \idx{\omega}{E} \otimes \idx{\rho}{R} \right\|}_1 \d{U} \leq
2^{-\frac{1}{2} H_2(A'|E)_{\omega} - \frac{1}{2} H_2(A|R)_{\rho} }$$
where the integration goes over all unitaries and with respect to the probability Haar measure $\d U$.
}\\
\\
The proof goes as follows. As we did in \eqref{intxinew}, we work with the integrand in terms of the decoupling state. We then use the H\"older inequality as stated in Appendix A with parameters $r=t=4$ and $s=2$ (or alternatively \textit{Lemma 4} in \cite{Semthesis}) to bound the Schatten 1-norm of the integrand in terms of the Schatten 2-norm:
\begin{align}
\Norm{ABC}{1}\leq\Norm{\abs{A}^4}{1}^{\frac{1}{4}}\Norm{\abs{B}^2}{1}^{\frac{1}{2}}\Norm{\abs{C}^4}{1}^{\frac{1}{4}}\label{hldrfk}
\end{align}
Note that the term $\Norm{\abs{B}^2}{1}^{\frac{1}{2}}$ is just the Schatten 2-norm of $B$.
Introducing the positive definite and normalized operators $\idx{\sigma}{E}$ and $\idx{\zeta}{R}$ with
\begin{align}
&A:=(\sigmaE\otimes\zetaR)^{\frac{1}{4}},\\
&B:=(\sigmaE\otimes\zetaR)^{-\frac{1}{4}}\left((\cT\otimes\cE)(\idx{U}{A} \otimes \idi{$\tilde{A}$}\: \idx{\xi}{A$\tilde{\textnormal{A}}$}\:\idx{U}{A}^\dagger \otimes \idi{$\tilde{A}$})\right)(\sigmaE\otimes\zetaR)^{-\frac{1}{4}},\\
&C:=(\sigmaE\otimes\zetaR)^{\frac{1}{4}}.
\end{align}
the above \eqref{hldrfk} specializes to
\begin{align}
&\Norm{(\cT\otimes\cE)(\idx{U}{A} \otimes \idi{$\tilde{A}$}\: \idx{\xi}{A$\tilde{\textnormal{A}}$}\:\idx{U}{A}^\dagger \otimes \idi{$\tilde{A}$})}{1}\nonumber\\
&\leq\Norm{(\sigmaE\otimes\zetaR)^{-\frac{1}{4}}\left((\cT\otimes\cE)(\idx{U}{A} \otimes \idi{$\tilde{A}$}\: \idx{\xi}{A$\tilde{\textnormal{A}}$}\:\idx{U}{A}^\dagger \otimes \idi{$\tilde{A}$})\right)(\sigmaE\otimes\zetaR)^{-\frac{1}{4}}}{2}.
\end{align}
One can abbreviate the notation introducing the completely positive map $\idx{\tilde{\mathcal{T}}}{$A \rightarrow E$}$ with $\tilde{\mathcal{T}}(\idx{\tau}{A$\tilde{\textnormal{A}}$})\ :=\ (\sigmaE \otimes \idi{$\tilde{\textnormal{A}}$})^{-1/4} \mathcal{T}(\idx{\tau}{A$\tilde{\textnormal{A}}$}) (\sigmaE \otimes \idi{$\tilde{\textnormal{A}}$})^{-1/4}$ for any $\idx{\tau}{A$\tilde{\textnormal{A}}$}\in\linops{\idx{\cH}{A$\tilde{\textnormal{A}}$}}$ and similarly the map $\idx{\tilde{\cE}}{$\tilde{\textnormal{A}} \rightarrow \textnormal{R}$}$ is defined to be $\tilde{\mathcal{E}}(\idx{\tau}{A$\tilde{\textnormal{A}}$})\ :=\ (\idi{A} \otimes \zetaR)^{-1/4} \cE(\idx{\tau}{A$\tilde{\textnormal{A}}$}) (\idi{A} \otimes \zetaR)^{-1/4}$ for any $\idx{\tau}{A$\tilde{\textnormal{A}}$}\in\linops{\idx{\cH}{A$\tilde{\textnormal{A}}$}}$. With the short notation we have
\begin{align}
&\int_{\mathbb{U}(A)}{\left\| (\cT\otimes\cE)((\idx{U}{A} \otimes \idi{$\tilde{A}$}) (\idx{\xi}{A$\tilde{\textnormal{A}}$})(\idx{U}{A}^\dagger \otimes \idi{$\tilde{A}$}))\right\|}_1^2\d{U}\nonumber\\
&\leq\int_{\mathbb{U}(A)}{\left\| (\tilde{\cT}\otimes\tilde{\cE})((\idx{U}{A} \otimes \idi{$\tilde{A}$}) (\idx{\xi}{A$\tilde{\textnormal{A}}$})(\idx{U}{A}^\dagger \otimes \idi{$\tilde{A}$}))\right\|}_2^2\d{U}\\
&=\frac{\idx{d}{A}^2}{\idx{d}{A}^2-1}\:\Norm{\tilde{\cT}(\idx{\xi}{A$\tilde{\textnormal{A}}$})}{2}^2\:\Norm{\tilde{\cE}(\idx{\xi}{A$\tilde{\textnormal{A}}$})}{2}^2,\label{nwwtmn}
\end{align}
where we applied the \textit{Decoupling Lemma} in the last step.
Going the steps from \eqref{shbet1} to \eqref{shbet2} backwards gives:
\begin{align}
&\frac{\idx{d}{A}^2}{\idx{d}{A}^2-1}\:\Norm{\tilde{\cT}(\idx{\xi}{A$\tilde{\textnormal{A}}$})}{2}^2\:\Norm{\tilde{\cE}(\idx{\xi}{A$\tilde{\textnormal{A}}$})}{2}^2\nonumber\\
&=(1-\frac{1}{\idx{d}{A}^2})\:\trace{\idx{\tilde{\omega}}{A'E}^2}\:\trace{\idx{\tilde{\rho}}{AR}^2}\left(\frac{\idx{d}{A}^2-\frac{\idx{d}{A}\:\Trace{\idx{\tilde{\omega}}{E}^2}}{\Trace{\idx{\tilde{\omega}}{A'E}^2}}}{\idx{d}{A}^2-1}\right)\left(\frac{\idx{d}{A}^2-\frac{\idx{d}{A}\:\Trace{\idx{\tilde{\rho}}{R}^2}}{\Trace{\idx{\tilde{\rho}}{AR}^2}}}{\idx{d}{A}^2-1}\right)\\
&\leq(1-\frac{1}{\idx{d}{A}^2})\:\trace{\idx{\tilde{\omega}}{A'E}^2}\:\trace{\idx{\tilde{\rho}}{AR}^2}\label{uslm}\\
&\leq\trace{\idx{\tilde{\omega}}{A'E}^2}\:\trace{\idx{\tilde{\rho}}{AR}^2}\\
&\leq\frac{1}{\tr{[\idx{\omega}{A'E}]}}\Trace{((\sigmaE^{-1/2}\otimes\idi{A'})\idx{\omega}{A'E})^2}\:\frac{1}{\tr{[\idx{\rho}{AR}]}}\Trace{((\idi{A}\otimes\zetaR^{-1/2})\idx{\rho}{AR})^2}\label{newwntmin}
\end{align}
In equation \eqref{uslm} we used \textit{Lemma 3.5} in \cite{Fred:PHD} (see also: \cite{Merging}) to obtain that both bracket terms are smaller than one.
The last inequality follows from the fact that both $\idx{\omega}{A'E}$ and $\idx{\rho}{AR}$ are sub normalized by the assumptions of the theorem. The whole derivation is valid for any positive and normalized operators $\idx{\sigma}{E}$ and $\zetaR$, therefore we can choose $\idx{\sigma}{E}^*$ and $\zetaR^*$
such that they minimize the expression in \eqref{newwntmin}. We get
\begin{align}
&\int_{\mathbb{U}(A)} {\left\| \mathcal{T}((\idx{U}{A} \otimes \idi{R}) \ \rhoAR \  (\idx{U}{A}^\dagger \otimes \idi{R})) - \idx{\omega}{E} \otimes \idx{\rho}{R} \right\|}_1^2 \d{U} \nonumber\\
&\leq\min_{\idx{\sigma}{E} \in \normstates{\idx{\cH}{E}}}\frac{1}{\tr{[\idx{\omega}{A'E}]}}\Trace{((\sigmaE^{-1/2}\otimes\idi{A'})\idx{\omega}{A'E})^2}\:\min_{\idx{\zeta}{R} \in \normstates{\idx{\cH}{R}}}\frac{1}{\tr{[\idx{\rho}{AR}]}}\Trace{((\idi{A}\otimes\zetaR^{-1/2})\idx{\rho}{AR})^2}\\
&=2^{-\chtwo{A'}{E}{\omega}\ -\ \chtwo{A}{R}{\rho}}.
\end{align}
After taking the square root and applying the Jensen inequality (Appendix B), we recapture the decoupling theorem.\\
We note that in the above calculation we had to go back the steps from \eqref{shbet1} to \eqref{shbet2} and to use several bounds to obtain expressions with the ``unphysical'' $\idx{H}{2}$-entropy. It therefore seems that one should try to go to the $\idx{H}{min}$-entropy directly. This is done in the following section.

\section{An improved bound for the decoupling theorem}
We reconsider formula \eqref{nwwtmn} and write it out using the hidden operators $\sigmaE$ and $\zetaR$.
\begin{align}
&\int_{\mathbb{U}(A)}{\left\| (\cT\otimes\cE)((\idx{U}{A} \otimes \idi{$\tilde{A}$}) (\idx{\xi}{A$\tilde{\textnormal{A}}$})(\idx{U}{A}^\dagger \otimes \idi{$\tilde{A}$}))\right\|}_1^2\d{U}\nonumber\\
&\leq\frac{\idx{d}{A}^2}{\idx{d}{A}^2-1}\:\Norm{\tilde{\cT}(\idx{\xi}{A$\tilde{\textnormal{A}}$})}{2}^2\:\Norm{\tilde{\cE}(\idx{\xi}{A$\tilde{\textnormal{A}}$})}{2}^2\\
&=\frac{\idx{d}{A}^2}{\idx{d}{A}^2-1}\:\Norm{\idi{A}\otimes\zetaR^{-\frac{1}{4}}\:(\idx{\rho}{AR}-\idx{\pi}{A}\otimes\rhoR)\:\idi{A}\otimes\zetaR^{-\frac{1}{4}}}{2}^2\cdot\nonumber\\
&\qquad\qquad\qquad\Norm{\idi{A'}\otimes\sigmaE^{-\frac{1}{4}}\:(  \idx{\omega}{A'E}-\idx{\pi}{A'}\otimes\idx{\omega}{E}   )\:\idi{A'}\otimes\sigmaE^{-\frac{1}{4}}}{2}^2\label{twplg}
\end{align}
Due to the similarity of the two terms with the Schatten 2-norm above it is sufficient to bound one of them. We choose the first and use the definition of the Schatten 2-norm to find
\begin{align}
&\Norm{\idi{A}\otimes\zetaR^{-\frac{1}{4}}\:(\idx{\rho}{AR}-\idx{\pi}{A}\otimes\rhoR)\:\idi{A}\otimes\zetaR^{-\frac{1}{4}}}{2}^2\\
&=\Trace{\idi{A}\otimes\zetaR^{-\frac{1}{2}}\:(\idx{\rho}{AR}-\idx{\pi}{A}\otimes\rhoR)\:\idi{A}\otimes\zetaR^{-\frac{1}{2}}\:(\idx{\rho}{AR}-\idx{\pi}{A}\otimes\rhoR)}\nonumber\\
&\leq\Norm{\idi{A}\otimes\zetaR^{-\frac{1}{2}}\:(\idx{\rho}{AR}-\idx{\pi}{A}\otimes\rhoR)\:\idi{A}\otimes\zetaR^{-\frac{1}{2}}\:(\idx{\rho}{AR}-\idx{\pi}{A}\otimes\rhoR)}{1}\\
&\leq\Norm{\idi{A}\otimes\zetaR^{-\frac{1}{2}}\:(\idx{\rho}{AR}-\idx{\pi}{A}\otimes\rhoR)\:\idi{A}\otimes\zetaR^{-\frac{1}{2}}}{\infty}\normBig{\idx{\rho}{AR}-\idx{\pi}{A}\otimes\rhoR}{1}.
\end{align}
The last inequality was obtained by an application of \textit{Lemma 2}. Now we rewrite the expression with the $\infty$-norm in a way that reveals its relation to the $\idx{H}{min}$-entropy. For this we choose from the beginning of our calculation on $\zetaR$ to minimize the term with the $\infty$-norm i.\:e.\:we evaluate
\begin{align}
&\min_{\zetaR\in\normstates{\idx{H}{R}}}\Norm{\idi{A}\otimes\zetaR^{-\frac{1}{2}}\:(\idx{\rho}{AR}-\idx{\pi}{A}\otimes\rhoR)\:\idi{A}\otimes\zetaR^{-\frac{1}{2}}}{\infty}\nonumber\\
&=\min\left\{\lambda\in\mathbb{R}\ |\ \lambda=\norm{\idi{A}\otimes\zetaR^{-\frac{1}{2}}\:(\idx{\rho}{AR}-\idx{\pi}{A}\otimes\rhoR)\:\idi{A}\otimes\zetaR^{-\frac{1}{2}}}{\infty};\ \zetaR\in\normstates{\idx{H}{R}}\right\}\\
&=\min\left\{\lambda\in\mathbb{R}\ |\ \idi{A}\otimes\zetaR^{-\frac{1}{2}}\:(\idx{\rho}{AR}-\idx{\pi}{A}\otimes\rhoR)\:\idi{A}\otimes\zetaR^{-\frac{1}{2}}\leq\lambda\:\idi{AR};\ \zetaR\in\normstates{\idx{H}{R}}\right\}\\
&=\min\left\{\lambda\in\mathbb{R}\ |\ \idx{\rho}{AR}-\idx{\pi}{A}\otimes\rhoR\leq\lambda\:\idi{A}\otimes\zetaR;\ \zetaR\in\normstates{\idx{H}{R}}\right\}\\
&=\min\left\{\Trace{\zetaR}\ |\ \idx{\rho}{AR}\leq\:\idi{A}\otimes(\zetaR+\frac{1}{\idx{d}{A}}\rhoR);\ \zetaR\in\posops{\idx{H}{R}}\right\}\\
&=\min\left\{\Trace{\tilde{\zetaR}-\frac{1}{\idx{d}{A}}\rhoR}\ |\ \idx{\rho}{AR}\leq\:\idi{A}\otimes\tilde{\zetaR};\ \tilde{\zetaR}\in\posops{\idx{H}{R}}+\frac{1}{\idx{d}{A}}\rhoR\right\}\label{samesets}\\
&=\min\left\{\Trace{\zetaR-\frac{1}{\idx{d}{A}}\rhoR}\ |\ \idx{\rho}{AR}\leq\:\idi{A}\otimes\zetaR;\ \zetaR\in\posops{\idx{H}{R}}\right\}\\
&=\min\left\{\Trace{\zetaR}\ |\ \idx{\rho}{AR}\leq\:\idi{A}\otimes\zetaR;\ \zetaR\in\posops{\idx{H}{R}}\right\}-\frac{1}{\idx{d}{A}}\Trace{\rhoR}\\
&=\min\left\{\lambda\in\mathbb{R}\ |\ \idx{\rho}{AR}\leq\lambda\:\idi{A}\otimes\zetaR;\ \zetaR\in\normstates{\idx{H}{R}}\right\}-\frac{1}{\idx{d}{A}}\Trace{\rhoR}\\
&=2^{\min\left\{\lambda\in\mathbb{R}\ |\ \idx{\rho}{AR}\leq2^\lambda\:\idi{A}\otimes\zetaR;\ \zetaR\in\normstates{\idx{H}{R}}\right\}}-\frac{1}{\idx{d}{A}}\Trace{\rhoR}\\
&=2^{-\chmin{A}{R}{\rho}}-\frac{1}{\idx{d}{A}}\Trace{\rhoR}.
\end{align}
For the last equality we used the definition of the $\idx{H}{min}$-entropy.
In equation \eqref{samesets} we exploited the fact that the sum of two positive-semidefinite matrices is given by a positive-semidefinite matrix again. Therefore the two groups (together with $+$) $\posops{\idx{H}{R}}+\frac{1}{\idx{d}{A}}\rhoR$ and $\posops{\idx{H}{R}}$ are identical. By analogy we can conclude that
\begin{align}
&\Norm{\idi{A'}\otimes\sigmaE^{-\frac{1}{4}}\:(  \idx{\omega}{A'E}-\idx{\pi}{A'}\otimes\idx{\omega}{E}   )\:\idi{A'}\otimes\sigmaE^{-\frac{1}{4}}}{2}^2\nonumber\\
&\leq\left(2^{-\chmin{A'}{E}{\omega}}-\frac{1}{\idx{d}{A}}\Trace{\idx{\omega}{E}}\right)\normBig{\idx{\omega}{A'E}-\idx{\pi}{A}\otimes\idx{\omega}{E}}{1}.
\end{align}
Plugging in both results into equation \eqref{twplg}, we conclude
\begin{align}
&\int_{\mathbb{U}(A)}{\left\| (\cT\otimes\cE)((\idx{U}{A} \otimes \idi{$\tilde{A}$}) (\idx{\xi}{A$\tilde{\textnormal{A}}$})(\idx{U}{A}^\dagger \otimes \idi{$\tilde{A}$}))\right\|}_1^2\d{U}\nonumber\\
&\leq\frac{1}{1-\frac{1}{\idx{d}{A}^2}}\left(2^{-\chmin{A'}{E}{\omega}}-\frac{1}{\idx{d}{A}}\Trace{\idx{\omega}{E}}\right)\left(2^{-\chmin{A}{R}{\rho}}-\frac{1}{\idx{d}{A}}\Trace{\rhoR}\right)\cdot\nonumber\\
&\qquad\qquad\qquad\qquad\normBig{\idx{\omega}{A'E}-\idx{\pi}{A}\otimes\idx{\omega}{E}}{1}\:\normBig{\idx{\rho}{AR}-\idx{\pi}{A}\otimes\idx{\rho}{R}}{1}\label{prfrmsq}.
\end{align}
Note that the $\idx{H}{min}$-entropy is upper bounded by the logarithm of the dimension i. e. $\chmin{A}{R}{\rho}\leq\log{\idx{d}{A}}$, for example. Since the states $\idx{\omega}{A'E}$ and $\idx{\rho}{AR}$ are sub normalized by the conditions of the theorem, both bracket terms must be positive. This implies that the subtrahends in the brackets may be left out to obtain a shorter formulation of the result.\\
As a final step we perform the square root on both sides of \eqref{prfrmsq} and then use Jensen inequality to be able to take the square root of the integrand. We find that
\begin{align}
&\int_{\mathbb{U}(A)}{\left\| (\cT\otimes\cE)(\idx{U}{A} \otimes \idi{$\tilde{A}$}) (\idx{\xi}{A$\tilde{\textnormal{A}}$})(\idx{U}{A}^\dagger \otimes \idi{$\tilde{A}$})\right\|}_1\d{U}\nonumber\\
&\leq\sqrt{\frac{1}{1-\frac{1}{\idx{d}{A}^2}}\left(2^{-\chmin{A'}{E}{\omega}}-\frac{1}{\idx{d}{A}}\Trace{\idx{\omega}{E}}\right)\left(2^{-\chmin{A}{R}{\rho}}-\frac{1}{\idx{d}{A}}\Trace{\rhoR}\right)}\cdot\nonumber\\
&\qquad\qquad\qquad\sqrt{\normBig{\idx{\omega}{A'E}-\idx{\pi}{A}\otimes\idx{\omega}{E}}{1}\:\normBig{\idx{\rho}{AR}-\idx{\pi}{A}\otimes\idx{\rho}{R}}{1}}
\end{align}
and arrive at a better bound for the decoupling theorem. We state our new version of the theorem for completeness:\\
\\
\textbf{Theorem:} (Decoupling theorem)\textit{
Let $\rhoAR\in\subnormstates{\idx{\cH}{A}\otimes\idx{\cH}{R}}$ be a sub normalized density operator and let $\cTAE$ be a completely positive linear map going from $\subnormstates{\idx{\cH}{A}\otimes\idx{\cH}{R}}$ to $\posops{\idx{\cH}{E} \otimes \idx{\cH}{R}}$ with Choi-Jamiolkowski representation $\idx{\omega}{A'E}\in\subnormstates{\idx{\cH}{E}\otimes\idx{\cH}{A'}}$, then
\begin{align}
&\int_{\mathbb{U}(A)} {\left\| \mathcal{T}((\idx{U}{A} \otimes \idi{R}) \ \rhoAR \  (\idx{U}{A}^\dagger \otimes \idi{R})) - \idx{\omega}{E} \otimes \idx{\rho}{R} \right\|}_1 \d{U}\nonumber\\
&\leq\sqrt{\frac{1}{1-\frac{1}{\idx{d}{A}^2}}\left(2^{-\chmin{A'}{E}{\omega}}-\frac{1}{\idx{d}{A}}\Trace{\idx{\omega}{E}}\right)\left(2^{-\chmin{A}{R}{\rho}}-\frac{1}{\idx{d}{A}}\Trace{\rhoR}\right)}\cdot\nonumber\\
&\qquad\qquad\qquad\sqrt{\normBig{\idx{\omega}{A'E}-\idx{\pi}{A}\otimes\idx{\omega}{E}}{1}\:\normBig{\idx{\rho}{AR}-\idx{\pi}{A}\otimes\idx{\rho}{R}}{1}}\nonumber
\end{align}
where the integration goes over all unitaries and with respect to the probability Haar measure $\d U$.
}\\
\\
In the sequel we will always work with the original decoupling theorem and we will not consider this version anymore. Thus when we refer to the ``Decoupling Theorem'', we always mean the original one as stated in the last section. All future results related to the original decoupling theorem can easily be generalized to formulations including the latest version. (This is for example the case for the main theorem about decoupling with almost 2-designs.) 
\chapter{Decoupling with almost 2-designs}

We consider the situation of a system $A$ on which some physical process described by a unitary operation occurs. Generally we allow for correlations of the system $A$ to a reference system $R$, on which no physical evolution takes place. Afterwards a TPCPM $\cTAE$ is applied to the $A$ subsystem again leaving the $R$ subsystem unaffected. The joint state of the system $AR$ before any process takes place is described by the density operator $\rhoAR$. Accordingly, the state of the system after the whole evolution is given by $\cTAE(\idx{U}{A}\otimes\idi{R}\:\rhoAR\:\idx{U}{A}^\dagger\otimes\idi{R})$. For fixed $\cT$ and $\rhoAR$ the Decoupling Theorem can be used to guarantee the existence of some unitary operator, such that applying that unitary and afterwards the map $\cT$ on the $A$ subsystem of $\rhoAR$ destroys almost all correlations between the two subsystems:
Since we know that the average distance is bounded by
\begin{align}
\int_{\mathbb{U}(A)} {\left\| \mathcal{T}((\idx{U}{A} \otimes \idi{R}) \ \rhoAR \  (\idx{U}{A}^\dagger \otimes \idi{R})) - \idx{\omega}{E} \otimes \idx{\rho}{R} \right\|}_1 \d{U} \leq
2^{-\frac{1}{2} H_2(A'|E)_{\omega} - \frac{1}{2} H_2(A|R)_{\rho} }
\end{align}
we can be sure that there exists a unitary $U^{*}$ that decouples well in the sense that
\begin{align}
{\left\| \mathcal{T}((\idx{U^{*}}{A} \otimes \idi{R}) \ \rhoAR \  ((\idx{U}{A}^{*})^\dagger \otimes \idi{R})) - \idx{\omega}{E} \otimes \idx{\rho}{R} \right\|}_1 \leq
2^{-\frac{1}{2} H_2(A'|E)_{\omega} - \frac{1}{2} H_2(A|R)_{\rho} }.
\end{align}
This is especially relevant for channel coding, since that unitary can be interpreted as an encoding operator \cite{Fred:PHD,mother,Merging}. In this chapter we go a step further and try to generalize the decoupling theorem to the case where we have an ``almost-integration'' only. As a motivation, we consider a concrete physical realization of the systems $A$ and $R$ assuming them to be made of interacting particles. We model the internal dynamics of the $A$ subsystem in terms of a random quantum circuit and address the question whether or not a \textit{physical} process occurring on the $A$ subsystem may be used for encoding purposes. This means that some TPCPM $\cTAE$ is fixed, and we analyze if a random quantum circuit does well in the sense of decoupling. This question is particularly relevant for applications in case that $\cT$ is given by the partial trace.
By the usual formulation of the decoupling theorem we already know that there exists some hypothetical physical process on the $A$ subsystem which decouples well. But here we have a concrete model of the internal dynamics and we would like to see if these dynamics actually can produce the desired process. And how long would this take?
We would like to give a precise meaning to this question using the concepts of \textit{unitary designs} and \textit{quantum circuits} in the following sections.

\section{Unitary 2-designs}

Suppose we are interested in taking the average of a function $f$ over the Lie group $\mathbb{U}$. To obtain a guess about this value it is helpful to take a finite set $\mathcal{D}$ of unitaries and instead evaluate $f$ at these points and average over $\cD$. This procedure is fairly similar to the the well known calculation of ``upper sums'' and ``lower sums'' in the definition of the Riemann integral over a subset or $\mathbb{R}$. (But here we don't take the limit to infinite partitions). Of course for any such set there are functions $f$ whose average over $\mathbb{U}$ is arbitrarily different from the one over $\mathcal{D}$. But if we fix a certain type of ``good'' functions and assume the set $\mathcal{D}$ to be ``large'' enough, we should obtain at least a good guess.
Heuristically, a unitary $k$-designs is a finite subset $\mathcal{D}$ of the Lie group $\mathbb{U}$ that has the property that integrating any polynomial of degree $k$ over the whole unitary group with respect to the Haar measure gives the same result as averaging over $\mathcal{D}$. We fix the vague statements from above in a definition \cite{RQC}:

\begin{definition}(Unitary design) Let $\mathcal{D}=\{U_i\}_{i=1,...,n}$ be a set of unitary matrices on a Hilbert space $\cH$. Attach to each $U_i$ a probability $p_i$ with $\sum_i p_i=1$.
Define the functions:
$$\mathcal{G}_W(\rho):=\sum_i{p_i U_i^{\otimes k}\rho (U_i^\dagger)^{\otimes k}}$$
$$\mathcal{G}_H(\rho):=\int_{\mathbb{U}}{{U^{\otimes k}\rho (U^\dagger)^{\otimes k}}}\d{U}$$
for $\rho\in\linops{\cH^{\otimes k}}$.
$\mathcal{D}$ is called a unitary $k$-design if and only if $\mathcal{G}_W=\mathcal{G}_H$.
\end{definition}
This implies, that any polynomial of degree $k$ in the matrix elements of a unitary $U$ and of degree $k$ in the matrix elements of $\bar{U}$ has the same expectation value with respect to the two different probability distribution underlying the expressions for $\mathcal{G}_W$ and $\mathcal{G}_H$.
To see this we evaluate the terms $\bra{i_1,...,i_k}\mathcal{G}_W(\proj{j_1,...,j_k}{j'_1,...,j'_k})\ket{i'_1,...,i'_k}$ and $\bra{i_1,...,i_k}\mathcal{G}_H(\proj{j_1,...,j_k}{j'_1,...,j'_k})\ket{i'_1,...,i'_k}$ and use the defining property of a k-design $\mathcal{G}_W=\mathcal{G}_H$ to see that the average of a monomial over $\mathbb{U}$ and $\mathcal{D}$ is always the same. Then any polynomial will also have the same average over $\mathbb{U}$ and $\mathcal{D}$, which justifies the above motivation.\\
Obviously 2-designs are relevant in our context. In equations \eqref{intgrlpr} we integrate over the unitary group. Since the state in the integrand is conjugated by a two-fold tensor product of a unitary, a 2-design would be sufficient to perform the integration. Thus in the statement of the decoupling theorem the integral can be replaced by a 2-design immediately.\\
\\
\textbf{Proposition:} (Decoupling with 2-designs)\textit{
Let $\cD$ be a 2-design, let $\rhoAR\in\subnormstates{\idx{\cH}{A}\otimes\idx{\cH}{R}}$ be a sub normalized density operator and let $\cTAE$ be a completely positive linear map going from $\subnormstates{\idx{\cH}{A}\otimes\idx{\cH}{R}}$ to $\posops{\idx{\cH}{E} \otimes \idx{\cH}{R}}$ with Choi-Jamiolkowski representation $\idx{\tilde{\omega}}{A'E}\in\subnormstates{\idx{\cH}{E}\otimes\idx{\cH}{A'}}$, then
$$\sum_{U^i\in\cD} {p_i\left\| \mathcal{T}((\idx{U^i}{A} \otimes \idi{R}) \ \rhoAR \  ((\idx{U^i}{A})^\dagger \otimes \idi{R})) - \idx{\omega}{E} \otimes \idx{\rho}{R} \right\|}_1 \leq
2^{-\frac{1}{2} H_2(A'|E)_{\omega} - \frac{1}{2} H_2(A|R)_{\rho} }.$$
}\\
\\
(We wrote the index $i$ as a superscript for typographical convenience, which will also be done in the future whenever it simplifies the notation.)
The main result of this chapter will be a decoupling formula with $\varepsilon$-almost 2-designs. Then the above proposition is an immediate corollary in the case $\varepsilon=0$.
For the proof we generalize the derivation of the usual decoupling theorem which only requires the evaluation of expressions involving 2-designs. 
Therefore although we stated the definition of general $k$-designs, for our decoupling results only 2-designs are relevant. We will only consider this special case in the future and refer to this case when we write $\mathcal{G}_W$ or $\mathcal{G}_H$.

\section{Random quantum circuits}
A quantum circuit is a sequence of wires and gates. To each wire corresponds some qubit, which is transported through space or time by this wire and to each gate corresponds some unitary operation. A $k$-qubit gate takes $k$ input qubits (i.\:e.\:$k$ wires) and performs some operation on them to give back $k$ qubits after its application. (I.\:e.\:it is given by an element of $\mathbb{U}(2^k)$.) Since the wires do not perform any operation it is sufficient to think of the circuit as a sequence of unitaries, that are applied in a certain order: $W=W_t\cdot...\cdot W_2\cdot W_1$, where we call $t$ the time of the circuit. As in the case of classical computation the most common gates are $1$ and $2$ bit gates. We call a set of gates \textit{universal} for $k$- qubits if any operation which can be performed on $k$ qubits can be approximated to arbitrary precision using operations from the universal gate set only. A more detailed introduction to the topic of quantum circuits may be found in \cite{NielsenChuang}.\\
Typically quantum computations are modeled using quantum circuits. But for the motivation of our theorem it is also interesting to model the randomization process of a many-particle physical system using quantum circuits. Such approaches were considered in \cite{ETE} and \cite{RQC} and here we will keep close to the second one.\\
There the authors start with a $k$-qubit Hilbert space $\cH$ describing some system whose state is given by $\ket{\psi}$. In nature the most common type of interaction is a two particle interaction. It corresponds to the application of a 2-qubit unitary gate. Thus a universal gate set of gates in $\mathbb{U}(4)$ is particularly interesting. A canonical example for such a universal set would be the set of all one qubit gates together with the CNOT gate. The circuit acts in the following way: At each step of the circuit two qubits and an element of the universal gate set are chosen uniformly at random. The gate is applied go the qubits and the circuit proceeds to the next step. Since the gate set is assumed to be universal any element of $\mathbb{U}(4)$ can be reached. This set is again universal for $\mathbb{U}(2^k)$, so that any unitary can be generated by the circuit.\\
The crucial property of the described circuit is its relation to 2-designs. In the next section we state some related results.

\section{Unitary almost 2-desings}

In this section we link the above topics of 2-designs and random circuits by introducing what is called an \textit{almost 2-design}. At the end of this section we state a pivotal theorem, which establishes the fact that random quantum circuits are approximate 2-designs.
But first we recapture the definition of the diamond norm of some linear map $\cTAE$ from $\linops{\idx{\cH}{A}}$ to $\linops{\idx{\cH}{E}}$ \cite{Watrous}.

\begin{definition}
Let $\cTAE$ be a linear map from $\linops{\idx{\cH}{A}}$ to $\linops{\idx{\cH}{E}}$ the diamond norm of $\cTAE$ is defined to be:
$$\norm{\cTAE}{\diamond}\ =\ \sup_{\idx{d}{R}}{\max_{\rhoAR\in\linops{\idx{\cH}{AR}}}{\frac{\norm{(\cTAE\otimes\opidi{R})({\rhoAR})}{1}}{\norm{\rhoAR}{1}}}}$$
\end{definition}
If calculated for a difference of two quantum channels, the diamond norm gives the maximum probability of being able to distinguish the two channels experimentally. This is because of the property of the trace distance between two quantum states that it quantifies how well these states can be distinguished with arbitrary measurements. Thus for two quantum channels $\cT$, $\cE$ the expression
\begin{align}
\norm{\cT-\cE}{1}:=\max_{\rho\in\normstates{\cH}}{\norm{\cT(\rho)-\cE(\rho)}{1}}
\end{align}
corresponds to the experimental situation, where an optimal input state $\rho^*$ is chosen and afterwards one tries to distinguish the states $\cT(\rho^*)$ and $\cE(\rho^*)$ with some measurement. But this is still not the best way to distinguish the quantum channels $\cT$ and $\cE$ in an experiment. The definition of $\norm{\cT-\cE}{1}$ does not include the possibility of choosing an initial state in some ``larger'' Hilbert space. In general one has that
\begin{align}
\norm{\cT-\cE}{1}\leq\norm{\cT-\cE}{\diamond},
\end{align}
which motivates the definition of the diamond norm as stated above.\\
Intuitively an almost 2-design is a finite set of unitary operators which approximates a 2-design. It is a priori not clear which of the many properties a 2-design has, apart from our defining property, should be used for comparison. And once a property is fixed it is also nontrivial to fix the norm in which this approximation should be measured. This results in inconsistent and different definitions in the current literature. We would like to apply the results obtained in \cite{RQC} for decoupling purposes and therefore abide by their definition.\\
\begin{definition}
Let $\mathcal{G}_W$ and $\mathcal{G}_H$ be as in the definition of the $k$-design (Definition 6).
$\mathcal{G}_W$ is called an $\varepsilon$-approximate unitary $k$-design, if
$$\norm{\mathcal{G}_W-\mathcal{G}_H}{\diamond}\leq\varepsilon$$
\end{definition}
Since the map $\mathcal{G}_W$ is entirely determined by the set of pairs $\{(p_i,U_i)\}_{i=1,...,n}$, we sometimes will also refer to that set as the almost 2-design.
We now consider the random circuit described above. For infinitely many time- steps the outcome after applying the circuit to some state will be independent of the state on which it is applied. This means that the measure on the set of unitaries reached by the circuit gets unitarily invariant. Since the Haar measure is the unique biinvariant measure on $\mathbb{U}$, we can conclude that in the limit of infinite time the distribution of unitaries generated by the circuit reaches the Haar distribution.\\
Let now $W$ be a unitary generated by our circuit after $t$ steps of time. Applying this circuit two times to the different subsystems of a bipartite state $\rhoAB$ gives a state $\rhoAB^W=W\otimes W \rhoAB W^\dagger\otimes W^\dagger$. If the average of all the $\rhoAB^W$ taken over the different circuits $W$ equals the average calculated over $\mathbb{U}$ with respect to the Haar measure, we say that the random circuits constitute a 2-design.
Unfortunately it turns out that the convergence rate of the random circuits towards the Haar distribution is exponentially slow in the number of qubits of the underlying system \cite{RQC}, \cite{ETE}, \cite{NielsenChuang}. Nevertheless, the authors of \cite{RQC} (Theorems 9 and 10) derive the following pivotal theorem:\\
\\
\textbf{Theorem:} (Random quantum circuits are approximate 2-designs) \textit{Let $\mu$ be the probability distribution corresponding to any universal gate set on $\mathbb{U}(4)$ and let $W$ be a random circuit on $n$ qubits obtained by drawing $t$ random unitaries according to $\mu$ and applying each of them to a random pair of qubits. Then there exists $C$ and ($C=C(\mu)$ only) 
such that for any $\varepsilon>0$ and any $t\geq C(n^2+n\log(1/\varepsilon))$, $\cG_W$ is an $\varepsilon$-approximate unitary 2-design.}\\
\\
This theorem implies that random two particle interactions as described above yield approximate 2-designs. We would like to understand, if the process is really decoupling in the sense that if it occurs on some system it destroys the correlations to its reference system. This is relevant from a physical point of view, because as already mentioned the time until the random circuits get close to being distributed with respect to Haar measure and thus the time until the usual decoupling theorem is applicable is exponentially large. So what happens with the system in the physical situation after polynomial circuit time? We consider a decoupling theorem with almost 2-designs.\\

\section{Decoupling with almost 2-designs}
In this section we formulate and prove the core theorem of this chapter. It generalizes the usual decoupling theorem in the sense that it is valid in the case where almost 2-designs are considered.\\
\\
\textbf{Theorem:} (Decoupling with $\varepsilon$-approximate unitary 2-designs)\textit{
Let $\rhoAR\in\subnormstates{\idx{\cH}{A}\otimes\idx{\cH}{R}}$ be a sub normalized density operator and let $\cTAE$ be a completely positive, linear map going from $\subnormstates{\idx{\cH}{A}\otimes\idx{\cH}{R}}$ to $\posops{\idx{\cH}{E} \otimes \idx{\cH}{R}}$ with Choi-Jamiolkowski representation $\idx{\omega}{A'E}\in\subnormstates{\idx{\cH}{E}\otimes\idx{\cH}{A'}}$, then
\begin{align}
&\sum_{(p_i,U_i)\in\cD}{p_i}{{\left\| \mathcal{T}((\idx{U^i}{A} \otimes \idi{R}) \ \rhoAR \  ((\idx{U^i}{A})^\dagger \otimes \idi{R})) - \idx{\omega}{E} \otimes \idx{\rho}{R} \right\|}_1}\nonumber\\
&\leq\sqrt{1+4\varepsilon\idx{d}{A}^4}\ 2^{-\frac{1}{2}\:(\chtwo{A'}{E}{\omega}\ +\ \chtwo{A}{R}{\rho})}\nonumber
\end{align}
where the summation goes over pairs $(p_i,U_i)$, such that $\cD$ constitutes an $\varepsilon$-approximate 2-design.
}\\
\\
For a proof we proceed similarly to the proof of the decoupling theorem in the last chapter.
Like before, we introduce the map $\idx{\cE}{$\tilde{\textnormal{A}}\rightarrow$E}$ which we define to be the unique Choi-Jamiolkowski preimage of the state $\idx{\rho}{AR}$ and write for any $i$:
\begin{align}
&\mathcal{T}((\idx{U^i}{A} \otimes \idi{R}) \ \rhoAR \  (\idx{U^i}{A}^\dagger \otimes \idi{R})) - \idx{\omega}{E} \otimes \idx{\rho}{R}\nonumber\\
&=(\cT\otimes\cE)((\idx{U^i}{A} \otimes \idi{$\tilde{A}$}) (\idx{\xi}{A$\tilde{\textnormal{A}}$})(\idx{U^i}{A}^\dagger \otimes \idi{$\tilde{A}$}))\label{intxi},
\end{align}
where $\idx{\xi}{A$\tilde{\textnormal{A}}$}=\idx{\Phi}{A$\tilde{\textnormal{A}}$}-\idx{\pi}{A}\otimes\idx{\pi}{$\tilde{\textnormal{A}}$}$ is the decoupling state.\\
The idea ot the derivation is to add and subtract an integral term which is "close" to the sum at a step where this is convenient and then use the defining property of the almost 2-design.\\
To go from the difficult Schatten 1-norm of the theorem to the manageable Schatten 2-norm we use H\"older inequality as stated in Appendix A (or \textit{Lemma 5}, \cite{Semthesis}) in exactly the same manner as it was done in the last chapter for the proof of the decoupling theorem. 
Again we introduce  positive semidefinite, normalized operators $\sigmaE$ and $\zetaR$ and the maps $\tilde{\cT}$ and $\tilde{\cE}$ with
\begin{align}
\tilde{\mathcal{T}}(\idx{\tau}{A$\tilde{\textnormal{A}}$})\ &:=\ (\sigmaE \otimes \idi{$\tilde{\textnormal{A}}$})^{-1/4} \mathcal{T}(\idx{\tau}{A$\tilde{\textnormal{A}}$}) (\sigmaE \otimes \idi{$\tilde{\textnormal{A}}$})^{-1/4}\qquad\ \forall\ \idx{\tau}{A$\tilde{\textnormal{A}}$}\in\linops{\idx{\cH}{A$\tilde{\textnormal{A}}$}}\\
\tilde{\mathcal{E}}(\idx{\tau}{A$\tilde{\textnormal{A}}$})\ &:=\ (\idi{A} \otimes \zetaR)^{-1/4} \cE(\idx{\tau}{A$\tilde{\textnormal{A}}$})(\idi{A} \otimes \zetaR)^{-1/4}\qquad\ \ \forall\ \idx{\tau}{A$\tilde{\textnormal{A}}$}\in\linops{\idx{\cH}{A$\tilde{\textnormal{A}}$}}
\end{align}
and find
\begin{align}
&{\left\| (\cT\otimes\cE)((\idx{U^i}{A} \otimes \idi{$\tilde{A}$}) (\idx{\xi}{A$\tilde{\textnormal{A}}$})(\idx{U^i}{A}^\dagger \otimes \idi{$\tilde{A}$}))\right\|}_1\nonumber\\
&\leq{\left\| (\tilde{\cT}\otimes\tilde{\cE})((\idx{U^i}{A} \otimes \idi{$\tilde{A}$}) (\idx{\xi}{A$\tilde{\textnormal{A}}$})(\idx{U^i}{A}^\dagger \otimes \idi{$\tilde{A}$}))\right\|}_2\\
&=\sqrt{\Trace{(\tilde{\cT}\otimes\tilde{\cE})(\idx{U^i}{A} \otimes \idi{$\tilde{A}$}  (\idx{\xi}{A$\tilde{\textnormal{A}}$}) \idx{U^i}{A}^\dagger \otimes \idi{$\tilde{A}$})^2}}.
\end{align}
The next step is to apply the swap trick (Appendix C) as was already done in the proof of the usual decoupling theorem. This gives
\begin{align}
&\sqrt{\Trace{(\tilde{\cT}\otimes\tilde{\cE})(\idx{U^i}{A} \otimes \idi{$\tilde{\textnormal{A}}$}\ \idx{\xi}{A$\tilde{\textnormal{A}}$}\ \idx{U^i}{A}^\dagger \otimes \idi{$\tilde{\textnormal{A}}$})^2}}\nonumber\\
&= \sqrt{\Trace{(\tilde{\mathcal{T}}\otimes\tilde{\cE})^{\otimes2}\left((\idx{U^i}{A} \otimes \idi{$\tilde{\textnormal{A}}$})^{\otimes2} \ (\idx{\xi}{A$\tilde{\textnormal{A}}$})^{\otimes2} \ (\idx{U^i}{A}^\dagger \otimes \idi{$\tilde{\textnormal{A}}$})^{\otimes 2}\right)  \idx{\mathcal{F}}{E} \otimes \idx{\mathcal{F}}{R}}} \label{swintrod}\\
&=\sqrt{\Trace{\left((\idx{U^i}{A} \otimes \idi{$\tilde{\textnormal{A}}$})^{\otimes2} \ (\idx{\xi}{A$\tilde{\textnormal{A}}$})^{\otimes2} \ (\idx{U^i}{A}^\dagger \otimes \idi{$\tilde{\textnormal{A}}$})^{\otimes 2}\right) (\tilde{\mathcal{T}}^{\dagger})^{\otimes2}[\idx{\mathcal{F}}{E}]\otimes (\tilde{\cE}^{\dagger})^{\otimes2}[\idx{\mathcal{F}}{R}]}}.\label{adjexp1}
\end{align}
In order to deal with the square root we use Jensen inequality, which gives
\begin{align}
&\sum_{i}{p_i\sqrt{\Trace{\left((\idx{U^i}{A} \otimes \idi{$\tilde{\textnormal{A}}$})^{\otimes2} \ (\idx{\xi}{A$\tilde{\textnormal{A}}$})^{\otimes2} \ (\idx{U^i}{A}^\dagger \otimes \idi{$\tilde{\textnormal{A}}$})^{\otimes 2}\right) (\tilde{\mathcal{T}}^{\dagger})^{\otimes2}[\idx{\mathcal{F}}{E}]\otimes (\tilde{\cE}^{\dagger})^{\otimes2}[\idx{\mathcal{F}}{R}]}}}\nonumber\\
&\leq\sqrt{\sum_{i}{p_i\Trace{\left((\idx{U^i}{A} \otimes \idi{$\tilde{\textnormal{A}}$})^{\otimes2} \ (\idx{\xi}{A$\tilde{\textnormal{A}}$})^{\otimes2} \ (\idx{U^i}{A}^\dagger \otimes \idi{$\tilde{\textnormal{A}}$})^{\otimes 2}\right) (\tilde{\mathcal{T}}^{\dagger})^{\otimes2}[\idx{\mathcal{F}}{E}]\otimes (\tilde{\cE}^{\dagger})^{\otimes2}[\idx{\mathcal{F}}{R}]}}}\\
&=\sqrt{\Trace{\left(\sum_i{p_i(\idx{U^i}{A} \otimes \idi{$\tilde{\textnormal{A}}$})^{\otimes2} \ (\idx{\xi}{A$\tilde{\textnormal{A}}$})^{\otimes2} \ (\idx{U^i}{A}^\dagger \otimes \idi{$\tilde{\textnormal{A}}$})^{\otimes 2}}\right) (\tilde{\mathcal{T}}^{\dagger})^{\otimes2}[\idx{\mathcal{F}}{E}]\otimes (\tilde{\cE}^{\dagger})^{\otimes2}[\idx{\mathcal{F}}{R}]}}\\
&=\sqrt{\Trace{\left(\sum_i{p_i(\idx{U^i}{A}^{\otimes2} \otimes \idi{$\tilde{\textnormal{A}}$}^{\otimes2}) \ (\idx{\xi}{A$\tilde{\textnormal{A}}$})^{\otimes2} \ ((\idx{U^i}{A}^\dagger)^{\otimes2} \otimes \idi{$\tilde{\textnormal{A}}$}^{\otimes2})}\right) (\tilde{\mathcal{T}}^{\dagger})^{\otimes2}[\idx{\mathcal{F}}{E}]\otimes (\tilde{\cE}^{\dagger})^{\otimes2}[\idx{\mathcal{F}}{R}]}}.\label{drpsqrt}
\end{align}
To apply the defining property of the almost 2-design we need to compare the term $\sum_i{p_i(\idx{U^i}{A}^{\otimes2} \otimes \idi{$\tilde{\textnormal{A}}$}^{\otimes2}) \ (\idx{\xi}{A$\tilde{\textnormal{A}}$})^{\otimes2} \ ((\idx{U^i}{A}^\dagger)^{\otimes2} \otimes \idi{$\tilde{\textnormal{A}}$}^{\otimes2})}$ with the corresponding integral. We therefore add $\int{(\idx{U}{A}^{\otimes2} \otimes \idi{$\tilde{\textnormal{A}}$}^{\otimes2}) \ (\idx{\xi}{A$\tilde{\textnormal{A}}$})^{\otimes2} \ ((\idx{U}{A}^\dagger)^{\otimes2} \otimes \idi{$\tilde{\textnormal{A}}$}^{\otimes2})}\d{U}$ to the argument of the square root and subtract it again. Note that we have the following two relations:
\begin{align}
\sum_i{p_i(\idx{U^i}{A}^{\otimes2} \otimes \idi{$\tilde{\textnormal{A}}$}^{\otimes2}) \ (\idx{\xi}{A$\tilde{\textnormal{A}}$})^{\otimes2} \ ((\idx{U^i}{A}^\dagger)^{\otimes2} \otimes \idi{$\tilde{\textnormal{A}}$}^{\otimes2})}&=(\cG_W\otimes\opidi{$\tilde{\textnormal{A}}\tilde{\textnormal{A'}}$})(\idx{\xi}{A$\tilde{\textnormal{A}}$}^{\otimes2}),\\
\int{(\idx{U}{A}^{\otimes2} \otimes \idi{$\tilde{\textnormal{A}}$}^{\otimes2}) \ (\idx{\xi}{A$\tilde{\textnormal{A}}$})^{\otimes2} \ ((\idx{U}{A}^\dagger)^{\otimes2} \otimes \idi{$\tilde{\textnormal{A}}$}^{\otimes2})}\d{U}&=(\cG_H\otimes\opidi{$\tilde{\textnormal{A}}\tilde{\textnormal{A'}}$})(\idx{\xi}{A$\tilde{\textnormal{A}}$}^{\otimes2}).
\end{align}
Where the $\cG_W$ and $\cG_H$ are as in 
Definition 6 for $k=2$ and $\opidi{$\tilde{\textnormal{A}}\tilde{\textnormal{A'}}$}$ denotes the operator identity on $\tilde{A}\tilde{A'}$. The above evidently would be valid if there were no correlations between the systems $A$, $A'$, $\tilde{A}$ and $\tilde{A'}$ (which is not true for $\idx{\xi}{A$\tilde{\textnormal{A}}$}$). But by linearity of all the considered maps the statements also follow for $\idx{\xi}{A$\tilde{\textnormal{A}}$}$.
For notational convenience we drop the square root in equation \eqref{drpsqrt} and consider its argument only. We have:
\begin{align}
&\Trace{\left(\sum_i{p_i(\idx{U^i}{A}^{\otimes2} \otimes \idi{$\tilde{\textnormal{A}}$}^{\otimes2}) \ (\idx{\xi}{A$\tilde{\textnormal{A}}$})^{\otimes2} \ ((\idx{U^i}{A}^\dagger)^{\otimes2} \otimes \idi{$\tilde{\textnormal{A}}$}^{\otimes2})}\right) (\tilde{\mathcal{T}}^{\dagger})^{\otimes2}[\idx{\mathcal{F}}{E}]\otimes (\tilde{\cE}^{\dagger})^{\otimes2}[\idx{\mathcal{F}}{R}]}\nonumber\\
&=\Trace{\left((\cG_W\otimes\opidi{$\tilde{\textnormal{A}}\tilde{\textnormal{A'}}$})(\idx{\xi}{A$\tilde{\textnormal{A}}$}^{\otimes2})-\left(\cG_H\otimes\opidi{$\tilde{\textnormal{A}}\tilde{\textnormal{A'}}$}\right)(\idx{\xi}{A$\tilde{\textnormal{A}}$}^{\otimes2})\right) (\tilde{\mathcal{T}}^{\dagger})^{\otimes2}[\idx{\mathcal{F}}{E}]\otimes (\tilde{\cE}^{\dagger})^{\otimes2}[\idx{\mathcal{F}}{R}]}\nonumber\\
&\quad+\Trace{\left(\cG_H\otimes\opidi{$\tilde{\textnormal{A}}\tilde{\textnormal{A'}}$}\right)(\idx{\xi}{A$\tilde{\textnormal{A}}$}^{\otimes2})\  (\tilde{\mathcal{T}}^{\dagger})^{\otimes2}[\idx{\mathcal{F}}{E}]\otimes (\tilde{\cE}^{\dagger})^{\otimes2}[\idx{\mathcal{F}}{R}]}\\
&\leq\Norm{\left((\cG_W\otimes\opidi{$\tilde{\textnormal{A}}\tilde{\textnormal{A'}}$})(\idx{\xi}{A$\tilde{\textnormal{A}}$}^{\otimes2})-\left(\cG_H\otimes\opidi{$\tilde{\textnormal{A}}\tilde{\textnormal{A'}}$}\right)(\idx{\xi}{A$\tilde{\textnormal{A}}$}^{\otimes2})\right) (\tilde{\mathcal{T}}^{\dagger})^{\otimes2}[\idx{\mathcal{F}}{E}]\otimes (\tilde{\cE}^{\dagger})^{\otimes2}[\idx{\mathcal{F}}{R}]}{1}\nonumber\\
&\quad+\Trace{\left(\cG_H\otimes\opidi{$\tilde{\textnormal{A}}\tilde{\textnormal{A'}}$}\right)(\idx{\xi}{A$\tilde{\textnormal{A}}$}^{\otimes2})\  (\tilde{\mathcal{T}}^{\dagger})^{\otimes2}[\idx{\mathcal{F}}{E}]\otimes (\tilde{\cE}^{\dagger})^{\otimes2}[\idx{\mathcal{F}}{R}]}\label{tterms}
\end{align}
The inequality is by the fact that the trace is given by the sum of the eigenvalues of some matrix in contrast to the Schatten 1-norm which is given by the sum of the absolute values of the eigenvalues. Thus the Schatten 1-norm of a matrix is always an upper bound on its trace.\\
The second term of equation \eqref{tterms} is calculated in an analogous way as in the proof of the original decoupling theorem. We postpone its evaluation to the end of our proof and first consider the term with the Schatten 1-norm. This term can be upper bounded with an application of \textit{Lemma 2}. We apply
\begin{align}
\Norm{ABC}{1}&\leq\Norm{A}{\infty}\Norm{B}{1}\Norm{C}{\infty}
\end{align}
with $A=\idi{AA'$\tilde{\textnormal{A}}\tilde{\textnormal{A'}}$}$ and find
\begin{align}
&\Norm{\left((\cG_W\otimes\opidi{$\tilde{\textnormal{A}}\tilde{\textnormal{A'}}$})(\idx{\xi}{A$\tilde{\textnormal{A}}$}^{\otimes2})-\left(\cG_H\otimes\opidi{$\tilde{\textnormal{A}}\tilde{\textnormal{A'}}$}\right)(\idx{\xi}{A$\tilde{\textnormal{A}}$}^{\otimes2})\right) (\tilde{\mathcal{T}}^{\dagger})^{\otimes2}[\idx{\mathcal{F}}{E}]\otimes (\tilde{\cE}^{\dagger})^{\otimes2}[\idx{\mathcal{F}}{R}]}{1}\nonumber\\
&\leq\Norm{\left((\cG_W\otimes\opidi{$\tilde{\textnormal{A}}\tilde{\textnormal{A'}}$})(\idx{\xi}{A$\tilde{\textnormal{A}}$}^{\otimes2})-\left(\cG_H\otimes\opidi{$\tilde{\textnormal{A}}\tilde{\textnormal{A'}}$}\right)(\idx{\xi}{A$\tilde{\textnormal{A}}$}^{\otimes2})\right)}{1}\Norm{ (\tilde{\mathcal{T}}^{\dagger})^{\otimes2}[\idx{\mathcal{F}}{E}]\otimes (\tilde{\cE}^{\dagger})^{\otimes2}[\idx{\mathcal{F}}{R}]}{\infty}\\
&=\Norm{\left((\cG_W\otimes\opidi{$\tilde{\textnormal{A}}\tilde{\textnormal{A'}}$})(\idx{\xi}{A$\tilde{\textnormal{A}}$}^{\otimes2})-\left(\cG_H\otimes\opidi{$\tilde{\textnormal{A}}\tilde{\textnormal{A'}}$}\right)(\idx{\xi}{A$\tilde{\textnormal{A}}$}^{\otimes2})\right)}{1}\Norm{ (\tilde{\mathcal{T}}^{\dagger})^{\otimes2}[\idx{\mathcal{F}}{E}]}{\infty}\Norm{(\tilde{\cE}^{\dagger})^{\otimes2}[\idx{\mathcal{F}}{R}]}{\infty}.\label{nnde}
\end{align}
The last equality is by the fact the the absolute value of the biggest eigenvalue of the tensor product is just the product of the absolute values of the biggest eigenvalues of the two components. By  the definition of the $\varepsilon$-almost 2-design and the definition of the diamond norm we have:
\begin{align}
\varepsilon&\geq\Norm{\cG_W-\cG_H}{\diamond}\\
&:=\sup_{\idx{d}{R}}{\max_{\rhoAR\in\linops{\idx{\cH}{AR}}}{\frac{\norm{(\cG_W\otimes\opidi{R})({\rhoAR})-(\cG_H\otimes\opidi{R})({\rhoAR})}{1}}{\norm{\rhoAR}{1}}}}\\
&\geq\frac{\Norm{\left((\cG_W\otimes\opidi{$\tilde{\textnormal{A}}\tilde{\textnormal{A'}}$})(\idx{\xi}{A$\tilde{\textnormal{A}}$}^{\otimes2})-\left(\cG_H\otimes\opidi{$\tilde{\textnormal{A}}\tilde{\textnormal{A'}}$}\right)(\idx{\xi}{A$\tilde{\textnormal{A}}$}^{\otimes2})\right)}{1}}{\Norm{\idx{\xi}{A$\tilde{\textnormal{A}}$}^{\otimes2}}{1}}\\
&\geq\frac{1}{4}\Norm{\left((\cG_W\otimes\opidi{$\tilde{\textnormal{A}}\tilde{\textnormal{A'}}$})(\idx{\xi}{A$\tilde{\textnormal{A}}$}^{\otimes2})-\left(\cG_H\otimes\opidi{$\tilde{\textnormal{A}}\tilde{\textnormal{A'}}$}\right)(\idx{\xi}{A$\tilde{\textnormal{A}}$}^{\otimes2})\right)}{1}\label{bndxi}
\end{align}
Inequality \eqref{bndxi} can be seen using the evident fact that ${\Norm{\idx{\xi}{A$\tilde{\textnormal{A}}$}^{\otimes2}}{1}}={\Norm{\idx{\xi}{A$\tilde{\textnormal{A}}$}}{1}}^2$ and by plugging in the definition $\idx{\xi}{A$\tilde{\textnormal{A}}$}:=\idx{\Phi}{A$\tilde{\textnormal{A}}$}-\idx{\pi}{A}\otimes\idx{\pi}{$\tilde{\textnormal{A}}$}$ into this expression. Using \eqref{bndxi} we obtain an upper bound for \eqref{nnde}:
\begin{align}
&\Norm{\left((\cG_W\otimes\opidi{$\tilde{\textnormal{A}}\tilde{\textnormal{A'}}$})(\idx{\xi}{A$\tilde{\textnormal{A}}$}^{\otimes2})-\left(\cG_H\otimes\opidi{$\tilde{\textnormal{A}}\tilde{\textnormal{A'}}$}\right)(\idx{\xi}{A$\tilde{\textnormal{A}}$}^{\otimes2})\right)}{1}\Norm{ (\tilde{\mathcal{T}}^{\dagger})^{\otimes2}[\idx{\mathcal{F}}{E}]}{\infty}\Norm{(\tilde{\cE}^{\dagger})^{\otimes2}[\idx{\mathcal{F}}{R}]}{\infty}\\
&\leq4\varepsilon\Norm{(\tilde{\mathcal{T}}^{\dagger})^{\otimes2}[\idx{\mathcal{F}}{E}]}{\infty}\Norm{(\tilde{\cE}^{\dagger})^{\otimes2}[\idx{\mathcal{F}}{R}]}{\infty}\label{zsmbstl}
\end{align}
Note that both terms with norms look almost identical so it is sufficient to find an upper bound for one of them only. We analyze the first term $\norm{(\tilde{\mathcal{T}}^{\dagger})^{\otimes2}[\idx{\mathcal{F}}{E}]}{\infty}$. Let $\idx{P}{AA'}^+$ be the projector corresponding to the biggest absolute eigenvalue of $(\tilde{\mathcal{T}}^{\dagger})^{\otimes2}[\idx{\mathcal{F}}{E}]$. Then the $\infty$-norm can be rewritten in the following way:
\begin{align}
\Norm{(\tilde{\mathcal{T}}^{\dagger})^{\otimes2}[\idx{\mathcal{F}}{E}]}{\infty}&=\Trace{\idx{P}{AA'}^+(\tilde{\mathcal{T}}^{\dagger})^{\otimes2}[\idx{\mathcal{F}}{E}]}\label{infnrmbnd}\\
&=\Trace{(\tilde{\mathcal{T}})^{\otimes2}[\idx{P}{AA'}^+]\idx{\mathcal{F}}{E}}
\end{align}
To be able to apply the swap trick and thus get rid of the operator $\cF_E$, we need to decompose $\idx{P}{AA'}^+$ into some basis: $\idx{P}{AA'}^+=\sum\limits_{i,j} c_{ij} \sigma^{A}_i \otimes \sigma^{A'}_j$. Without loss of generality we choose the coefficients $c_{ij}$ to be real. (For example, $\{\sigma^{A}_i\}_{i=1,...,d_A}$ might be an orthonormal basis of the space $\hermops{\idx{\cH}{A}}$.)  This gives:
\begin{align}
&\Trace{(\tilde{\mathcal{T}})^{\otimes2}[\idx{P}{AA'}^+]\idx{\mathcal{F}}{E}}\\
&=\sum\limits_{i,j} c_{ij}\Trace{(\tilde{\mathcal{T}}(\sigma^{A}_i)\otimes\tilde{\mathcal{T}}(\sigma^{A'}_j))\idx{\mathcal{F}}{E}}\\
&=\sum\limits_{i,j} c_{ij}\Trace{\tilde{\mathcal{T}}(\sigma^{A}_i)\tilde{\mathcal{T}}(\sigma^{A'}_j)}
\end{align}
We rewrite $\tilde{\mathcal{T}}(\sigma^{A}_i)$ using its Choi-Jamiolkowski representation.
\begin{align}
\sum\limits_{i,j} c_{ij}\Trace{(\tilde{\mathcal{T}}(\sigma^{A}_i)\tilde{\mathcal{T}}(\sigma^{A'}_j))}
&=\idx{d}{A}^2\sum\limits_{i,j}c_{ij}\Trace{\Ptrace{A}{\idx{\tilde{\omega}}{AE}\idi{E}\otimes\idx{\sigma^i}{A}^\intercal}\Ptrace{A'}{\idx{\tilde{\omega}}{A'E}\idi{E}\otimes\idx{\sigma^j}{A'}^\intercal}}\\
&=\idx{d}{A}^2\sum\limits_{i,j}c_{ij}\Trace{\idi{A'}\otimes(\idx{\tilde{\omega}}{AE}\idi{E}\otimes\idx{\sigma^i}{A}^\intercal)\idi{A}\otimes(\idx{\tilde{\omega}}{A'E}\idi{E}\otimes\idx{\sigma^j}{A'}^\intercal)}\\
&=\idx{d}{A}^2\sum\limits_{i,j}c_{ij}\Trace{(\idi{A'}\otimes\idx{\tilde{\omega}}{AE})\:(\idi{A}\otimes\idx{\tilde{\omega}}{A'E})\: (\idi{E}\otimes\idx{\sigma^i}{A}^\intercal\otimes\idx{\sigma^j}{A'}^\intercal)}\\
&=\idx{d}{A}^2\Trace{(\idi{A'}\otimes\idx{\tilde{\omega}}{AE})\:(\idi{A}\otimes\idx{\tilde{\omega}}{A'E})\: (\idi{E}\otimes(\idx{P}{AA'}^+)^\intercal)}\label{ttexpd}
\end{align}
Untill now no inequalities were used and the calculation following equation \eqref{infnrmbnd} is still exact. We need to find the entropies again. For this we introduce a basis $\{\sigma^{A}_i\}_{i=1,...,d_A}$ for $\hermops{\idx{\cH}{A}}$ and a basis $\{\sigma^{E}_i\}_{i=1,...,d_E}$ for $\hermops{\idx{\cH}{E}}$. Moreover we choose them to be orthonormal with respect to the scalar product:
\begin{align}
{\braket{\idx{\mu}{X}}{\idx{\nu}{X}}}_{\hermops{\idx{\cH}{X}}}\ :=\ \frac{1}{\idx{d}{X}}\ \Trace{\idx{\mu}{X}\cdot\idx{\nu}{X}}\qquad\forall\idx{\mu}{X},\idx{\nu}{X}\in\hermops{\idx{\cH}{X}}\label{scpr}
\end{align}
i.\:e.\:we have
\begin{align}
\frac{1}{\idx{d}{A}}\ \Trace{\sigma^{A}_i\sigma^{A}_j}=\delta_{ij}.
\end{align}
Now the product states $\{\sigma^{A}_i\otimes\sigma^{E}_j\}_{i=1,...,d_A;\: j=1,...,d_E}$ form an orthonormal basis for $\hermops{\idx{\cH}{AE}}$ with respect to the scalar product introduced in equation \eqref{scpr}:
\begin{align}
\frac{1}{\idx{d}{A}\idx{d}{E}}\ \Trace{\sigma^{A}_i\otimes\sigma^{E}_j\sigma^{A}_k\otimes\sigma^{E}_l}&=\frac{1}{\idx{d}{A}}\:\Trace{\sigma^{A}_i\sigma^{A}_k}\cdot\frac{1}{\idx{d}{E}}\:\Trace{\sigma^{E}_j\sigma^{E}_l}\\
&=\delta_{ik}\delta_{jl}
\end{align}
We now write the states $\idx{\tilde{\omega}}{AE}$, $\idx{\tilde{\omega}}{A'E}$ and $(\idx{P}{AA'}^+)^\intercal$ in that basis:
\begin{align}
\idx{\tilde{\omega}}{AE}\ &:=\ \sum\limits_{i,j}a_{ij}\sigma^{A}_i\otimes\sigma^{E}_j\qquad\wedge\qquad a_{ij}\ :=\ \frac{1}{\idx{d}{A}\idx{d}{E}}\ \Trace{\sigma^{A}_i\otimes\sigma^{E}_j\idx{\tilde{\omega}}{A'E}}\\
\idx{\tilde{\omega}}{A'E}\ &:=\ \sum\limits_{i,j}a_{ij}\sigma^{A'}_i\otimes\sigma^{E}_j\qquad\wedge\qquad a_{ij}\ :=\ \frac{1}{\idx{d}{A}\idx{d}{E}}\ \Trace{\sigma^{A}_i\otimes\sigma^{E}_j\idx{\tilde{\omega}}{A'E}}\\
(\idx{P}{AA'}^+)^\intercal\ &:=\ \sum\limits_{i,j}c_{ij}\sigma^{A}_i\otimes\sigma^{A'}_j\qquad\wedge\qquad c_{ij}\ :=\ \frac{1}{\idx{d}{A}\idx{d}{A}}\ \Trace{\sigma^{A}_i\otimes\sigma^{A'}_j(\idx{P}{AA'}^+)^\intercal}
\end{align}
Since all matrices in the above statements are hermitian the coefficients $a_{ij}$ and $c_{ij}$ are real. Moreover the coefficients in the expansion of $\idx{\tilde{\omega}}{AE}$ and $\idx{\tilde{\omega}}{A'E}$ are the same, because the corresponding matrices are the same.
Plugging in the expansions into equation \eqref{ttexpd} yields:
\begin{align}
&\idx{d}{A}^2\Trace{\idi{A'}\otimes\idx{\tilde{\omega}}{AE}\:\idi{A}\otimes\idx{\tilde{\omega}}{A'E}\: \idi{E}\otimes(\idx{P}{AA'}^+)^\intercal}\\
&=\idx{d}{A}^2\sum\limits_{i,j,k,l,m,n}{a_{ij}a_{kl}c_{mn}\Trace{\idi{A'}\otimes\sigma^{A}_i\otimes\sigma^{E}_j\              \idi{A}\otimes\sigma^{A'}_k\otimes\sigma^{E}_l\ \idi{E}\otimes\sigma^{A}_m\otimes\sigma^{A'}_n}}\\
&=\idx{d}{A}^2\sum\limits_{i,j,k,l,m,n}{a_{ij}a_{kl}c_{mn}\Trace{\sigma^{A}_i\sigma^{A}_m}\Trace{\sigma^{A'}_k\sigma^{A'}_n}\Trace{\sigma^{E}_j\sigma^{E}_l}}\\
&=\idx{d}{A}^4\idx{d}{E}\sum\limits_{i,j,k,l,m,n}{a_{ij}a_{kl}c_{mn}\delta_{im}\delta_{kn}\delta_{jl}}\\
&=\idx{d}{A}^4\idx{d}{E}\sum\limits_{i,j,k}{a_{ij}a_{kj}c_{ik}}\label{intrmat}
\end{align}
We now introduce the matrices:
\begin{align}
A&:=(a_{ij})\\
C&:=(c_{ij})
\end{align}
By the definition of the transpose of a matrix we have: $A^\intercal:=(a_{ji})$ for $A$ defined as above. Then \eqref{intrmat} becomes:
\begin{align}
\idx{d}{A}^4\idx{d}{E}\sum\limits_{i,j,k}{a_{ij}a_{kj}c_{ik}}&=\idx{d}{A}^4\idx{d}{E}\Trace{A^\intercal C A}\\
&=\idx{d}{A}^4\idx{d}{E}\ \Trace{AA^\intercal C}\\
&\leq \idx{d}{A}^4\idx{d}{E}\ \Norm{AA^\intercal C}{1}\\
&=\idx{d}{A}^4\idx{d}{E}\ \Norm{AA^\dagger C}{1}\label{rplcintcl}\\
&\leq \idx{d}{A}^4\idx{d}{E}\ \Norm{AA^\dagger}{1}\Norm{C}{\infty}\\
&\leq \idx{d}{A}^4\idx{d}{E}\ \Norm{AA^\dagger}{1}\Norm{C}{2}\\
&=\idx{d}{A}^4\idx{d}{E}\ \Trace{AA^\dagger}\Norm{C}{F}\label{frend}
\end{align}
The replacement of the transpose of $A$ with the hermitian conjugate $A^\dagger$ in \eqref{rplcintcl} is possible because of the fact that all entries of $A$ are real.
In equation \eqref{frend} we rewrote the Schatten 2-norm in terms of the Frobenius norm $\Norm{C}{F}=\sqrt{\sum_{ij}{\abs{c_{ij}}^2}}$ \cite{Bhatia} and used the fact that all eigenvalues of $AA^\dagger$ are real and positive.
Using the explicit formula for the coefficients $c_{ij}$ we calculate the Frobenius norm of $C$, where we use that all of its entries are real.
\begin{align}
\Norm{C}{F}^2&=\sum_{ij}{\abs{c_{ij}}^2}\\
&=\sum_{ij}{c_{ij}^2}\\
&=\frac{1}{\idx{d}{A}^4}\sum_{ij}{\Trace{\sigma^{A}_i\otimes\sigma^{A'}_j(\idx{P}{AA'}^+)^\intercal}\Trace{\sigma^{A}_i\otimes\sigma^{A'}_j(\idx{P}{AA'}^+)^\intercal}}\\
&=\frac{1}{\idx{d}{A}^4}\Trace{\left(\sum_{ij}{\Trace{\sigma^{A}_i\otimes\sigma^{A'}_j(\idx{P}{AA'}^+)^\intercal}\sigma^{A}_i\otimes\sigma^{A'}_j}\right)(\idx{P}{AA'}^+)^\intercal}\\
&=\frac{1}{\idx{d}{A}^2}\Trace{(\idx{P}{AA'}^+)^\intercal(\idx{P}{AA'}^+)^\intercal}\\
&=\frac{1}{\idx{d}{A}^2}\Trace{(\idx{P}{AA'}^+)^\intercal}\label{prjd1}\\
&=\frac{1}{\idx{d}{A}^2}\label{prjd2}
\end{align}
So we have
\begin{align}
\Norm{C}{F}&=\frac{1}{\idx{d}{A}}.\label{fronbnd}
\end{align}
In equation \eqref{prjd1} the fact that $(\idx{P}{AA'}^+)^\intercal$ is a projector was used. And in equation \eqref{prjd2} we exhibited the fact that $(\idx{P}{AA'}^+)^\intercal$ corresponds to some fixed eigenvalue of $(\tilde{\mathcal{T}}^{\dagger})^{\otimes2}[\idx{\mathcal{F}}{E}]$ and therefore has eigenvalues which are all zero beside one which is one.\\
The trace term in \eqref{frend} can be calculated similarly. We use the explicit formula for the coefficients:
\begin{align}
\Trace{AA^\dagger}&=\sum_{ij}{a_{ij}a_{ij}}\\
&=\frac{1}{\idx{d}{A}^2\idx{d}{E}^2}\sum_{ij}{\Trace{\sigma^{A}_i\otimes\sigma^{E}_j\idx{\tilde{\omega}}{A'E}}\Trace{\sigma^{A}_i\otimes\sigma^{E}_j\idx{\tilde{\omega}}{A'E}}}\\
&=\frac{1}{\idx{d}{A}^2\idx{d}{E}^2}\Trace{\left(\sum_{ij}\Trace{\sigma^{A}_i\otimes\sigma^{E}_j\idx{\tilde{\omega}}{A'E}}\sigma^{A}_i\otimes\sigma^{E}_j\right)\idx{\tilde{\omega}}{A'E}}\\
&=\frac{1}{\idx{d}{A}\idx{d}{E}}\Trace{\idx{\tilde{\omega}}{A'E}^2}
\end{align}
Plugging this expression together with \eqref{fronbnd} into equation \eqref{frend} yields:
\begin{align}
\idx{d}{A}^4\idx{d}{E}\sum\limits_{i,j,k}{a_{ij}a_{kj}c_{ik}}\leq\idx{d}{A}^2\Trace{\idx{\tilde{\omega}}{A'E}^2},
\end{align}
and with it the desired upper bound for $\Norm{(\tilde{\mathcal{T}}^{\dagger})^{\otimes2}[\idx{\mathcal{F}}{E}]}{\infty}$:
\begin{align}
\Norm{(\tilde{\mathcal{T}}^{\dagger})^{\otimes2}[\idx{\mathcal{F}}{E}]}{\infty}\leq\idx{d}{A}^2\Trace{\idx{\tilde{\omega}}{A'E}^2}.
\end{align}
An identical calculation reveals that
\begin{align}
\Norm{(\tilde{\cE}^{\dagger})^{\otimes2}[\idx{\mathcal{F}}{R}]}{\infty}\leq\idx{d}{A}^2\Trace{\idx{\tilde{\rho}}{AR}^2}.
\end{align}
Thus we finally obtain the bound for \eqref{tterms} using \eqref{zsmbstl}:
\begin{align}
&\Norm{\left((\cG_W\otimes\opidi{$\tilde{\textnormal{A}}\tilde{\textnormal{A'}}$})(\idx{\xi}{A$\tilde{\textnormal{A}}$}^{\otimes2})-\left(\cG_H\otimes\opidi{$\tilde{\textnormal{A}}\tilde{\textnormal{A'}}$}\right)(\idx{\xi}{A$\tilde{\textnormal{A}}$}^{\otimes2})\right) (\tilde{\mathcal{T}}^{\dagger})^{\otimes2}[\idx{\mathcal{F}}{E}]\otimes (\tilde{\cE}^{\dagger})^{\otimes2}[\idx{\mathcal{F}}{R}]}{1}\nonumber\\
&\leq 4\varepsilon\idx{d}{A}^4\Trace{\idx{\tilde{\omega}}{A'E}^2}\Trace{\idx{\tilde{\rho}}{AR}^2}
\end{align}
The only thing left to do is to evaluate the trace term in \eqref{tterms} but this is equivalent to proving the usual decoupling theorem in the way it was done in chapter 2. We have in accordance with the steps following \eqref{dfndgr} to \eqref{newwntmin}:
\begin{align}
&\Trace{\left(\cG_H\otimes\opidi{$\tilde{\textnormal{A}}\tilde{\textnormal{A'}}$}\right)(\idx{\xi}{A$\tilde{\textnormal{A}}$}^{\otimes2})\  (\tilde{\mathcal{T}}^{\dagger})^{\otimes2}[\idx{\mathcal{F}}{E}]\otimes (\tilde{\cE}^{\dagger})^{\otimes2}[\idx{\mathcal{F}}{R}]}\nonumber\\
&=\Trace{(\idx{\xi}{A$\tilde{\textnormal{A}}$}^{\otimes2})\int{[(\idx{U}{A})^{\dagger\otimes2}(\tilde{\mathcal{T}}^{\dagger})^{\otimes2}[\idx{\mathcal{F}}{E}](\idx{U}{A})^{\otimes2}]}\d{U} \otimes (\tilde{\cE}^{\dagger})^{\otimes2}[\idx{\mathcal{F}}{R}]}\\
&\leq\frac{1}{\tr{[\idx{\omega}{A'E}]}}\Trace{((\sigmaE^{-1/2}\otimes\idi{A'})\idx{\omega}{A'E})^2}\:\frac{1}{\tr{[\idx{\rho}{AR}]}}\Trace{((\idi{A}\otimes\zetaR^{-1/2})\idx{\rho}{AR})^2}\label{wntmin}
\end{align}
Adjusting the operators $\sigmaE$ and $\zetaR$ in a way such that the above expressions correspond to $\idx{H}{2}$-entropies, we could recapture after taking the square root and applying Jensen's inequality the decoupling theorem.\\
Here, we are interested in decoupling with almost 2-designs and thus we need to consider the first term of \eqref{tterms}, too. Taking the obtained upper bounds on the two terms together with the correct adjustment of $\sigmaE$ and $\zetaR$ yields
\begin{align}
&\sum_i{p_i{\left\| \mathcal{T}((\idx{U^i}{A} \otimes \idi{R}) \ \rhoAR \  (\idx{U^i}{A}^\dagger \otimes \idi{R})) - \idx{\omega}{E} \otimes \idx{\rho}{R} \right\|}_1}\nonumber\\
&\leq\sqrt{2^{-\chtwo{A'}{E}{\omega}\ -\ \chtwo{A}{R}{\rho}}+4\varepsilon\idx{d}{A}^4\:2^{-\chtwo{A'}{E}{\omega}\ -\ \chtwo{A}{R}{\rho}}}\\
&=\sqrt{1+4\varepsilon\idx{d}{A}^4}\ 2^{-\frac{1}{2}\:(\chtwo{A'}{E}{\omega}\ +\ \chtwo{A}{R}{\rho})},
\end{align}
which concludes the proof of the decoupling theorem with almost 2-designs.

\section{Analysis of the decoupling formula}
In this section we would like to shortly go back to our original motivation and analyze whether the discussed circuits do well in the sense of decoupling. From the fact that quantum circuits constitute approximate 2-designs we already now that in a many qubit system with dynamics described with the above quantum circuit, the different possible unitary evolutions are given by elements of an almost 2-designs. Since typical dynamics in nature are given by local two particle interactions and the corresponding circuits (as discussed in \cite{ETE}) are of the above type (but less general), we conclude that our model can roughly be used to describe the internal dynamics of a many qubit system. Moreover we know that in order to reach an $\varepsilon$-almost 2-design we need at least $t\geq C(n^2+n\log{\frac{1}{\varepsilon}})$ time steps in the circuit, with $C$ being some constant that only depends on the concrete circuit used.\\
To say that some process occurring on the system is ``decoupling'' we need to be sure that the required unitary is reached by the circuit in polynomial time only. This is certainly not true for any unitary with distribution according to Haar measure and thus the original decoupling theorem cannot be used to make a statement about this question. In contrast the derived decoupling formula states that there exists some process $W$ on the system that reaches decoupling in the sense that
\begin{align}
{{\left\| \mathcal{T}((\idx{W}{A} \otimes \idi{R}) \ \rhoAR \  (\idx{W}{A}^\dagger \otimes \idi{R})) - \idx{\omega}{E} \otimes \idx{\rho}{R} \right\|}_1}
\leq\sqrt{1+4\varepsilon\idx{d}{A}^4}\ 2^{-\frac{1}{2}\:(\chtwo{A'}{E}{\omega}\ +\ \chtwo{A}{R}{\rho})}.
\end{align}
This formula implies that the time that one has to wait until one reaches a certain quality of decoupling does not depend on the dimensions of the reference system $R$ and the output system $E$ of the channel $\cT$. Note, moreover, that the factor $\idx{d}{A}^4$ does not significantly increase the time that is required until decoupling is reached. To reach an $\bar{\varepsilon}$-approximate 2-design with $\bar{\varepsilon}:= \frac{\varepsilon}{\idx{d}{A}^4}$ the circuit requires at least a time $\bar{t}$ with $\bar{t}$ given by:
\begin{align}
\bar{t}&=C\left(n^2+n\log{\left(\frac{\idx{d}{A}^4}{\varepsilon}\right)}\right)\\
&=C\left(n^2+n\log{\left(\frac{2^{4n}}{\varepsilon}\right)}\right)
\end{align}
Assuming that $t$ is the minimum time required for the circuit to reach an $\varepsilon$-almost 2-design a short calculation reveals:
\begin{align}
\bar{t}&=C\left(n^2+4n^2+n\log{\left(\frac{1}{\varepsilon}\right)}\right)\\
&\leq C\left(5n^2+5n\log{\left(\frac{1}{\varepsilon}\right)}\right)\\
&=5t
\end{align}
This means that once the circuit has reached an $\varepsilon$-almost 2-design, one has to wait only five times longer until the circuits form an $\bar{\varepsilon}$-almost 2-design. This additional time certainly does not affect the physical realizability of the required unitary evolution.

\chapter{A smoothed version of the decoupling formula for $\varepsilon$-almost 2-designs}

As in the case of the original Decoupling Theorem, we would like to obtain a smoothed version of the decoupling formula with almost 2-designs i.\:e.\:we would like to replace the occurring $H_{\textnormal{2}}$-entropies using smooth $H_{\textnormal{min}}$-entropies.
This is done conceptually in exactly the same way in both cases. We therefore keep our discussions short and refer to \cite{Semthesis} Chapter 3.
First we rewrite the formula for almost 2-designs using the fact that $\chmin{A}{B}{\rho}\leq\chtwo{A}{B}{\rho}$ and then we use the unsmoothed formula to find its smoothed counterpart.\\
\\
\textbf{Theorem:} (Smoothed decoupling formula for $\varepsilon$-approximate 2-designs)\textit{
Let $\rhoAR\in\subnormstates{\idx{\cH}{A}\otimes\idx{\cH}{R}}$ be a sub normalized density operator and let $\cTAE$ be a completely positive linear map going from $\subnormstates{\idx{\cH}{A}\otimes\idx{\cH}{R}}$ to $\posops{\idx{\cH}{E} \otimes \idx{\cH}{R}}$ with Choi-Jamiolkowski representation $\idx{\omega}{A'E}\in\subnormstates{\idx{\cH}{E}\otimes\idx{\cH}{A'}}$ and let $\delta>0$ be small enough, then
\begin{align}
&\sum_{(p_i,U_i)\in\cD}{p_i\left\|\mathcal{T}((\idx{U^i}{A} \otimes \idi{R}) \idx{\rho}{AR} (\idx{(U^i)^{\dagger}}{A} \otimes \idi{R}))-\idx{\omega}{E}\otimes\idx{\rho}{R}\right\|}_1\nonumber\\
&\qquad\qquad\qquad\qquad\leq\sqrt{1+4\varepsilon\idx{d}{A}^4}\:2^{-\frac{1}{2}\:\chmineeps{\delta}{A'}{E}{\omega}-\frac{1}{2}\:\chmineeps{\delta}{A}{R}{\rho}}+8\idx{d}{A}\varepsilon\:\delta+12\delta\nonumber
\end{align}
where the summation goes over pairs $(p_i,U_i)$, such that $\cD$ constitutes an $\varepsilon$-approximate 2-design.
}\\
\\
For the proof let us introduce the state $\idx{\hat{\omega}}{A'E}\in\subnormstates{\idx{\cH}{A'E}}$ (which we sometimes denote shortly as $\hat{\omega}$) with the properties $\bar{P}(\idx{\omega}{A'E},\idx{\hat{\omega}}{A'E})\leq\delta$ and $\chmin{A'}{E}{\hat{\omega}} = \chmineeps{\delta}{A'}{E}{\omega}$. This state hits the bound in the definition of  $H_{\textnormal{min}}^\delta$ i.\:e.\:it maximizes $\chmin{A'}{E}{}$ over $\cB^{\delta}(\omega)$. Analogously $\idx{\hat{\rho}}{AR}$ is defined to be an operator with $\bar{P}(\idx{\hat{\rho}}{AR},\idx{\rho}{AR})\leq\delta$ and $\chmin{A}{R}{\hat{\rho}} = \chmineeps{\delta}{A}{R}{\rho}$.\\
Using the generalized Fuchs- van der Graaf inequalities (as in \cite{Semthesis} Chapter 3) we find that:
\begin{align}
\left\|\idx{\omega}{A'E}-\idx{\hat{\omega}}{A'E}\right\|_1\leq2\delta\ \wedge\ \left\|\idx{\rho}{AR}-\idx{\hat{\rho}}{AR}\right\|_1\leq2\delta\label{omrobnd}
\end{align}
Now we decompose $\hat{\omega}-\omega$ and $\hat{\rho}-\rho$ into positive operators with orthogonal support.
We write
\begin{align}
\hat{\omega}-\omega=\Delta_+-\Delta_-\ \wedge \ \hat{\rho}-\rho=\Gamma_+-\Gamma_-
\end{align}
and conclude from \eqref{omrobnd} that
\begin{align}
\Norm{\Delta_+}{1}\leq2\delta\ \wedge\ \Norm{\Delta_-}{1}\leq2\delta\ \wedge\ \Norm{\Gamma_+}{1}\leq2\delta\ \wedge\ \Norm{\Gamma_-}{1}\leq2\delta.
\end{align}
Moreover we introduce the completely positive maps $\hat{\cT}$, $\cD_+$ and $\cD_-$ which we define to be the unique Choi-Jamiolkowski preimages of  $\idx{\hat{\omega}}{A'E}$, $\Delta_+$ and $\Delta_-$ respectively.
Now we are in the position to apply the $\varepsilon$-almost decoupling theorem on $\hat{\rho}$ and $\hat{\omega}$ to find
\begin{align}
&\sqrt{1+4\varepsilon\idx{d}{A}^4}\:2^{-\frac{1}{2}\:\chmineeps{\delta}{A'}{E}{\omega}-\frac{1}{2}\:\chmineeps{\delta}{A}{R}{\rho}}\nonumber\\ &=\sqrt{1+4\varepsilon\idx{d}{A}^4}\:2^{-\frac{1}{2}\:\chmineeps{}{A'}{E}{\hat{\omega}}-\frac{1}{2}\:\chmineeps{}{A}{R}{\hat{\rho}}}\\
&\geq\sqrt{1+4\varepsilon\idx{d}{A}^4}\:2^{-\frac{1}{2}\:\chtwo{A'}{E}{\hat{\omega}}-\frac{1}{2}\:\chtwo{A}{R}{\hat{\rho}}}\\
&\geq\sum_{i}{p_i}{{\left\| \mathcal{\hat{T}}((\idx{U^i}{A} \otimes \idi{R}) \ \idx{\hat{\rho}}{AR} \  ((\idx{U^i}{A})^\dagger \otimes \idi{R})) - \idx{\hat{\omega}}{E} \otimes \idx{\hat{\rho}}{R} \right\|}_1}.
\end{align}
To obtain a smoothed version of the $\varepsilon$-almost decoupling formula we need to get rid of the hat-terms in the above expression using $\delta$-bounds only. This is realized with several applications of the triangle inequality. For any $i$, we have
\begin{align}
&{\left\| \mathcal{\hat{T}}((\idx{U^i}{A} \otimes \idi{R}) \ \idx{\hat{\rho}}{AR} \  ((\idx{U^i}{A})^\dagger \otimes \idi{R})) - \idx{\hat{\omega}}{E} \otimes \idx{\hat{\rho}}{R} \right\|}_1\nonumber\\
&\geq{\left\| \mathcal{\hat{T}}((\idx{U^i}{A} \otimes \idi{R}) \ \idx{\hat{\rho}}{AR} \  ((\idx{U^i}{A})^\dagger \otimes \idi{R})) - \idx{\omega}{E} \otimes \idx{\hat{\rho}}{R} \right\|}_1-{\left\|\idx{\omega}{A'E}-\idx{\hat{\omega}}{A'E}\right\|}_1\\
&\geq{\left\| \mathcal{\hat{T}}((\idx{U^i}{A} \otimes \idi{R}) \ \idx{\hat{\rho}}{AR} \  ((\idx{U^i}{A})^\dagger \otimes \idi{R})) - \idx{\omega}{E} \otimes \idx{\hat{\rho}}{R} \right\|}_1-2\delta.
\end{align}
In the same way $\idx{\hat{\rho}}{R}$ is eliminated from the product term and we get in total
\begin{align}
&{\left\| \mathcal{\hat{T}}((\idx{U^i}{A} \otimes \idi{R}) \ \idx{\hat{\rho}}{AR} \  ((\idx{U^i}{A})^\dagger \otimes \idi{R})) - \idx{\hat{\omega}}{E} \otimes \idx{\hat{\rho}}{R} \right\|}_1\nonumber\\
&\geq{\left\| \mathcal{\hat{T}}((\idx{U^i}{A} \otimes \idi{R}) \ \idx{\hat{\rho}}{AR} \  ((\idx{U^i}{A})^\dagger \otimes \idi{R})) - \idx{\omega}{E} \otimes \idx{\rho}{R} \right\|}_1-4\delta.
\end{align}
Still $\hat{\cT}$ and $\idx{\hat{\rho}}{AR}$ occur in the above expression. To recover the original expression, over which the summation takes place, we need to rewrite the above using $\cT$ and $\rhoAR$ only. This is achieved with two further applications of the triangle inequality.
\begin{align}
&{\left\| \hat{\mathcal{T}}((\idx{U^i}{A} \otimes \idi{R}) \ \idx{\hat{\rho}}{AR} \  (\idx{(U^i)^{\dagger}}{A} \otimes \idi{R})) -\idx{\omega}{E}\otimes\idx{\rho}{R}\right\|}_1-4\delta\nonumber\\
&\geq {\left\|\mathcal{T}((\idx{U^i}{A} \otimes \idi{R}) \idx{\hat{\rho}}{AR} (\idx{(U^i)^{\dagger}}{A} \otimes \idi{R}))-\idx{\omega}{E}\otimes\idx{\rho}{R}\right\|}_1\nonumber\\
&\quad- {\left\|\hat{\mathcal{T}}((\idx{U^i}{A} \otimes \idi{R}) \idx{\hat{\rho}}{AR} (\idx{(U^i)^{\dagger}}{A} \otimes \idi{R}))-\mathcal{T}((\idx{U^i}{A} \otimes \idi{R}) \idx{\hat{\rho}}{AR} (\idx{(U^i)^{\dagger}}{A} \otimes \idi{R}))\right\|}_1-4\delta\\
&\geq{\left\|\mathcal{T}((\idx{U^i}{A} \otimes \idi{R}) \idx{\rho}{AR} (\idx{(U^i)^{\dagger}}{A} \otimes \idi{R}))-\idx{\omega}{E}\otimes\idx{\rho}{R}\right\|}_1\nonumber\\
&\quad -{\left\|\mathcal{T}((\idx{U^i}{A} \otimes \idi{R}) \idx{\rho}{AR} (\idx{(U^i)^{\dagger}}{A} \otimes \idi{R}))-\mathcal{T}((\idx{U^i}{A} \otimes \idi{R}) \idx{\hat{\rho}}{AR} (\idx{(U^i)^{\dagger}}{A} \otimes \idi{R}))\right\|}_1\nonumber\\
&\quad- {\left\|\hat{\mathcal{T}}((\idx{U^i}{A} \otimes \idi{R}) \idx{\hat{\rho}}{AR} (\idx{(U^i)^{\dagger}}{A} \otimes \idi{R}))-\mathcal{T}((\idx{U^i}{A} \otimes \idi{R}) \idx{\hat{\rho}}{AR} (\idx{(U^i)^{\dagger}}{A} \otimes \idi{R}))\right\|}_1-4\delta\label{finbrg}
\end{align}
The first term of equation \eqref{finbrg} corresponds to the unsmoothened decoupling formula. For the remaining two terms
\begin{align}
\sum_{i}{p_i{\left\|\mathcal{T}((\idx{U^i}{A} \otimes \idi{R}) \idx{\rho}{AR} (\idx{(U^i)^{\dagger}}{A} \otimes \idi{R}))-\mathcal{T}((\idx{U^i}{A} \otimes \idi{R}) \idx{\hat{\rho}}{AR} (\idx{(U^i)^{\dagger}}{A} \otimes \idi{R}))\right\|}_1}\label{frstsma}
\end{align} and
\begin{align}
\sum_{i}{p_i{\left\|\hat{\mathcal{T}}((\idx{U^i}{A} \otimes \idi{R}) \idx{\hat{\rho}}{AR} (\idx{(U^i)^{\dagger}}{A} \otimes \idi{R}))-\mathcal{T}((\idx{U^i}{A} \otimes \idi{R}) \idx{\hat{\rho}}{AR} (\idx{(U^i)^{\dagger}}{A} \otimes \idi{R}))\right\|}_1}\label{scndsma}
\end{align}
we need to find upper bounds. We treat them separately beginning with the first one.
To preform the calculation we write $\hat{\rho}-\rho=\Gamma_+-\Gamma_-$ and use the linearity of $\cT$. We get
\begin{align}
&\sum_{i}{p_i{\left\|\mathcal{T}((\idx{U^i}{A} \otimes \idi{R}) \idx{\rho}{AR} (\idx{(U^i)^{\dagger}}{A} \otimes \idi{R}))-\mathcal{T}((\idx{U^i}{A} \otimes \idi{R}) \idx{\hat{\rho}}{AR} (\idx{(U^i)^{\dagger}}{A} \otimes \idi{R}))\right\|}_1}\\
&=\sum_{i}{p_i{\left\|\mathcal{T}((\idx{U^i}{A} \otimes \idi{R}) \Gamma_+ (\idx{(U^i)^{\dagger}}{A} \otimes \idi{R}))-\mathcal{T}((\idx{U^i}{A} \otimes \idi{R}) \Gamma_- (\idx{(U^i)^{\dagger}}{A} \otimes \idi{R}))\right\|}_1}\\
&\leq\sum_{a\in\{+,-\}}{\sum_{i}{p_i{\left\|\mathcal{T}((\idx{U^i}{A} \otimes \idi{R}) \Gamma_a (\idx{(U^i)^{\dagger}}{A} \otimes \idi{R}))\right\|}_1}}\\
&=\sum_{a\in\{+,-\}}{\Trace{\mathcal{T}\left(\sum_{i}{p_i\idx{U^i}{A} \otimes \idi{R}\: \Gamma_a\: \idx{(U^i)^{\dagger}}{A} \otimes \idi{R}}-\int_{\mathbb{U}}{\idx{U}{A} \otimes \idi{R}\:\Gamma_a\:\idx{U^{\dagger}}{A} \otimes \idi{R}}\d{U}\right)}}\nonumber\\
&\quad+\sum_{a\in\{+,-\}}{\Trace{\cT\Big(\int_{\mathbb{U}}{\idx{U}{A} \otimes \idi{R}\:\Gamma_a\:\idx{U^{\dagger}}{A} \otimes \idi{R}}\d{U}\Big)}}\\
&\leq\sum_{a\in\{+,-\}}{\left\|\sum_{i}{p_i\idx{U^i}{A} \otimes \idi{R}\: \Gamma_a\: \idx{(U^i)^{\dagger}}{A} \otimes \idi{R}}-\int_{\mathbb{U}}{\idx{U}{A} \otimes \idi{R}\:\Gamma_a\:\idx{U^{\dagger}}{A} \otimes \idi{R}}\d{U}\right\|_1}{\left\|\cT^\dagger(\idi{E})\right\|}_{\infty}\nonumber\\
&\quad+\sum_{a\in\{+,-\}}\Trace{\mathcal{T}(\idx{\pi}{A})\otimes\ptr{A}{\Gamma_a}}\\
&\leq\sum_{a\in\{+,-\}}{\varepsilon\:\left\|\Gamma_a\right\|_1}{\left\|\cT^\dagger(\idi{E})\right\|}_{\infty} +\sum_{a\in\{+,-\}}\Trace{\idx{\omega}{A'E}}\Trace{\Gamma_a}.
\end{align}
In the last inequality we used the property of the almost 2-design that it constitutes an almost 1-design automatically. This can be seen straight from the definition considering states that are given by the identity operator on one of the systems on which the unitaries act. Now choose the eigenvalue of $\cT^\dagger(\idi{E})$ which is biggest in absolute value. Let $\idx{P}{A}$ be the projector corresponding to this eigenvalue. Note, moreover, that $\trace{\Gamma_a}\leq2\delta$ as we have already shown. Using all this, we get
\begin{align}
&\sum_{i}{p_i{\left\|\mathcal{T}((\idx{U^i}{A} \otimes \idi{R}) \idx{\rho}{AR} (\idx{(U^i)^{\dagger}}{A} \otimes \idi{R}))-\mathcal{T}((\idx{U^i}{A} \otimes \idi{R}) \idx{\hat{\rho}}{AR} (\idx{(U^i)^{\dagger}}{A} \otimes \idi{R}))\right\|}_1}\nonumber\\
&\leq4\varepsilon\:\delta{\left\|\cT^\dagger(\idi{E})\right\|}_{\infty}+4\delta\\
&\leq4\varepsilon\:\delta\:\Trace{\cT(\idx{P}{A})}+4\delta\\
&=4\idx{d}{A}\varepsilon\delta\:\Trace{\idx{\omega}{A'E}\:(\idx{P}{A})^\intercal\otimes\idi{E}}+4\delta\\
&\leq4\idx{d}{A}\varepsilon\delta\:\Trace{\idx{\omega}{A'E}}\:\Norm{(\idx{P}{A})^\intercal\otimes\idi{E}}{\infty}+4\delta\\
&\leq4\idx{d}{A}\varepsilon\delta+4\delta.\label{rstabsch}
\end{align}
The last step makes use of the fact that $(\idx{P}{A})^\intercal$ being a projector has positive eigenvalues smaller or equal than one. 
For the evaluation of the second term \eqref{scndsma} we decompose $\hat{\cT}-\cT\: =\: \cD_+-\cD_-$ in accordance with the above decomposition $\hat{\omega}-\omega=\Delta_+-\Delta_-$. We then get
\begin{align}
&\sum_{i}{p_i{\left\|\hat{\mathcal{T}}((\idx{U^i}{A} \otimes \idi{R}) \idx{\hat{\rho}}{AR} (\idx{(U^i)^{\dagger}}{A} \otimes \idi{R}))-\mathcal{T}((\idx{U^i}{A} \otimes \idi{R}) \idx{\hat{\rho}}{AR} (\idx{(U^i)^{\dagger}}{A} \otimes \idi{R}))\right\|}_1}\nonumber\\
&=\sum_i{{p_i\left\|\left(\cD_+-\cD_-\right)(\idx{U^i}{A} \otimes \idi{R}\:\idx{\hat{\rho}}{AR}\:\idx{(U^i)^{\dagger}}{A} \otimes \idi{R})\right\|}_1}\\
&\leq\sum_{a\in\{+,-\}}{\sum_i{{p_i\left\|\cD_a(\idx{U^i}{A} \otimes \idi{R}\:\idx{\hat{\rho}}{AR}\:\idx{(U^i)^{\dagger}}{A} \otimes \idi{R})\right\|}_1}}\\
&\leq\sum_{a\in\{+,-\}}{\Trace{\cD_a\Big(\sum_i{p_i\:(\idx{U^i}{A} \otimes \idi{R})\:\idx{\hat{\rho}}{AR}\:(\idx{(U^i)^{\dagger}}{A} \otimes \idi{R})}\Big)}}\\
&=\sum_{a\in\{+,-\}}{\Trace{\cD_a(\sum_i{p_i\idx{U^i}{A} \otimes \idi{R}\:\idx{\hat{\rho}}{AR}\:\idx{(U^i)^{\dagger}}{A} \otimes \idi{R}}-\int_{\mathbb{U}}{\idx{U}{A} \otimes \idi{R}\:\idx{\hat{\rho}}{AR}\:\idx{U^{\dagger}}{A} \otimes \idi{R}}\d{U})}}\nonumber\\
&+\sum_{a\in\{+,-\}}{\Trace{\cD_a(\int_{\mathbb{U}}{\idx{U}{A} \otimes \idi{R}\:\idx{\hat{\rho}}{AR}\:\idx{U^{\dagger}}{A} \otimes \idi{R}}\d{U})}}\\
&\leq\sum_{a\in\{+,-\}}{\left\|\sum_i{p_i\idx{U^i}{A} \otimes \idi{R}\idx{\hat{\rho}}{AR}\idx{(U^i)^{\dagger}}{A} \otimes \idi{R}}-
\int_{\mathbb{U}}{\idx{U}{A} \otimes \idi{R}\idx{\hat{\rho}}{AR}\idx{U^{\dagger}}{A} \otimes \idi{R}}\d{U}\right\|}_1{\left\|\cD_a^\dagger(\idi{E})\right\|}_{\infty}\nonumber\\
&+\sum_{a\in\{+,-\}}{\Trace{\cD_a(\idx{\pi}{A}\otimes\idx{\hat{\rho}}{R})}}\\
&\leq\sum_{a\in\{+,-\}}\varepsilon\:{\big\|\idx{\hat{\rho}}{AR}\big\|}_{1}\:{\left\|\cD_a^\dagger(\idi{E})\right\|}_{\infty}
+\sum_{a\in\{+,-\}}{\Trace{\Delta_a\otimes\idx{\hat{\rho}}{R}}}.\label{leqedsg}
\end{align}
Again, we used the fact that an almost 2-design \eqref{leqedsg} is an almost 1-design automatically. Thus we can use the bound with the diamond norm here, too. Furthermore, note that $\idx{\hat{\rho}}{AR}$ is sub-normalized by definition and that we found that $\Norm{\Delta_a}{1}\leq2\delta$. We define $\idx{P}{A}^a$ to be the projector corresponding to the biggest absolute eigenvalue of $\cD_a^\dagger(\idi{E})$ and we get that
\begin{align}
&\sum_{i}{p_i{\left\|\hat{\mathcal{T}}((\idx{U^i}{A} \otimes \idi{R}) \idx{\hat{\rho}}{AR} (\idx{(U^i)^{\dagger}}{A} \otimes \idi{R}))-\mathcal{T}((\idx{U^i}{A} \otimes \idi{R}) \idx{\hat{\rho}}{AR} (\idx{(U^i)^{\dagger}}{A} \otimes \idi{R}))\right\|}_1}\nonumber\\
&\leq\sum_{a\in\{+,-\}}\varepsilon\:{\left\|\cD_a^\dagger(\idi{E})\right\|}_{\infty}+4\delta\\
&=\sum_{a\in\{+,-\}}\varepsilon\:\Trace{\idx{P}{A}^a\:\cD_a^\dagger(\idi{E})}+4\delta\\
&=\sum_{a\in\{+,-\}}\varepsilon\:\Trace{\cD_a(\idx{P}{A}^a)}+4\delta\\
&=\idx{d}{A}\sum_{a\in\{+,-\}}\varepsilon\:\Trace{\Delta_a\:\right((\idx{P}{A}^a)^\intercal\otimes\idi{E}\left)}+4\delta\\
&\leq\idx{d}{A}\sum_{a\in\{+,-\}}{\varepsilon \Norm{\Delta_a}{1}\Norm{(\idx{P}{A}^a)^\intercal\otimes\idi{E}}{\infty}}+4\delta\\
&=4\idx{d}{A}\varepsilon\delta +4\delta.\label{scndtrm}
\end{align}
In the last step we used the fact that the biggest eigenvalue of $(\idx{P}{A}^a)^\intercal$ is one.\\
Taking together the expressions \eqref{rstabsch} and \eqref{scndtrm} and plugging them into \eqref{finbrg}, we obtain
\begin{align}
&\sum_{i}{p_i\left\| \mathcal{\hat{T}}((\idx{U^i}{A} \otimes \idi{R}) \ \idx{\hat{\rho}}{AR} \  ((\idx{U^i}{A})^\dagger \otimes \idi{R})) - \idx{\hat{\omega}}{E} \otimes \idx{\hat{\rho}}{R} \right\|}_1\nonumber\\
&\geq\sum_{i}{p_i\left\|\mathcal{T}((\idx{U^i}{A} \otimes \idi{R}) \idx{\rho}{AR} (\idx{(U^i)^{\dagger}}{A} \otimes \idi{R}))-\idx{\omega}{E}\otimes\idx{\rho}{R}\right\|}_1-4\idx{d}{A}\varepsilon\delta-4\delta-4\idx{d}{A}\varepsilon\delta-4\delta-4\delta.
\end{align}
Finally this yields 
\begin{align}
&\sum_{i}{p_i\left\|\mathcal{T}((\idx{U^i}{A} \otimes \idi{R}) \idx{\rho}{AR} (\idx{(U^i)^{\dagger}}{A} \otimes \idi{R}))-\idx{\omega}{E}\otimes\idx{\rho}{R}\right\|}_1\nonumber\\
&\leq\sqrt{1+4\varepsilon\idx{d}{A}^4}\:2^{-\frac{1}{2}\:\chmineeps{\delta}{A'}{E}{\omega}-\frac{1}{2}\:\chmineeps{\delta}{A}{R}{\rho}}+8\idx{d}{A}\varepsilon\:\delta+12\delta,
\end{align}
which proves the smoothed decoupling formula for $\varepsilon$-approximate 2-designs.
\chapter{A classical analogue of the decoupling formula}
The decoupling theorem bounds the average distance of some quantum state after a random unitary operation and some quantum channel have been applied to it from a fully uncorrelated state. Generally a unitary operation on a quantum system $A$ corresponds to a quantum-mechanical evolution of this system. Thus the decoupling theorem is only relevant under the assumption that the evolution is governed by the laws of quantum mechanics. In this and the following chapters we would like to understand in how far classical operations can be used for decoupling purposes. Instead of averaging over the group of all unitary matrices $\mathbb{U}(A)$, we restrict the problem to the case where only classical operations are allowed, i.\:e.\:we average over the group of all permutation operators on the system $A$, $\mathbb{P}(A)$. This corresponds physically to a situation where a system is subject to an evolution which is purely classical. Or, from the point of view of the computational sciences, we restrict our analysis to operations that can be performed on a classical computer in contrast to a quantum computer. While it turns out that this problem is difficult to solve in the general case, several special cases of particular interest will be studied in the sequel. In this chapter we assume that the state of the $A$ system is classical, but still we allow that it has correlations with a reference quantum-system $R$. A typical situation, where this is the case is that the $A$ system represents the outcome of a random variable and an adversary possesses quantum side information about $A$. The question we are interested in is whether using some classical operation it is possible to extract from $A$ a part that is almost completely unknown to the adversary. Our discussion will give results strongly resembling the ``General Leftover Hash Lemma'' as derived in \cite{Renner:PHD}, which has large-scale implications in the area of quantum cryptography.

\section{A ``classicalized'' Decoupling Lemma }
The aim of this section is to derive a decoupling lemma as in Section 2.1 for our new setup, where the integration over the unitary group is replaced by an average over all permutation operators and the quantum state under consideration has CQ-structure. For completeness we shortly recover some terminology and basic definitions:\\
The \textit{symmetric group} $S_n$ is the set of all bijections of $\{1,...,n\}$ to itself together with the concatenation of maps as the group multiplication. Elements $\pi$ of $S_n$ are called \textit{permutations}.
\begin{definition}(Permutation operator \cite{SymGroup}) Let  $\idx{\cH}{A}$ be a Hilbert space together with a fixed  basis $\{\ket{i}\}_{i=1,...,d_A}$. For $\pi$ in $S_{\idx{d}{A}}$, we define the permutation operator $P(\pi)$ to be the operator which has the matrix representation $(P_{ij})$, where
\[(P_{ij})=\begin{cases}
1 & \text{if}\ \pi(j)=i\\
0& \text{otherwise}
\end{cases}\]
with respect to the given basis.
\end{definition}
For a given Hilbert space $\idx{\cH}{A}$, we denote by $\mathbb{P}(A)$ the set of all permutation operators for some fixed basis. This set has $\idx{d}{A}!$ elements. Since the above definition includes a group homomorphism from $S_{\idx{d}{A}}$ to the group of automorphisms of $\idx{\cH}{A}$, it defines a representation (\cite{SymGroup}) of the symmetric group $S_{\idx{d}{A}}$ called the \textit{defining representation}.\\
Furthermore we define what is meant by a CQ-state.
\begin{definition}(CQ-state \cite{Renner:Skript}) Let $\idx{\cH}{A}$ and $\idx{\cH}{R}$ be Hilbert spaces and let $\{\ket{i}\}_{i=1,...,d_A}$ be a fixed orthonormal basis of $\idx{\cH}{A}$. A density operator $\rhoAR\in\subnormstates{\idx{\cH}{A}\otimes\idx{\cH}{R}}$ is said to be classical on $\idx{\cH}{A}$ with respect to the basis $\{\ket{i}\}_{i=1,...,d_A}$ if
$$\rhoAR\in span\{\proj{i}{i}_{i=1,...,d_A}\}\otimes\hermops{\idx{\cH}{R}}.$$
If in addition $\rhoAR$ is non-classical on $\idx{\cH}{R}$, we call it a hybrid classical-quantum state or shortly CQ-state.
\end{definition}
Typically CQ-states are used in situations, where the quantum state of some system (here $R$) is not known with certainty. Instead it depends on the outcome of some classical random event (described by system $A$ in our case). Note that a CQ-state $\rhoAR$ can be written in the basis $\{\ket{i}\}_{i=1,...,d_A}$ introduced in the definition to be
\begin{align}
\rhoAR=\sum_{i}\proji{i}{i}{A}\otimes\rhoR^{[i]},
\end{align}
where the $\rhoR^{[i]}$ are sub-normalized density operators with $\sum_{i}{\rhoR^{[i]}}=\rhoR$.\\
Finally for a linear map $\cTAE\in\textnormal{Hom}(\linops{\HA},\linops{\HE})$, we define its classicalized version $\cTAE^{cl}$.
\begin{definition}(Classicalized map) 
Let $\idx{\cH}{A}$ be a Hilbert space with a fixed orthonormal basis $\{\ket{i}\}_{i=1,...,d_A}$ and let $\cTAE$ be a linear map in $\textnormal{Hom}(\linops{\HA},\linops{\HE})$. We define the map $\cTAE^{cl}\in\textnormal{Hom}(\linops{\HA},\linops{\HE})$ by
$$\cTAE^{cl}(\rhoA):=\cTAE\left(\sum_{i}{\proji{i}{i}{A}\rhoA\proji{i}{i}{A}}\right)\quad\forall\rhoA\in\linops{\HA}$$
and call $\cTAE^{cl}$ the classicalized version of $\cTAE$.
\end{definition}
The Choi-Jamiolkowski representation of $\cTAE^{cl}$ will be denoted by $\idx{\omega}{A'E}^{cl}$, i.\:e.\: 
\begin{align}
\idx{\omega}{A'E}^{cl}&=\cTAE^{cl}(\idx{\Phi}{AA'})\\
&=\cTAE\left(\sum_{i}{\proji{i}{i}{A}\idx{\Phi}{AA'}\proji{i}{i}{A}}\right)\\
&=\cTAE\left(\frac{1}{\idx{d}{A}}\sum_{i}{\proji{i}{i}{A}\otimes\proji{i}{i}{A'}}\right)\label{cqcjr}\\
&=:\cTAE(\idx{T}{AA'})\label{intrT},
\end{align}
where in the last equation we introduced the state $\idx{T}{AA'}:=\frac{1}{\idx{d}{A}}\sum_{i}{\proji{i}{i}{A}\otimes\proji{i}{i}{A'}}$, which includes the classical correlations between the systems $A$ and $A'$. From equation \eqref{cqcjr} one can see that the Choi-Jamiolkowski representation of a classicalized map always has CQ-structure. A short calculation reveals that 
\begin{align}
\idx{\omega}{E}=\idx{\omega}{E}^{cl}\label{clecl}.
\end{align}
The term ``classicalized map'' is motivated by the fact that applying such a map to a state $\rhoAR$ always produces a CQ-state. Moreover it does not matter whether one applies a map $\cTAE$ or its classicalized version $\cTAE^{cl}$ on a CQ-state. We get assuming that $\rhoAR$ has CQ-structure that
\begin{align}
\cTAE^{cl}(\rhoAR)&=\cTAE\left(\sum_{j}{\proji{j}{j}{A}\left(\sum_{i}\proji{i}{i}{A}\otimes\rhoR^{[i]}\right)\proji{j}{j}{A}}\right)\\
&=\cTAE(\rhoAR)\label{clcqinv}.
\end{align}
In fact this is intuitively clear. The state $\rhoAR$ is already classical on $A$; further classicalization of this state has no effect anymore.
We are now in the position to state the decoupling lemma for CQ-states.\\
\\
\textbf{Lemma:} (Decoupling Lemma for CQ-states)\textit{
Let $\rhoAR$ be classical on $\HA$ with respect to $\{\ket{i}\}_{i=1,...,d_A}$ and let $\cTAE\in\textnormal{Hom}(\linops{\HA},\linops{\HE})$ be a linear map with Choi-Jamiolkowski representation $\idx{\omega}{A'E}\in\hermops{\idx{\cH}{E}\otimes\idx{\cH}{A'}}$, then
\begin{align}
\frac{1}{\idx{d}{A}!}\sum_{\idx{P}{A}\in\mathbb{P}(A)} &{\left\| \mathcal{T}(\idx{P}{A} \otimes \idi{R} \ \rhoAR \  \idx{P}{A}^\dagger \otimes \idi{R}) - \idx{\omega}{E} \otimes \idx{\rho}{R} \right\|}_2^2 \nonumber\\
&=\frac{\idx{d}{A}^2}{\idx{d}{A}-1}\:{\left\|\idx{\rho}{AR}-\idx{\pi}{A}\otimes\rhoR\right\|}_2^2\: {\left\|\idx{\omega}{A'E}^{cl}-\idx{\pi}{A'}\otimes\idx{\omega}{E}^{cl}\right\|}_2^2,\nonumber
\end{align}
where the summation goes over all permutation operators, which act by permuting the basis vectors of $\{\ket{i}\}_{i=1,...,d_A}$.
}\\
\\
Since conjugating $\rhoAR$ with a permutation operator acting on $A$ does not affect its CQ-structure, one can apply equation \eqref{clcqinv} directly to the statement of the decoupling lemma for CQ-states. Together with equation \eqref{clecl} this results in
\begin{align}
&\frac{1}{\idx{d}{A}!}\sum_{\idx{P}{A}\in\mathbb{P}(A)} {\left\| \mathcal{T}(\idx{P}{A} \otimes \idi{R} \ \rhoAR \  \idx{P}{A}^\dagger \otimes \idi{R}) - \idx{\omega}{E} \otimes \idx{\rho}{R} \right\|}_2^2\nonumber\\
&=\frac{1}{\idx{d}{A}!}\sum_{\idx{P}{A}\in\mathbb{P}(A)} {\left\| \mathcal{T}^{cl}(\idx{P}{A} \otimes \idi{R} \ \rhoAR \  \idx{P}{A}^\dagger \otimes \idi{R}) - \idx{\omega}{E}^{cl} \otimes \idx{\rho}{R} \right\|}_2^2,
\end{align}
which explains the occurrence of the Choi-Jamiolkowski representation of $\cTAE^{cl}$ on the right hand side of the lemma.\\
The rest of this section is devoted to a proof of the Decoupling Lemma for CQ-states. This proof will be organized in three subsections. In the first and second we proof two general claims about the action of permutation operators and apply them in the third subsection to conclude the proof.
\subsection{Counting permutation operators}
The following claim will be useful in our proof. It shows how twofold tensor products of permutation operators act on classical states.\\
\\
\textbf{Claim 1:} Let $\{\ket{i}\}_{i=1,...,d_A}$ with $\idx{d}{A}\geq2$ be some basis and $\mathbb{P}(A)$ be the corresponding set of permutation operators. Then for any $i,j$
\begin{align}
\frac{1}{\idx{d}{A}!}\sum_{\idx{P}{A}\in\mathbb{P}(A)}{ (\idx{P}{A}\otimes\idx{P}{A'})\:(\proji{i}{i}{A}\otimes\proji{j}{j}{A'})\:(\idx{P}{A}^\dagger\otimes\idx{P}{A}^\dagger)}
=\frac{1-\delta_{ij}}{\idx{d}{A}^2-\idx{d}{A}}\idi{AA'}-\frac{1-\delta_{ij}}{\idx{d}{A}-1}\idx{T}{AA'}+\delta_{ij}\idx{T}{AA'}\nonumber.
\end{align}
\\
This formula can be seen as follows. Consider first the case when $i=j$ then
\begin{align}
\frac{1}{\idx{d}{A}!}\sum_{\idx{P}{A}\in\mathbb{P}(A)}{ (\idx{P}{A}\otimes\idx{P}{A'})\:(\proji{i}{i}{A}\otimes\proji{i}{i}{A'})\:(\idx{P}{A}^\dagger\otimes\idx{P}{A'}^\dagger)}
&=\frac{(\idx{d}{A}-1)!}{\idx{d}{A}!}\sum_{k}{\proji{k}{k}{A}\otimes\proji{k}{k}{A'}}\label{explvrltl}\\
&=\idx{T}{AA'},
\end{align}
The crucial step in equation \eqref{explvrltl} can be seen from an easy counting argument. Since the permutations are bijective maps for any $\ket{j},\ket{k}\in\{\ket{i}\}_{i=1,...,d_A}$ there are $(\idx{d}{A}-1)!$ permutation operators with $P\ket{j}=\ket{k}$. Thus when summing over all permutation operators as in equation \eqref{explvrltl} for any fixed $\ket{i}$ every basis vector in $\{\ket{i}\}_{i=1,...,d_A}$ contributes $(\idx{d}{A}-1)!$ times to the whole sum.
The above implies that Claim 1 is valid for $i=j$. Now assume that $i\neq j$. For fixed $k\neq l$ there are $(\idx{d}{A}-2)!$ permutation operators that map $\proji{i}{i}{A}\otimes\proji{j}{j}{A'}$ to $\proji{k}{k}{A}\otimes\proji{l}{l}{A'}$ but there is no permutation operator mapping $\proji{i}{i}{A}\otimes\proji{j}{j}{A'}$ to $\proji{k}{k}{A}\otimes\proji{l}{l}{A'}$ with $k=l$. We conclude that in the case $i\neq j$ we have
\begin{align}
&\frac{1}{\idx{d}{A}!}\sum_{\idx{P}{A}\in\mathbb{P}(A)}{ (\idx{P}{A}\otimes\idx{P}{A'})\:(\proji{i}{i}{A}\otimes\proji{j}{j}{A'})\:(\idx{P}{A}^\dagger\otimes\idx{P}{A}^\dagger)}\nonumber\\
&=\frac{(\idx{d}{A}-2)!}{\idx{d}{A}!}\sum_{k\neq l}{\proji{k}{k}{A}\otimes\proji{l}{l}{A'}}\\
&=\frac{1}{\idx{d}{A}(\idx{d}{A}-1)}\sum_{k,l}{\proji{k}{k}{A}\otimes\proji{l}{l}{A'}}-\frac{1}{\idx{d}{A}(\idx{d}{A}-1)}\sum_{k}{\proji{k}{k}{A}\otimes\proji{k}{k}{A'}}\\
&=\frac{1}{\idx{d}{A}(\idx{d}{A}-1)}\idi{AA'}-\frac{1}{\idx{d}{A}-1}\idx{T}{AA'}.
\end{align}
So Claim 1 is also valid in the case that $i\neq j$, which concludes the proof. Finally we note that Claim 1 is intuitively clear: If we choose a permutation operator uniformly at random from $\mathbb{P}$ and apply $P\otimes P$ to any state of the type $\ket{i}\otimes\ket{i}$ we expect to get the uniform distribution over the set of such states. Since there are $\idx{d}{A}$ such states, the result of the summation should be $T$ for $i=j$. If $i\neq j$ we still expect that applying $P\otimes P$ to $\ket{i}\otimes\ket{j}$ yields the uniform distribution on the set of $\ket{i}\otimes\ket{j}$. Since there are $\idx{d}{A}(\idx{d}{A}-1)$ such states, the result of the summation in this case is 
\begin{align}
\frac{1}{\idx{d}{A}(\idx{d}{A}-1)}\sum_{k\neq l}{\proj{k}{k}\otimes\proj{l}{l}}.
\end{align}
\subsection{Action of twofold tensor products of permutation operators on a CQ-decoupling state.}
In the next subsection we will see that in order to proof the CQ-Decoupling Lemma one can proceed similarly as was done in the proof of the usual Decoupling Lemma with unitary operations in Section 2.1. There it was convenient to introduce the Decoupling State $\idx{\xi}{A$\tilde{\textnormal{A}}$}:=\idx{\Phi}{A$\tilde{\textnormal{A}}$} - \idx{\pi}{A}\otimes\idx{\pi}{$\tilde{\textnormal{A}}$}$. Here a similar approach will be used but this time we work with the CQ-Decoupling State $\idx{\lambda}{A$\tilde{\textnormal{A}}$}:=\idx{T}{A$\tilde{\textnormal{A}}$} - \idx{\pi}{A}\otimes\idx{\pi}{$\tilde{\textnormal{A}}$}$. This subsection shows a mathematical claim about that state. In the next subsection we will see why $\idx{\lambda}{A$\tilde{\textnormal{A}}$}$ is of relevance and use our claim.\\
\\
\textbf{Claim 2:} Let $\idx{\lambda}{A$\tilde{\textnormal{A}}$}=\idx{T}{A$\tilde{\textnormal{A}}$} - \idx{\pi}{A}\otimes\idx{\pi}{$\tilde{\textnormal{A}}$}$ be the CQ-Decoupling State. In the setup of the previous section with $\idx{d}{A}\geq2$, we have that
\begin{align}
\frac{1}{\idx{d}{A}!}\sum_{\idx{P}{A}\in\mathbb{P}(A)}{(\idx{P}{A} \otimes \idi{$\tilde{A}$})^{\otimes2}\  (\idx{\lambda}{A$\tilde{\textnormal{A}}$})^{\otimes2}\ (\idx{P}{A}^\dagger \otimes \idi{$\tilde{A}$})^{\otimes2}}
=\frac{1}{\idx{d}{A}-1}\:(\idx{\lambda}{AA'}\otimes\idx{\lambda}{$\tilde{\textnormal{A}}\tilde{\textnormal{A'}}$})\nonumber.
\end{align}
A conceptually easy way of proving this is by writing out $\idx{\lambda}{A$\tilde{\textnormal{A}}$}=\idx{T}{A$\tilde{\textnormal{A}}$} - \idx{\pi}{A}\otimes\idx{\pi}{$\tilde{\textnormal{A}}$}$, which gives
\begin{align}
\idx{\lambda}{A$\tilde{\textnormal{A}}$}^{\otimes2}
&=\idx{T}{A$\tilde{\textnormal{A}}$}\otimes\idx{T}{A'$\tilde{\textnormal{A'}}$}\nonumber\\
&-\idx{T}{A$\tilde{\textnormal{A}}$}\otimes\idx{\pi}{A'$\tilde{\textnormal{A'}}$}\nonumber\\
&-\idx{\pi}{A$\tilde{\textnormal{A}}$}\otimes\idx{T}{A'$\tilde{\textnormal{A'}}$}\nonumber\\
&+\idx{\pi}{A$\tilde{\textnormal{A}}$}\otimes\idx{\pi}{A'$\tilde{\textnormal{A'}}$}\label{4terms}
\end{align}
and by applying the summation to the four different terms separately.
The last term is invariant under the summation. For the second term we get
\begin{align}
&\frac{1}{\idx{d}{A}!}\sum_{\idx{P}{A}\in\mathbb{P}(A)}{(\idx{P}{A} \otimes \idi{$\tilde{A}$})^{\otimes2}\  (\idx{T}{A$\tilde{\textnormal{A}}$}\otimes\idx{\pi}{A'$\tilde{\textnormal{A'}}$})\ (\idx{P}{A}^\dagger \otimes \idi{$\tilde{A}$})^{\otimes2}}\nonumber\\
&=\frac{1}{\idx{d}{A}!}\sum_{\idx{P}{A}\in\mathbb{P}(A)}{(\idx{P}{A} \otimes \idi{$\tilde{A}$}) \idx{T}{A$\tilde{\textnormal{A}}$}(\idx{P}{A}^\dagger \otimes \idi{$\tilde{A}$})\otimes\idx{\pi}{A'$\tilde{\textnormal{A'}}$}}\\
&=\frac{1}{\idx{d}{A}\:\idx{d}{A}!}\sum_{i}{\sum_{\idx{P}{A}\in\mathbb{P}(A)}{(\idx{P}{A}\proji{i}{i}{A}\idx{P}{A}^\dagger)\otimes\proji{i}{i}{$\tilde{\textnormal{A}}$}\otimes\idx{\pi}{A'$\tilde{\textnormal{A'}}$}}}\\
&=\frac{(\idx{d}{A}-1)!}{\idx{d}{A}\:\idx{d}{A}!}\sum_{i}{\idi{A}\otimes\proji{i}{i}{$\tilde{\textnormal{A}}$}\otimes\idx{\pi}{A'$\tilde{\textnormal{A'}}$}}\label{explpermsum1}\\
&=\idx{\pi}{A$\tilde{\textnormal{A}}$}\otimes\idx{\pi}{A'$\tilde{\textnormal{A'}}$}\label{uslat}.
\end{align}
In \eqref{explpermsum1} we again used the fact that for any $\ket{j},\ket{k}\in\{\ket{i}\}_{i=1,...,d_A}$ there are $(\idx{d}{A}-1)!$ permutation operators with $P\ket{j}=\ket{k}$. Therefore in the sum over all permutation operators in equation \eqref{explpermsum1} for any fixed $\ket{i}$ every basis vector in $\{\ket{i}\}_{i=1,...,d_A}$ contributes $(\idx{d}{A}-1)!$ times to the whole sum.\\
Due to the symmetry of the problem the third an the second term in equation \eqref{4terms} yield the same result after summation, such that we only have to evaluate the sum over the first term.
\begin{align}
&\frac{1}{\idx{d}{A}!}\sum_{\idx{P}{A}\in\mathbb{P}(A)}{(\idx{P}{A} \otimes \idi{$\tilde{A}$})^{\otimes2}\  (\idx{T}{A$\tilde{\textnormal{A}}$}\otimes\idx{T}{A'$\tilde{\textnormal{A'}}$})\ (\idx{P}{A}^\dagger \otimes \idi{$\tilde{A}$})^{\otimes2}}\nonumber\\
&=\frac{1}{\idx{d}{A}^2\:\idx{d}{A}!}     \sum_{i,j}{\sum_{\idx{P}{A}\in\mathbb{P}(A)}{(\idx{P}{A}\proji{i}{i}{A}\idx{P}{A}^\dagger)\otimes\proji{i}{i}{$\tilde{\textnormal{A}}$}\otimes(\idx{P}{A'}\proji{j}{j}{A'}\idx{P}{A'}^\dagger)\otimes\proji{j}{j}{$\tilde{\textnormal{A'}}$}}}\label{useclaim2}
\end{align}
We now use Claim 1 and plug this result into equation \eqref{useclaim2}.
\begin{align}
&\frac{1}{\idx{d}{A}^2\:\idx{d}{A}!}     \sum_{i,j}{\sum_{\idx{P}{A}\in\mathbb{P}(A)}{(\idx{P}{A}\proji{i}{i}{A}\idx{P}{A}^\dagger)\otimes\proji{i}{i}{$\tilde{\textnormal{A}}$}\otimes(\idx{P}{A'}\proji{j}{j}{A'}\idx{P}{A'}^\dagger)\otimes\proji{j}{j}{$\tilde{\textnormal{A'}}$}}}\nonumber\\
&=\frac{1}{\idx{d}{A}^2}     \sum_{i,j}{\left(\left(\frac{1-\delta_{ij}}{\idx{d}{A}^2-\idx{d}{A}}\idi{AA'}-\frac{1-\delta_{ij}}{\idx{d}{A}-1}\idx{T}{AA'}+\delta_{ij}\idx{T}{AA'}\right)\otimes\proji{i}{i}{$\tilde{\textnormal{A}}$}\otimes\proji{j}{j}{$\tilde{\textnormal{A'}}$}\right)}\\
&=\frac{1}{\idx{d}{A}^2}     \sum_{i}{\left(\idx{T}{AA'}\otimes\proji{i}{i}{$\tilde{\textnormal{A}}$}\otimes\proji{i}{i}{$\tilde{\textnormal{A'}}$}\right)}\nonumber\\
&\quad+\frac{1}{\idx{d}{A}^2}\sum_{i\neq j}{\left(\left(\frac{1}{\idx{d}{A}^2-\idx{d}{A}}\idi{AA'}-\frac{1}{\idx{d}{A}-1}\idx{T}{AA'}\right)\otimes\proji{i}{i}{$\tilde{\textnormal{A}}$}\otimes\proji{j}{j}{$\tilde{\textnormal{A'}}$}\right)}\\
&=\frac{1}{\idx{d}{A}}{\idx{T}{AA'}\otimes\idx{T}{$\tilde{\textnormal{A}}\tilde{\textnormal{A'}}$}}\nonumber\\
&\quad+\frac{1}{\idx{d}{A}^2}\sum_{i,j}{\left(\left(\frac{1}{\idx{d}{A}^2-\idx{d}{A}}\idi{AA'}-\frac{1}{\idx{d}{A}-1}\idx{T}{AA'}\right)\otimes\proji{i}{i}{$\tilde{\textnormal{A}}$}\otimes\proji{j}{j}{$\tilde{\textnormal{A'}}$}\right)}\nonumber\\
&\quad-\frac{1}{\idx{d}{A}^2}\sum_{i}{\left(\left(\frac{1}{\idx{d}{A}^2-\idx{d}{A}}\idi{AA'}-\frac{1}{\idx{d}{A}-1}\idx{T}{AA'}\right)\otimes\proji{i}{i}{$\tilde{\textnormal{A}}$}\otimes\proji{i}{i}{$\tilde{\textnormal{A'}}$}\right)}\\
&=\frac{1}{\idx{d}{A}}{\idx{T}{AA'}\otimes\idx{T}{$\tilde{\textnormal{A}}\tilde{\textnormal{A'}}$}}\nonumber\\
&\quad+{\left(\frac{1}{\idx{d}{A}^2-\idx{d}{A}}\idi{AA'}-\frac{1}{\idx{d}{A}-1}\idx{T}{AA'}\right)\otimes\idx{\pi}{$\tilde{\textnormal{A}}$}\otimes\idx{\pi}{$\tilde{\textnormal{A'}}$}}\nonumber\\
&\quad-\frac{1}{\idx{d}{A}}{\left(\frac{1}{\idx{d}{A}^2-\idx{d}{A}}\idi{AA'}-\frac{1}{\idx{d}{A}-1}\idx{T}{AA'}\right)\otimes\idx{T}{$\tilde{\textnormal{A}}\tilde{\textnormal{A'}}$}}\\
&=\frac{1}{\idx{d}{A}-1}\left(\idx{\pi}{AA'}-\idx{T}{AA'}\right)\otimes\left(\idx{\pi}{$\tilde{\textnormal{A}}\tilde{\textnormal{A'}}$}-\idx{T}{$\tilde{\textnormal{A}}\tilde{\textnormal{A'}}$}\right)+\idx{\pi}{AA'}\otimes\idx{\pi}{$\tilde{\textnormal{A}}\tilde{\textnormal{A'}}$}
\end{align}
Taking this together with \eqref{uslat} and \eqref{4terms} yields that
\begin{align}
\frac{1}{\idx{d}{A}!}\sum_{\idx{P}{A}\in\mathbb{P}(A)}{(\idx{P}{A} \otimes \idi{$\tilde{A}$})^{\otimes2}\  (\idx{\lambda}{A$\tilde{\textnormal{A}}$})^{\otimes2}\ (\idx{P}{A}^\dagger \otimes \idi{$\tilde{A}$})^{\otimes2}}=\frac{1}{\idx{d}{A}-1}\left(\idx{\pi}{AA'}-\idx{T}{AA'}\right)\otimes\left(\idx{\pi}{$\tilde{\textnormal{A}}\tilde{\textnormal{A'}}$}-\idx{T}{$\tilde{\textnormal{A}}\tilde{\textnormal{A'}}$}\right),
\end{align}
which concludes the proof of Claim 2.
\subsection{Proof of the Decoupling Lemma for CQ-states}
In this subsection we use Claim 2 to proof the Decoupling Lemma for CQ-states. For this we follow the discussion of Chapter 2.1. We introduce the map $\idx{\cE}{$\tilde{\textnormal{A}}\rightarrow$R}^{cl}$, which we define to be the unique Choi-Jamiolkowski preimage of the state $\rhoAR$ i.e. $\idx{\cE}{$\tilde{\textnormal{A}}\rightarrow$R}^{cl}(\idx{\Phi}{A$\tilde{\textnormal{A}}$})=\rhoAR$, where $\tilde{A}$ is just a copy of $A$. It suffices to consider classicalized maps $\idx{\cE}{$\tilde{\textnormal{A}}\rightarrow$R}^{cl}$ because in our setup the state $\rhoAR$ has CQ-structure. By our definition of a classicalized map, we conclude that there exists a map $\idx{\cE}{$\tilde{\textnormal{A}}\rightarrow$R}$ with (compare \eqref{intrT})
\begin{align}
\cE(\idx{T}{A$\tilde{\textnormal{A}}$})=\rhoAR.
\end{align}
In addition we have that
\begin{align}
\rhoR&=\Ptrace{A}{\cE(\idx{T}{A$\tilde{\textnormal{A}}$})}\\
&=\cE(\Ptrace{A}{\idx{T}{A$\tilde{\textnormal{A}}$}})\\
&=\cE(\pi_{\tilde{\textnormal{A}}}).
\end{align}
Following the discussion of Chapter 2.1 equations \eqref{advxi}, we can rewrite the term in the Schatten 2-norm for any permutation operator $\idx{P}{A}$ as follows.

\begin{align}
&\mathcal{T}((\idx{P}{A} \otimes \idi{R}) \ \rhoAR \  (\idx{P}{A}^\dagger \otimes \idi{R})) - \idx{\omega}{E} \otimes \idx{\rho}{R} \nonumber\\
&= \mathcal{T}((\idx{P}{A} \otimes \idi{R}) \cE(\idx{T}{A$\tilde{\textnormal{A}}$})(\idx{P}{A}^\dagger \otimes \idi{R})) - \cT(\idx{\pi}{A})\otimes \cE(\idx{\pi}{$\tilde{\textnormal{A}}$})\\
&=(\cT\otimes\cE)((\idx{P}{A} \otimes \idi{$\tilde{A}$}) (\idx{T}{A$\tilde{\textnormal{A}}$} - \idx{\pi}{A}\otimes\idx{\pi}{$\tilde{\textnormal{A}}$})(\idx{P}{A}^\dagger \otimes \idi{$\tilde{A}$}))\\
&=(\cT\otimes\cE)((\idx{P}{A} \otimes \idi{$\tilde{A}$}) (\idx{\lambda}{A$\tilde{\textnormal{A}}$})(\idx{P}{A}^\dagger \otimes \idi{$\tilde{A}$})),
\end{align}
where in the last step we used the CQ-Decoupling State to simplify the notation. Then the left hand side of the CQ-Decoupling Lemma becomes
\begin{align}
&\frac{1}{\idx{d}{A}!}\sum_{\idx{P}{A}\in\mathbb{P}(A)} {\left\| \mathcal{T}(\idx{P}{A} \otimes \idi{R} \ \rhoAR \  \idx{P}{A}^\dagger \otimes \idi{R}) - \idx{\omega}{E} \otimes \idx{\rho}{R} \right\|}_2^2 \nonumber\\
&=\frac{1}{\idx{d}{A}!}\sum_{\idx{P}{A}\in\mathbb{P}(A)} \Trace{(\cT\otimes\cE)(\idx{P}{A} \otimes \idi{$\tilde{A}$} \:\idx{\lambda}{A$\tilde{\textnormal{A}}$}\:\idx{P}{A}^\dagger \otimes \idi{$\tilde{A}$})^2}\\
&=\frac{1}{\idx{d}{A}!}\sum_{\idx{P}{A}\in\mathbb{P}(A)}{\Trace{\left((\idx{P}{A} \otimes \idi{$\tilde{A}$})^{\otimes2}\  (\idx{\lambda}{A$\tilde{\textnormal{A}}$})^{\otimes2}\   (\idx{P}{A}^\dagger \otimes \idi{$\tilde{A}$})^{\otimes2}\right)\:(\cT^\dagger)^{\otimes2}[\idx{\cF}{E}]\otimes(\cE^\dagger)^{\otimes2}[\idx{\cF}{R}]}}\label{sndswp}\\
&=\Trace{\left(\frac{1}{\idx{d}{A}!}\sum_{\idx{P}{A}\in\mathbb{P}(A)}(\idx{P}{A} \otimes \idi{$\tilde{A}$})^{\otimes2}\  (\idx{\lambda}{A$\tilde{\textnormal{A}}$})^{\otimes2}\  (\idx{P}{A}^\dagger \otimes \idi{$\tilde{A}$})^{\otimes2}\right)\:(\cT^\dagger)^{\otimes2}[\idx{\cF}{E}]\otimes(\cE^\dagger)^{\otimes2}[\idx{\cF}{R}]},
\end{align}
where equation \eqref{sndswp} makes use of the swap trick (Appendix B) and is analogous to equation \eqref{dfndgr}. We evaluate the sum over all permutation operators using Claim 2 which gives that 
\begin{align}
&\frac{1}{\idx{d}{A}!}\sum_{\idx{P}{A}\in\mathbb{P}(A)} {\left\| \mathcal{T}(\idx{P}{A} \otimes \idi{R} \ \rhoAR \  \idx{P}{A}^\dagger \otimes \idi{R}) - \idx{\omega}{E} \otimes \idx{\rho}{R} \right\|}_2^2 \nonumber\\
&=\frac{1}{\idx{d}{A}-1}\Trace{\left(\idx{\lambda}{AA'}\otimes\idx{\lambda}{$\tilde{\textnormal{A}}\tilde{\textnormal{A'}}$}\right)\:(\cT^\dagger)^{\otimes2}[\idx{\cF}{E}]\otimes(\cE^\dagger)^{\otimes2}[\idx{\cF}{R}]}\\
&=\frac{1}{\idx{d}{A}-1}\Trace{\cT^{\otimes2}(\idx{\lambda}{AA'})\otimes\cE^{\otimes2}(\idx{\lambda}{$\tilde{\textnormal{A}}\tilde{\textnormal{A'}}$})\ \idx{\cF}{E}\otimes\idx{\cF}{R}}\\
&=\frac{1}{\idx{d}{A}-1}\Trace{\cT^{\otimes2}(\idx{\lambda}{AA'})\idx{\cF}{E}}\:\Trace{\cE^{\otimes2}(\idx{\lambda}{$\tilde{\textnormal{A}}\tilde{\textnormal{A'}}$})\ \idx{\cF}{R}}\label{usrsym}.
\end{align}
Due to the symmetry of the occurring trace terms it is sufficient to evaluate one of them. We get
\begin{align}
\Trace{\cT^{\otimes2}(\idx{\lambda}{AA'})\idx{\cF}{E}}&=\Trace{\cT^{\otimes2}(\idx{T}{AA'}-\idx{\pi}{AA'})\idx{\cF}{E}}\\
&=\frac{1}{\idx{d}{A}}\sum_{i}{\Trace{\cT(\proji{i}{i}{A})^2}}-\Trace{\idx{\omega}{E}^2}\label{classmpviaclassCJ}\\
&=\frac{1}{\idx{d}{A}}\sum_{i,j}{\delta_{ij}\Trace{\cT(\proji{i}{i}{A})\cT(\proji{j}{j}{A})}}-\Trace{\idx{\omega}{E}^2}\\
&=\frac{1}{\idx{d}{A}}\sum_{i,j}{\Trace{\proji{i}{i}{A'}\proji{j}{j}{A'}\otimes\cT(\proji{i}{i}{A})\cT(\proji{j}{j}{A})}}-\Trace{\idx{\omega}{E}^2}\\
&=\frac{1}{\idx{d}{A}}\Trace{\sum_{i}{\left(\proji{i}{i}{A'}\otimes\cT(\proji{i}{i}{A})\right)}\sum_{j}{\left(\proji{j}{j}{A'}\otimes\cT(\proji{j}{j}{A})\right)}}-\Trace{\idx{\omega}{E}^2}\\
&=\idx{d}{A}\Trace{\cT(\idx{T}{AA'})\cT(\idx{T}{AA'})}-\Trace{\idx{\omega}{E}^2}\\
&=\idx{d}{A}\left(\Trace{(\idx{\omega}{A'E}^{cl})^2}-\frac{1}{\idx{d}{A}}\tracebig{\idx{\omega}{E}^2}\right),
\end{align}
where the last step is by equation \eqref{intrT}. An analogous calculation upon exploitation of the CQ-structure of $\rhoAR$ yields that
\begin{align}
\Trace{\cE^{\otimes2}(\idx{\lambda}{$\tilde{\textnormal{A}}\tilde{\textnormal{A'}}$})\idx{\cF}{R}}
=\idx{d}{A}\left(\Trace{\idx{\rho}{AR}^2}-\frac{1}{\idx{d}{A}}\tracebig{\idx{\rho}{R}^2}\right).
\end{align}
We find plugging into equation \eqref{usrsym} that
\begin{align}
&\frac{1}{\idx{d}{A}!}\sum_{\idx{P}{A}\in\mathbb{P}(A)} {\left\| \mathcal{T}(\idx{P}{A} \otimes \idi{R} \ \rhoAR \  \idx{P}{A}^\dagger \otimes \idi{R}) - \idx{\omega}{E} \otimes \idx{\rho}{R} \right\|}_2^2\nonumber\\&=
\frac{\idx{d}{A}^2}{\idx{d}{A}-1}\left(\Trace{(\idx{\omega}{A'E}^{cl})^2}-\frac{1}{\idx{d}{A}}\tracebig{\idx{\omega}{E}^2}\right)\left(\Trace{\idx{\rho}{AR}^2}-\frac{1}{\idx{d}{A}}\tracebig{\idx{\rho}{R}^2}\right)\\
&=\frac{\idx{d}{A}^2}{\idx{d}{A}-1}\:{\left\|\idx{\rho}{AR}-\idx{\pi}{A}\otimes\rhoR\right\|}_2^2\: {\left\|\idx{\omega}{A'E}^{cl}-\idx{\pi}{A'}\otimes\idx{\omega}{E}^{cl}\right\|}_2^2,
\end{align}
which concludes the proof of the Decoupling Lemma for CQ-states. In the next sections we derive three theorems from this lemma. The theorems get more general but partially we have to pay for this with worse bounds. The first one will be analogous to the ``Leftover Hash Lemma'' as derived in \cite{Renner:PHD} while the second and third resemble the decoupling theorem.

\section{A ``Hash Lemma'' like result}
\textbf{Theorem:} (CQ-Decoupling Theorem for the Partial Trace)\textit{
Let $\rhoAR$ be classical on $\HA$ with respect to $\{\ket{i}\}_{i=1,...,d_A}$ and let $A=(A_1A_2)$ be a composite system, then the average distance from uniform is given by
\begin{align}
\frac{1}{\idx{d}{A}!}\sum_{\idx{P}{A}\in\mathbb{P}(A)}\Norm{\textnormal{tr}_{\textnormal{\tiny{A}}_2}{(\idx{P}{A} \otimes \idi{R} \ \rhoAR \  \idx{P}{A}^\dagger \otimes \idi{R})} - \pi_{\textnormal{\tiny{A}}_1}\otimes \idx{\rho}{R}}{1}
\leq\sqrt{d_{\textnormal{\tiny{A}}_1}\:\frac{d_{\textnormal{\tiny{A}}}-d_{\textnormal{\tiny{A}}_2}}{\idx{d}{A}-1}\:2^{-\chmin{A}{R}{\rho}}},\nonumber
\end{align}
where the average goes over all permutation operators, which act by permuting the basis vectors of $\{\ket{i}\}_{i=1,...,d_A}$.
}\\
\\
This in particular implies that
\begin{align}
\frac{1}{\idx{d}{A}!}\sum_{\idx{P}{A}\in\mathbb{P}(A)}\Norm{\textnormal{tr}_{\textnormal{\tiny{A}}_2}{(\idx{P}{A} \otimes \idi{R} \ \rhoAR \  \idx{P}{A}^\dagger \otimes \idi{R})} - \pi_{\textnormal{\tiny{A}}_1}\otimes \idx{\rho}{R}}{1}
\leq\sqrt{d_{\textnormal{\tiny{A}}_1}\:2^{-\chmin{A}{R}{\rho}}}\label{tllhsh}.
\end{align}
In \cite{Renner:PHD}, \cite{TSSR10} the ``General Leftover Hash Lemma'' is derived. There the summation takes place over all elements from some 2-universal family of hash functions from $A$ to $A_1$ \cite{CarterWegmann} but the resulting bound is the same as in equation \eqref{tllhsh} i.\:e.\:slightly worse than in our theorem. Given a fixed state $\rhoAR$ both the general leftover hash lemma and our theorem imply that there exists a function $f: A\rightarrow A_1$ that extracts almost uniform randomness. We postpone a further analysis of this formula to the next chapter and for the moment turn our attention to the proof.
We proceed as in Chapter 2.2 and apply the H\"older inequality \eqref{hldrfk} to bound the trace distance. Introducing the positive definite and normalized operator $\zetaR$, we write:
\begin{align}
&\Norm{\textnormal{tr}_{\textnormal{\tiny{A}}_2}{(\idx{P}{A} \otimes \idi{R} \ \rhoAR \  \idx{P}{A}^\dagger \otimes \idi{R})} - \pi_{\textnormal{\tiny{A}}_1} \otimes \idx{\rho}{R}}{1}\nonumber\\
&\leq\Norm{(\pi_{\textnormal{\tiny{A}}_1}\otimes\zetaR)^{-\frac{1}{4}}\left(\textnormal{tr}_{\textnormal{\tiny{A}}_2}{(\idx{P}{A} \otimes \idi{R} \ \rhoAR \  \idx{P}{A}^\dagger \otimes \idi{R})} - \pi_{\textnormal{\tiny{A}}_1} \otimes \idx{\rho}{R}\right)(\pi_{\textnormal{\tiny{A}}_1}\otimes\zetaR)^{-\frac{1}{4}}}{2}\\
&=\sqrt{d_{\textnormal{\tiny{A}}_1}}\:\Norm{\textnormal{tr}_{\textnormal{\tiny{A}}_2}{(\idx{P}{A} \otimes \idi{R} \ \idx{\tilde{\rho}}{AR} \  \idx{P}{A}^\dagger \otimes \idi{R})} - \pi_{\textnormal{\tiny{A}}_1} \otimes \idx{\tilde{\rho}}{R}}{2}\label{applyCQdec}
\end{align}
To keep the notation simple we introduced a state $\idx{\tilde{\rho}}{AR}:=(\id_{\textnormal{\tiny{A}}_1}\otimes\zetaR)^{-\frac{1}{4}}\rhoAR(\id_{\textnormal{\tiny{A}}_1}\otimes\zetaR)^{-\frac{1}{4}}$. Applying equation \eqref{applyCQdec} first and afterwards the decoupling lemma for CQ-states (Section 5.1) yields
\begin{align}
&\frac{1}{\idx{d}{A}!}\sum_{\idx{P}{A}\in\mathbb{P}(A)}\Norm{\textnormal{tr}_{\textnormal{\tiny{A}}_2}{(\idx{P}{A} \otimes \idi{R} \ \rhoAR \  \idx{P}{A}^\dagger \otimes \idi{R})} - \pi_{\textnormal{\tiny{A}}_1}\otimes \idx{\rho}{R}}{1}^2\nonumber\\
&\leq d_{\textnormal{\tiny{A}}_1}\:\frac{1}{\idx{d}{A}!}\sum_{\idx{P}{A}\in\mathbb{P}(A)}\Norm{\textnormal{tr}_{\textnormal{\tiny{A}}_2}{(\idx{P}{A} \otimes \idi{R} \ \idx{\tilde{\rho}}{AR} \  \idx{P}{A}^\dagger \otimes \idi{R})} - \pi_{\textnormal{\tiny{A}}_1} \otimes \idx{\tilde{\rho}}{R}}{2}^2\label{usemeintheend}\\
&=d_{\textnormal{\tiny{A}}_1}\:\frac{\idx{d}{A}^2}{\idx{d}{A}-1}\:{\left\|\idx{\tilde{\rho}}{AR}-\idx{\pi}{A}\otimes\idx{\tilde{\rho}}{R}\right\|}_2^2\cdot\frac{1}{\idx{d}{A}}\cdot\left(1-\frac{1}{d_{\textnormal{\tiny{A}}_1}}\right)\label{thrfg}.
\end{align}
The last equation is an application of the CQ decoupling lemma, where we used the explicit form of the Choi-Jamiolkowski representation for the partial trace
\begin{align}
\omega_{\textnormal{\tiny{A'A}}_1}^{cl}= \textnormal{tr}_{\textnormal{\tiny{A}}_2}(\idx{T}{AA'})
\end{align}
to calculate ${\left\|\idx{\omega}{A'E}^{cl}-\idx{\pi}{A'}\otimes\idx{\omega}{E}^{cl}\right\|}_2^2$. We conclude rewriting \eqref{thrfg}
\begin{align}
&\frac{1}{\idx{d}{A}!}\sum_{\idx{P}{A}\in\mathbb{P}(A)}\Norm{\textnormal{tr}_{\textnormal{\tiny{A}}_2}{(\idx{P}{A} \otimes \idi{R} \ \rhoAR \  \idx{P}{A}^\dagger \otimes \idi{R})} - \pi_{\textnormal{\tiny{A}}_1}\otimes \idx{\rho}{R}}{1}^2\nonumber\\
&\leq (d_{\textnormal{\tiny{A}}_1}-1)\:\frac{\idx{d}{A}}{\idx{d}{A}-1}\:\left(\Trace{\idx{\tilde{\rho}}{AR}^2}-\frac{1}{\idx{d}{A}}\tracebig{\idx{\tilde{\rho}}{R}^2}\right)\\
&\leq (d_{\textnormal{\tiny{A}}_1}-1)\:\frac{\idx{d}{A}}{\idx{d}{A}-1}\:\Trace{\idx{\tilde{\rho}}{AR}^2}\\
&= d_{\textnormal{\tiny{A}}_1}\:\frac{d_{\textnormal{\tiny{A}}}-d_{\textnormal{\tiny{A}}_2}}{\idx{d}{A}-1}\:\Trace{\idx{\tilde{\rho}}{AR}^2}\label{muell}.
\end{align}
We choose now $\zetaR$ to minimize the above expression and use the fact that the $\idx{H}{min}$-entropy always constitutes a lower bound on the $\idx{H}{2}$-entropy (\textit{Lemma 5}) to arrive at
\begin{align}
\frac{1}{\idx{d}{A}!}\sum_{\idx{P}{A}\in\mathbb{P}(A)}\Norm{\textnormal{tr}_{\textnormal{\tiny{A}}_2}{(\idx{P}{A} \otimes \idi{R} \ \rhoAR \  \idx{P}{A}^\dagger \otimes \idi{R})} - \pi_{\textnormal{\tiny{A}}_1}\otimes \idx{\rho}{R}}{1}^2
\leq d_{\textnormal{\tiny{A}}_1}\cdot\frac{d_{\textnormal{\tiny{A}}}-d_{\textnormal{\tiny{A}}_2}}{\idx{d}{A}-1}\cdot2^{-\chmin{A}{R}{\rho}}.
\end{align}
The proof is concluded by taking the square root on both sides and applying the Jensen inequality. It is also possible to follow the derivation of Section 2.3, where an ``improved'' decoupling theorem is derived to get bounds involving the $\idx{H}{min}$-entropy and  $\sqrt{\norm{\idx{\rho}{AR}-\idx{\pi}{A}\otimes\idx{\rho}{R}}{1}}$.

\section{A decoupling theorem for CQ-states and TPCP maps}
\textbf{Theorem:} (Decoupling Theorem for CQ-states and TPCP maps)\textit{
Let $\rhoAR\in\subnormstates{\idx{\cH}{A}\otimes\idx{\cH}{R}}$ be classical on $\idx{\cH}{A}$ with respect to $\{\ket{i}\}_{i=1,...,d_A}$ and let $\cTAE$ be a completely positive and trace preserving, linear map then
$$\frac{1}{\idx{d}{A}!}\sum_{\idx{P}{A}\in\mathbb{P}(A)} {\left\| \mathcal{T}(\idx{P}{A} \otimes \idi{R} \ \rhoAR \  \idx{P}{A}^\dagger \otimes \idi{R}) - \idx{\omega}{E} \otimes \idx{\rho}{R} \right\|}_1
\leq\sqrt{d_{E}\:\frac{\idx{d}{A}-\frac{\idx{d}{A}}{\idx{d}{E}}}{\idx{d}{A}-1}\:2^{-\chtwo{A}{R}{\rho}}},$$
where the average is taken over all permutation operators, which act by permuting the basis vectors of $\{\ket{i}\}_{i=1,...,d_A}$.
}\\
\\
The proof is almost identical to the proof shown in the previous section. We apply the H\"older inequality as in \eqref{applyCQdec} and the CQ-decoupling yields
\begin{align}
&\frac{1}{\idx{d}{A}!}\sum_{\idx{P}{A}\in\mathbb{P}(A)}\Norm{\cT(\idx{P}{A} \otimes \idi{R} \ \rhoAR \  \idx{P}{A}^\dagger \otimes \idi{R}) - \idx{\omega}{E}\otimes \idx{\rho}{R}}{1}^2\nonumber\\
&\leq\idx{d}{E}\:\frac{1}{\idx{d}{A}!}\sum_{\idx{P}{A}\in\mathbb{P}(A)}\Norm{\cT(\idx{P}{A} \otimes \idi{R} \ \idx{\tilde{\rho}}{AR} \  \idx{P}{A}^\dagger \otimes \idi{R}) - \idx{\omega}{E} \otimes \idx{\tilde{\rho}}{R}}{2}^2\\
&=\idx{d}{E}\:\frac{\idx{d}{A}^2}{\idx{d}{A}-1}\:{\left\|\idx{\tilde{\rho}}{AR}-\idx{\pi}{A}\otimes\idx{\tilde{\rho}}{R}\right\|}_2^2{\left\|\idx{\omega}{A'E}^{cl}-\idx{\pi}{A'}\otimes\idx{\omega}{E}^{cl}\right\|}_2^2\\
&=\idx{d}{E}\:\frac{\idx{d}{A}^2}{\idx{d}{A}-1}\:{\left\|\idx{\tilde{\rho}}{AR}-\idx{\pi}{A}\otimes\idx{\tilde{\rho}}{R}\right\|}_2^2\left(\Trace{(\idx{\omega}{A'E}^{cl})^2}-\frac{1}{\idx{d}{A}}\tracebig{\idx{\omega}{E}^2}\right)\\
&=\idx{d}{E}\:\frac{1}{\idx{d}{A}-1}\:{\left\|\idx{\tilde{\rho}}{AR}-\idx{\pi}{A}\otimes\idx{\tilde{\rho}}{R}\right\|}_2^2  \left(\sum_{i}{\Trace{\cT(\proji{i}{i}{A})^2}}-\idx{d}{A}\:\Trace{\idx{\omega}{E}^2}\right)\label{followme}.
\end{align}
The last step makes use of the calculation following equation \eqref{classmpviaclassCJ}. By assumption $\cT$ is a TPCPM. Due to the trace preserving property, for any $i$ we have $\Trace{\cT(\proji{i}{i}{A})}=1$ and we know that all eigenvalues of $\cT(\proji{i}{i}{A})$ are nonnegative by the positivity of $\cT$. We conclude that the eigenvalues of any matrix $\cT(\proji{i}{i}{A})$ are between zero and one. Thus 
\begin{align}
\Trace{\cT(\proji{i}{i}{A})^2}&\leq\Trace{\cT(\proji{i}{i}{A})}\\
&=1.
\end{align}
Moreover an application of the Cauchy Schwarz inequality shows
\begin{align}
1&=\trace{\idx{\omega}{E}}\\
&\leq\sqrt{\trace{\idi{E}}\:\Trace{\idx{\omega}{E}^2}}.
\end{align}
Following \eqref{followme} and using the found bounds we get that
\begin{align}
&d_{E}\:\frac{1}{\idx{d}{A}-1}\:{\left\|\idx{\tilde{\rho}}{AR}-\idx{\pi}{A}\otimes\idx{\tilde{\rho}}{R}\right\|}_2^2  \left(\sum_{i}{\Trace{\cT(\proji{i}{i}{A})^2}}-\idx{d}{A}\:\Trace{\idx{\omega}{E}^2}\right)\nonumber\\
&\leq d_{E}\:\frac{\idx{d}{A}-\idx{d}{A}\:\Trace{\idx{\omega}{E}^2}}{\idx{d}{A}-1}\:{\left\|\idx{\tilde{\rho}}{AR}-\idx{\pi}{A}\otimes\idx{\tilde{\rho}}{R}\right\|}_2^2\\
&\leq d_{E}\:\frac{\idx{d}{A}-\frac{\idx{d}{A}}{\idx{d}{E}}}{\idx{d}{A}-1}\:{\left\|\idx{\tilde{\rho}}{AR}-\idx{\pi}{A}\otimes\idx{\tilde{\rho}}{R}\right\|}_2^2\\
&\leq d_{E}\:\frac{\idx{d}{A}-\frac{\idx{d}{A}}{\idx{d}{E}}}{\idx{d}{A}-1}\:\Trace{\idx{\tilde{\rho}}{AR}^2}.
\end{align}
The arguments following \eqref{muell}
conclude the proof.
\section{A decoupling theorem for CQ-states}
In this section a short derivation of a decoupling theorem for CQ-states is presented. The obtained theorem can be seen as a classical analogue of the decoupling theorem as derived in Section 2.2.\\
\\
\textbf{Theorem:} (Decoupling Theorem for CQ-states)\textit{
Let $\rhoAR\in\subnormstates{\idx{\cH}{A}\otimes\idx{\cH}{R}}$ be classical on $\idx{\cH}{A}$ with respect to $\{\ket{i}\}_{i=1,...,d_A}$ and let $\cTAE$ be a completely positive linear map going from $\subnormstates{\idx{\cH}{A}\otimes\idx{\cH}{R}}$ to $\posops{\idx{\cH}{E} \otimes \idx{\cH}{R}}$ with Choi-Jamiolkowski representation $\idx{\omega}{A'E}\in\subnormstates{\idx{\cH}{E}\otimes\idx{\cH}{A'}}$, then
$$\frac{1}{\idx{d}{A}!}\sum_{\idx{P}{A}\in\mathbb{P}(A)} {\left\| \mathcal{T}(\idx{P}{A} \otimes \idi{R} \ \rhoAR \  \idx{P}{A}^\dagger \otimes \idi{R}) - \idx{\omega}{E} \otimes \idx{\rho}{R} \right\|}_1
\leq\sqrt{(\idx{d}{A}+1)\:2^{-\chtwo{A}{R}{\rho}-\chtwo{A'}{E}{({\omega}^{cl})}}},$$
where the average is taken over all permutation operators, which act by permuting the basis vectors of $\{\ket{i}\}_{i=1,...,d_A}$.
}\\
\\
The proof is closely analogous to the proof of the decoupling theorem as stated in Section 2.2. We therefore can keep our discussion short and refer to Section 2.2 for explanations. We apply the H\"older inequality as in \eqref{hldrfk} to get
\begin{align}
&{\left\| \mathcal{T}(\idx{P}{A} \otimes \idi{R} \ \rhoAR \  \idx{P}{A}^\dagger \otimes \idi{R}) - \idx{\omega}{E} \otimes \idx{\rho}{R} \right\|}_1\nonumber\\
&\leq\Norm{(\sigmaE\otimes\zetaR)^{-\frac{1}{4}}\left(\mathcal{T}(\idx{P}{A} \otimes \idi{R} \ \rhoAR \  \idx{P}{A}^\dagger \otimes \idi{R}) - \idx{\omega}{E} \otimes \idx{\rho}{R}\right)(\sigmaE\otimes\zetaR)^{-\frac{1}{4}}}{2}\\
&=\Norm{\tilde{\cT}(\idx{P}{A} \otimes \idi{R} \ \idx{\tilde{\rho}}{AR} \  \idx{P}{A}^\dagger \otimes \idi{R}) - \idx{\tilde{\omega}}{E} \otimes \idx{\tilde{\rho}}{R}}{2},
\end{align}
where $\sigmaE\in\normstates{\idx{\cH}{E}}$ and $\zetaR\in\normstates{\idx{\cH}{R}}$. The map $\tilde{\cT}$ is defined as in Section 2.2 and $\idx{\tilde{\omega}}{A'E}$ is the Choi-Jamiolkowski representation of $\tilde{\cT}$. Furthermore we wrote $\idx{\tilde{\rho}}{AR}:=(\idx{\id}{A}\otimes\zetaR)^{-\frac{1}{4}}\rhoAR(\idx{\id}{A}\otimes\zetaR)^{-\frac{1}{4}}$. Using the above inequality and  applying the decoupling lemma for CQ-states yields
\begin{align}
\frac{1}{\idx{d}{A}!}\sum_{\idx{P}{A}\in\mathbb{P}(A)} &{\left\| \mathcal{T}(\idx{P}{A} \otimes \idi{R} \ \rhoAR \  \idx{P}{A}^\dagger \otimes \idi{R}) - \idx{\omega}{E} \otimes \idx{\rho}{R} \right\|}_1^2 \nonumber\\
&\leq\frac{1}{\idx{d}{A}!}\sum_{\idx{P}{A}\in\mathbb{P}(A)}\Norm{\tilde{\cT}(\idx{P}{A} \otimes \idi{R} \ \idx{\tilde{\rho}}{AR} \  \idx{P}{A}^\dagger \otimes \idi{R}) - \idx{\tilde{\omega}}{E} \otimes \idx{\tilde{\rho}}{R}}{2}^2\\
&=\frac{\idx{d}{A}^2}{\idx{d}{A}-1}\:{\left\|\idx{\tilde{\rho}}{AR}-\idx{\pi}{A}\otimes\idx{\tilde{\rho}}{R}\right\|}_2^2\: {\left\|\idx{\tilde{\omega}}{A'E}^{cl}-\idx{\pi}{A'}\otimes\idx{\tilde{\omega}}{E}^{cl}\right\|}_2^2\\
&=\frac{\idx{d}{A}^2}{\idx{d}{A}-1}\left(\Trace{\idx{\tilde{\rho}}{AR}^2}-\frac{1}{\idx{d}{A}}\tracebig{\idx{\tilde{\rho}}{R}^2}\right)\left(\Trace{(\idx{\tilde{\omega}}{A'E}^{cl})^2}-\frac{1}{\idx{d}{A}}\tracebig{\idx{\tilde{\omega}}{E}^2}\right)\\
&\leq(\idx{d}{A}+1)\:\trace{(\idx{\tilde{\omega}}{A'E}^{cl})^2}\:\trace{\idx{\tilde{\rho}}{AR}^2}\left(\frac{\idx{d}{A}^2-\frac{\idx{d}{A}\:\Trace{\idx{\tilde{\omega}}{E}^2}}{\Trace{(\idx{\tilde{\omega}}{A'E}^{cl})^2}}}{\idx{d}{A}^2-1}\right)\left(\frac{\idx{d}{A}^2-\frac{\idx{d}{A}\:\Trace{\idx{\tilde{\rho}}{R}^2}}{\Trace{\idx{\tilde{\rho}}{AR}^2}}}{\idx{d}{A}^2-1}\right).
\end{align}
As discussed in Section 2.2 both bracket terms are smaller or equal than one and we get
\begin{align}
\frac{1}{\idx{d}{A}!}\sum_{\idx{P}{A}\in\mathbb{P}(A)} {\left\| \mathcal{T}(\idx{P}{A} \otimes \idi{R} \ \rhoAR \  \idx{P}{A}^\dagger \otimes \idi{R}) - \idx{\omega}{E} \otimes \idx{\rho}{R} \right\|}_1^2
\leq(\idx{d}{A}+1)\:\trace{(\idx{\tilde{\omega}}{A'E}^{cl})^2}\:\trace{\idx{\tilde{\rho}}{AR}^2}.
\end{align}
We choose the $\sigmaE$ and $\zetaR$ hidden in the tildes of the above expression to minimize the terms $\trace{(\idx{\tilde{\omega}}{A'E}^{cl})^2}$ and
$\trace{\idx{\tilde{\rho}}{AR}^2}$ to obtain a bound in terms of the $\idx{H}{2}$-entropy. Finally we take the square root on both sides and apply the Jensen inequality. This reveals that
\begin{align}
\frac{1}{\idx{d}{A}!}\sum_{\idx{P}{A}\in\mathbb{P}(A)} {\left\| \mathcal{T}(\idx{P}{A} \otimes \idi{R} \ \rhoAR \  \idx{P}{A}^\dagger \otimes \idi{R}) - \idx{\omega}{E} \otimes \idx{\rho}{R} \right\|}_1
\leq\sqrt{(\idx{d}{A}+1)\:2^{-\chtwo{A}{R}{\rho}-\chtwo{A'}{E}{({\omega}^{cl})}}}.
\end{align}
\chapter{Decoupling with 2-wise almost independent families of permutations}
In the last chapter we dealt with an analogue of the decoupling theorem valid for CQ-states. It is now interesting to obtain a classical version of the results of Chapter 3. There the decoupling behavior of unitary almost 2-designs is discussed. But what is the classical analogue of a unitary almost 2-design? In this chapter's first section we introduce 2-wise almost independent families of permutations and show how they can be understood as a classical analogue of a unitary almost 2-design. Afterwards we generalize the versions of the decoupling theorem for CQ-states obtained in the previous chapter and formulate a result similar to the ``General Leftover Hash Lemma'' (\textit{Lemma 2} in \cite{TSSR10}). But instead of taking the average over a family of almost hash functions as is done in the latter, we average over almost independent permutations.\\
\section{Almost independent families of permutations}
The concept of $k$-wise independent permutations has been receiving an increasing amount of interest in the computer science literature (see \cite{LubyRackoffRev} for an overview). Loosely speaking a family of permutations is called $k$-wise almost independent if for any element $P$ chosen uniformly at random from that family and any $k$ different input values $x_1,...,x_k$ the distribution $Px_1,...,Px_k$ is almost uniform. For our purposes it will be interesting to consider pairwise independent permutations, nevertheless we state the definitions for any $k$ to emphasize the relation to unitary $k$-designs. In \cite{almostperm} a construction of a family of $k$-wise almost independent permutations is given for any $k$.
\begin{definition}(Statistical Distance \cite{almostperm}) 
Let $D_1$ and $D_2$ be two probability distributions defined over a finite set $\Omega$. The statistical distance of $D_1$ and $D_2$ is defined to be
$$\Norm{D_1-D_2}{1}:=\sum_{\omega\in\Omega}{\abs{D_1(\omega)-D_2(\omega)}}$$
$D_1$ and $D_2$ are called $\varepsilon$-close in statistical distance if $\Norm{D_1-D_2}{1}\leq\varepsilon$.
\end{definition}
Note that for two classical density operator $\rho=\sum_{i}D_1(i)\proj{i}{i}$ and $\sigma=\sum_{i}D_2(i)\proj{i}{i}$ the trace distance introduced in the preliminaries and the statistical distance of the corresponding distributions $D_1$ and $D_2$ coincide. Thus the trace distance can be viewed as a quantum generalization of the statistical distance, which justifies the above notation for the statistical distance with the ``Schatten 1-norm''. If two probability distributions are $\varepsilon$-close in statistical distance the maximum difference between the probabilities of an arbitrary event with respect to the different distributions is smaller or equal than $\varepsilon$. We denote with $\mathfrak{P}$ the family of all bijective functions
\begin{align}
P:\{0,1\}^n\rightarrow\{0,1\}^n.
\end{align}
\begin{definition}($k$-wise $\varepsilon$-dependent family of permutations \cite{almostperm})
Let $n,k\in\mathbb{N}$ and let $[N_k]$ be the set of all $k$-tuples of \textit{distinct} $n$-bit strings. Furthermore let $\cF\subset\mathfrak{P}$ be a family of permutations and $\varepsilon\geq0$. The family $\cF$ is called \textit{$k$-wise $\varepsilon$-dependent} if for every $k$-tuple $(x_1,...,x_k)\in[N_k]$ and $P$ chosen uniformly at random from $\cF$, the distribution on $[N_k]$ induced via $(Px_1,...,Px_k)$ is $\varepsilon$-close to the uniform distribution on $[N_k]$. A $k$-wise $0$-dependent family of permutations is called $k$-wise independent.
\end{definition}
As already mentioned, in our setup pairwise almost independent permutations will be most relevant; similarly to the fact that we dealt with unitary almost 2-designs in Chapter 3. To reveal the relation between the definitions of almost independent permutations and unitary almost 2-designs (Definition 8) it is convenient to reformulate the above Definition 13 in terms of density operators. For this we first introduce some new notation. In analogy to Definition 6 we have:
\begin{definition}
For $\cF\subset\mathbb{P}$ and any $\rho\in\linops{\cH^{\otimes k}}$ we define the functions:
\begin{align}
&\cP_W(\rho):=\sum_{P\in\cF}{\frac{1}{\abs{\cF}}P^{\otimes k}\rho (P^\dagger)^{\otimes k}}\nonumber\\
&\cP_H(\rho):=\sum_{P\in\mathbb{P}}{\frac{1}{\abs{\mathbb{P}}}P^{\otimes k}\rho (P^\dagger)^{\otimes k}}.\nonumber
\end{align}
\end{definition}
Note that in the above definition $\cF$ is a subset of $\mathbb{P}$ and not of $\mathfrak{P}$ as in Definition 13. We will see after some further definitions how this can be understood. To compare the distance of two maps we introduce a classical version of the diamond norm:
\begin{definition}
Fix an orthonormal basis $\{\ket{i}\}_{i=1,...,d_A}$ for $\HA$ and an orthonormal basis for $\linops{\idx{\cH}{E}}$. Let $\rhoA\in\linops{\HA}$ be classical with respect to the fixed basis of $\HA$. Let $\cTAE$ be a linear map from $\linops{\HA}$ to $\linops{\idx{\cH}{E}}$ that preserves the diagonal structure of the input state (i.e. it is classical). The classical diamond norm of $\cTAE$ is defined to be:
$$\Norm{\cTAE}{\diamond}^{cl}:=\max_{\rhoA}{\frac{\Norm{\cTAE{(\rhoA)}}{1}}{\Norm{\rhoA}{1}}}.$$
\end{definition}
The notion of ``classicality'' is basis dependent. Thus the above is not a good norm in the sense of operator norms. Nevertheless once a basis is chosen it can be seen as a classical analogue of the diamond norm. Moreover note that there is no notion of ``entanglement'' in the classical case, such that one can leave out the tensoring with the operator identity typical for the diamond norm in this case without changing the value of the norm. Finally we state an alternative definition of a $k$-wise almost independent family of permutations.
\begin{definition}
Let $\cP_W$ and $\cP_H$ be as in Definition 14. $\cP_W$ is called an $\varepsilon$-almost $k$-wise independent family of permutations
if and only if
$$\Norm{\cP_W-\cP_H}{1}^{cl}\leq\varepsilon.$$
\end{definition}
This definition strongly resembles the definition of a unitary $\varepsilon$-almost $k$-design. Indeed one can see it as a classical motivation  for the quantum mechanical definition of an almost $k$-design. In the next subsection we proof the consistency of the two different definitions of almost independent families of permutations (Definitions 13 and 16), i.\:e.\:we show that $\cP_W$ is an $\varepsilon$-almost $k$-wise independent family of permutations (with respect to Defition 16) if and only if the underlying set $\cF$ is $k$-wise almost independent (according to Defintion 13). Afterwards in a short subsection we give a concrete example of pairwise independent family of permutations.
\subsection{Proving the equivalence of the two different views on almost independent families of permutations}

For the proof we consider the case of pairwise independent families of permutations but the discussion can be generalized to $k>2$ in a straight forward manner.\\
Assume that $\cF$ is an $\varepsilon$-almost independent family of permutations according to Definition 13. We now that choosing a permutation uniformly at random from $\cF$ induces a probability distribution on $[N_2]$ that is close to the uniform one. Formalizing this statement, we get
\begin{align}
\sum_{\omega\in[N_2]}{\abs{\underset{P\in\cF}{\textnormal{Prob}}\left[(Px_1,Px_2)=\omega\right]-\underset{P\in\mathfrak{P}}{\textnormal{Prob}}\left[(Px_1,Px_2)=\omega\right]}}\leq\varepsilon\qquad\ \forall\ (x_1,x_2)\in[N_2]\label{stdsttotrdst},
\end{align}
where $\underset{P\in\cF}{\textnormal{Prob}}\left[(Px_1,Px_2)=\omega\right]$ denotes the probability that choosing $P\in\cF$ uniformly at random for a fixed element $(x_1,x_2)\in[N_2]$ we get $(Px_1,Px_2)=\omega$. Similarly $\underset{P\in\mathfrak{P}}{\textnormal{Prob}}\left[(Px_1,Px_2)=\omega\right]$ is defined but in this case $P$ is chosen uniformly at random from $\mathfrak{P}$. Note that $\underset{P\in\mathfrak{P}}{\textnormal{Prob}}\left[(Px_1,Px_2)=\omega\right]$ is constant and thus corresponds to the uniform distribution over $[N_2]$. The above can be translated into the language of quantum mechanics introducing the classical density operators $\rho$ and $\sigma$ with
\begin{align}
&\rho:=\sum_{\omega\in[N_2]}{\underset{P\in\cF}{\textnormal{Prob}}\left[(Px_1,Px_2)=\omega\right]}\:\proj{\omega}{\omega}\label{explr}\\
&\sigma:=\sum_{\omega\in[N_2]}{\underset{P\in\mathfrak{P}}{\textnormal{Prob}}\left[(Px_1,Px_2)=\omega\right]}\:\proj{\omega}{\omega}.
\end{align}
and equation \eqref{stdsttotrdst} is equivalently rewritten as
\begin{align}
\Norm{\rho-\sigma}{1}\leq\varepsilon\qquad\ \forall\ (x_1,x_2)\in[N_2]
\end{align}
since the Schatten 1-norm of some matrix is given by the sum of the absolute eigenvalues of this matrix.
To represent a classical $n$-bit string in a quantum mechanical language, we enumerate each bit string of length $n$ by a number $i\in\{1,...,2^n\}$. Then we choose the canonical basis $\{e_i\}_{i\in\{1,...,2^n\}}$ of a $2^n$-dimensional Hilbert space $\cH$ and identify the $i$-th bit-string with the $i$-th basis vector. As a result to each of the $(2^n)!$ permutations in $\mathfrak{P}$ corresponds a unique permutation operator in $\mathbb{P}$.
The vector $\ket{\omega}$ is an element of a bipartite Hilbert space, because $\omega=(y_1,y_2)\in[N_2]$ and each $y_i$ corresponds to an element of the canonical basis of $\cH$.\\
Strictly speaking the density operator $\rho$ defined in equation \eqref{explr} should have an index $(x_1,x_2)$ since it is dependent on the ``input'' $(x_1,x_2)$. We will keep this fact in mind. Let $e_1$ and $e_2$ be the quantum mechanical representations of the strings $x_1,x_2$. We can rewrite $\rho$ with a slight abuse of notation as
\begin{align}
\rho&=\sum_{\omega\in[N_2]}{\underset{P\in\cF}{\textnormal{Prob}}\left[(Px_1,Px_2)=\omega\right]}\:\proj{\omega}{\omega}\\
&=\sum_{\omega\in[N_2]}{\underset{P\in\cF}{\textnormal{Prob}}\left[(P\otimes P)\ket{e_1}\otimes\ket{e_2}=\ket{\omega}\right]}\:\proj{\omega}{\omega}.
\end{align}
The set $\cF$ is meant to be once a subset of $\mathfrak{P}$ and once of $\mathbb{P}$. But since there is a one to one correspondence between the elements of these sets as described above, the notation is still well defined. Furthermore we have
\begin{align}
&\sum_{\omega\in[N_2]}{\underset{P\in\cF}{\textnormal{Prob}}\left[(P\otimes P)\ket{e_1}\otimes\ket{e_2}=\ket{\omega}\right]}\:\proj{\omega}{\omega}\nonumber\\
&=\sum_{\omega\in[N_2]}{\left(\sum_{{P\in\cF}\atop{(P\otimes P)\ket{e_1}\otimes\ket{e_2}=\ket{\omega}}}\frac{1}{\abs{\cF}}\right)\:\proj{\omega}{\omega}}\\
&=\sum_{\omega\in[N_2]}{\left(\sum_{{P\in\cF}\atop{(P\otimes P)\ket{e_1}\otimes\ket{e_2}=\ket{\omega}}}\frac{1}{\abs{\cF}}\:(P\otimes P)\proj{e_1\otimes e_2}{e_1\otimes e_2}(P\otimes P)^{\dagger}\right)}\\
&=\sum_{P\in\cF}{\frac{1}{\abs{\cF}}\:(P\otimes P)\proj{e_1\otimes e_2}{e_1\otimes e_2}(P\otimes P)^{\dagger}}.
\end{align}
An identical calculation shows that
\begin{align}
\sigma=\sum_{P\in\mathbb{P}}{\frac{1}{\abs{\mathbb{P}}}\:(P\otimes P)\proj{e_1\otimes e_2}{e_1\otimes e_2}(P\otimes P)^{\dagger}}
\end{align}
and reveals that the condition \eqref{stdsttotrdst} is equivalent to
\begin{align}
\Norm{\cP_W(\proj{e_1\otimes e_2}{e_1\otimes e_2})-\cP_H(\proj{e_1\otimes e_2}{e_1\otimes e_2})}{1}\leq\varepsilon\qquad\forall\ \ket{e_1\otimes e_2};\:e_1\neq e_2\label{gdcond},
\end{align}
where the functions $\cP_W$ and $\cP_H$ are as in Definition 14 for $k=2$. If $\cF$ (or in other words $\cP_W$) is a pairwise almost independent family of permutations according to Definition 16, then the derivation of equation \eqref{gdcond} implies that it is also such a family with respect to Definition 13. On the other hand if $\cF$ is a pairwise almost independent family of permutations according to Definition 13 statement \eqref{gdcond} is always valid. Then for any bipartite classical density operator with $\sum_{i,j}{p_{ij}}\leq1$
\begin{align}
\zeta:=\sum_{i,j}{p_{ij}\proj{ij}{ij}}
\end{align}
by several applications of the triangle inequality one gets that
\begin{align}
&\Norm{\cP_W(\zeta)-\cP_H(\zeta)}{1}\nonumber\\
&=\Norm{\sum_{i,j}{p_{ij}\big(\cP_W(\proj{ij}{ij})-\cP_H(\proj{ij}{ij})\big)}}{1}\\
&\leq\Norm{\sum_{i\neq j}{p_{ij}\big(\cP_W(\proj{ij}{ij})-\cP_H(\proj{ij}{ij})\big)}}{1}
+\Norm{\sum_{i}{p_{ii}\big(\cP_W(\proj{ii}{ii})-\cP_H(\proj{ii}{ii})\big)}}{1}\\
&\leq\sum_{i\neq j}{p_{ij}\Norm{\cP_W(\proj{ij}{ij})-\cP_H(\proj{ij}{ij})}{1}}+\sum_{i}{p_{ii}\Norm{\cP_W(\proj{ii}{ii})-\cP_H(\proj{ii}{ii})}{1}}\label{2trter}\\
&\leq\sum_{i\neq j}{p_{ij}\:\varepsilon}+\sum_{i}{p_{ii}\Norm{\cP_W(\proj{ii}{ii})-\cP_H(\proj{ii}{ii})}{1}}\label{nwus}
\end{align}
To bound the first term of \eqref{2trter} we used the inequality \eqref{gdcond} directly. But the second summand has to be bound separately since for \eqref{gdcond} we require the condition $i\neq j$. We do this with a short reformulation, whereby we introduce an arbitrary element of the basis $\ket{j}\neq\ket{i}$:
\begin{align}
&\Norm{\cP_W(\proj{ii}{ii})-\cP_H(\proj{ii}{ii})}{1}\nonumber\\
&=\Norm{\sum_{P\in\cF}{\frac{1}{\abs{\cF}}\:(P\otimes P)\proj{ii}{ii}(P\otimes P)^{\dagger}}-\sum_{P\in\mathbb{P}}{\frac{1}{\abs{\mathbb{P}}}\:(P\otimes P)\proj{ii}{ii}(P\otimes P)^{\dagger}}}{1}\\
&=\Norm{\sum_{P\in\cF}{\frac{1}{\abs{\cF}}\:P\proj{i}{i}P^{\dagger}}-\sum_{P\in\mathbb{P}}{\frac{1}{\abs{\mathbb{P}}}\:P\proj{i}{i}P^{\dagger}}}{1}\\
&=\Norm{\sum_{P\in\cF}{\frac{1}{\abs{\cF}}\:P\proj{i}{i}P^{\dagger}\left(\sum_{k}{\bracket{k}{P}{j}\bracket{j}{P^\dagger}{k}}\right)}-\sum_{P\in\mathbb{P}}{\frac{1}{\abs{\mathbb{P}}}\:P\proj{i}{i}P^{\dagger}\left(\sum_{k}{\bracket{k}{P}{j}\bracket{j}{P^\dagger}{k}}\right)}}{1}\label{intrbracket}\\
&=\Norm{\Ptrace{2}{\sum_{P\in\cF}{\frac{1}{\abs{\cF}}\:(P\otimes P)\proj{ij}{ij}(P\otimes P)^{\dagger}}-\sum_{P\in\mathbb{P}}{\frac{1}{\abs{\mathbb{P}}}\:(P\otimes P)\proj{ij}{ij}(P\otimes P)^{\dagger}}}}{1}\label{explptr2}\\
&\leq\Norm{\sum_{P\in\cF}{\frac{1}{\abs{\cF}}\:(P\otimes P)\proj{ij}{ij}(P\otimes P)^{\dagger}}-\sum_{P\in\mathbb{P}}{\frac{1}{\abs{\mathbb{P}}}\:(P\otimes P)\proj{ij}{ij}(P\otimes P)^{\dagger}}}{1}\\
&\leq\varepsilon
\end{align}
In equation \eqref{intrbracket} both bracket terms are equal one and there exists always a vector $\ket{j}\neq\ket{i}$ (in all nontrivial cases).
With the partial trace in \eqref{explptr2} we mean that we trace out the second system and we use the monotonicity of the trace distance under TPCPM for the following inequality. We plug this result into equation \eqref{nwus} to find
\begin{align}
\Norm{\cP_W(\zeta)-\cP_H(\zeta)}{1}&\leq\sum_{i\neq j}{p_{ij}\:\varepsilon}+\sum_{i}{p_{ii}\:\varepsilon}\\
&\leq\varepsilon.\label{findef16}
\end{align}
If $\cF$ is a pairwise $\varepsilon$-almost independent family of permutations according to Definition 13 inequality \eqref{gdcond} is satisfied and subsequently \eqref{findef16} is valid for any classical $\zeta$. This implies that $\cF$ is an $\varepsilon$-almost independent family with respect to Definition 16, which concludes the proof of the equivalence of the two different definitions. 
\subsection{An exemplary pairwise independent family of permutations}
We would like to give a concrete example of a family of pairwise independent permutations. Choose the set $\{0,1\}^{n}$ of $n$-bit strings. Give it the structure of a field, $GF(2^n)$, introducing the entrywise addition and multiplication operations (mod 2).\\
\\
\textbf{Proposition:}\textit{
The family $\cF:=\{P_{a,b}\ |\ a,b\in GF(2^n)\ \wedge\ a\neq0\}$ of permutations defined via
\begin{align}
P_{a,b}:  GF(2^n)&\rightarrow GF(2^n)\nonumber\\
x&\mapsto P_{a,b}(x):= a\cdot x+ b\nonumber
\end{align}
is pairwise independent.
}\\
\\
First note that if choosing $x_1,x_2\in GF(2^n)$ with $x_1\neq x_2$ they are mapped to distinct elements $y_1:=P_{a,b}(x_1)$ and $y_2:=P_{a,b}(x_2)$ by any element of the family $\cF$. This is because we have by assumption that $a\neq0$ and the addition in a field is a bijective map. Thus the maps $P_{a,b}$ can be seen as maps from $[N_2]$ into itself. Choosing $(a,b)$ uniformly at random from $(GF(2^n)-\{0\})\times GF(2^n)$ corresponds to the choice of a permutation uniformly at random from $\cF$. To see that the induced distribution on $[N_2]$ is uniform, we check that for any input $(x_1,x_2)\in[N_2]$ each pair $(y_1,y_2)$ is hit with equal probability:
\begin{align}
&a\cdot x_1+b=y_1\\
&a\cdot x_2+b=y_2
\end{align}
This is equivalent to
\begin{align}
\begin{pmatrix} x_1 & 1 \\ x_2 & 1 \end{pmatrix}\cdot \left( \begin{array}{c} a \\ b \end{array} \right) = \left( \begin{array}{c} y_1 \\ y_2 \end{array} \right),
\end{align}
where the condition $x_1\neq x_2$ implies that
\begin{align}
\det\begin{pmatrix} x_1 & 1 \\ x_2 & 1 \end{pmatrix}\neq0.
\end{align}
Thus for any $(y_1,y_2)\in[N_2]$ there exists a unique $P_{a,b}\in\cF$ with
\begin{align}
&P_{a,b}(x_1)=y_1\\
&P_{a,b}(x_2)=y_2
\end{align}
which implies that the induced distribution is uniform.
\section{CQ-decoupling theorem for pairwise $\varepsilon$-dependent families of permutations}
We generalize the CQ-decoupling theorem for the partial trace derived in Section 5.3 in the sense that we only average over a $\varepsilon$-almost pairwise independent family of permutations. This can be seen as an example. It is also possible to generalize the results of the Sections 5.4 and 5.5 in the same sense but we won't do that explicitly since the derivations strongly resemble this chapter's discussion.\\
\\
\textbf{Theorem:} (CQ-Decoupling Theorem for the Partial Trace)\textit{
Let $\rhoAR$ be classical on $\HA$ with respect to $\{\ket{i}\}_{i=1,...,d_A}$ and let $A=(A_1A_2)$ be a composite system, then the average distance from uniform is given by
\begin{align}
\frac{1}{\abs{\cF}}\sum_{\idx{P}{A}\in\cF}&\Norm{\textnormal{tr}_{\textnormal{\tiny{A}}_2}{(\idx{P}{A} \otimes \idi{R} \ \rhoAR \  \idx{P}{A}^\dagger \otimes \idi{R})} - \pi_{\textnormal{\tiny{A}}_1}\otimes \idx{\rho}{R}}{1}\nonumber\\
&\leq\sqrt{ d_{\textnormal{\tiny{A}}_1}\left(\frac{d_{\textnormal{\tiny{A}}}-d_{\textnormal{\tiny{A}}_2}}{\idx{d}{A}-1}+4\varepsilon\idx{d}{A} \right)\:2^{-\chtwo{A}{R}{\rho}}},\nonumber
\end{align}
where the average goes over a family $\cF$ of pairwise $\varepsilon$-almost independent permutation operators, which act by permuting the basis vectors of $\{\ket{i}\}_{i=1,...,d_A}$.
}\\
\\
By an application of \eqref{applyCQdec} we get that
\begin{align}
&\frac{1}{\abs{\cF}}\sum_{\idx{P}{A}\in\cF}\Norm{\textnormal{tr}_{\textnormal{\tiny{A}}_2}{(\idx{P}{A} \otimes \idi{R} \ \rhoAR \  \idx{P}{A}^\dagger \otimes \idi{R})} - \pi_{\textnormal{\tiny{A}}_1}\otimes \idx{\rho}{R}}{1}^2\nonumber\\
&\leq d_{\textnormal{\tiny{A}}_1}\:\frac{1}{\abs{\cF}}\sum_{\idx{P}{A}\in\cF}\Norm{\textnormal{tr}_{\textnormal{\tiny{A}}_2}{(\idx{P}{A} \otimes \idi{R} \ \idx{\tilde{\rho}}{AR} \  \idx{P}{A}^\dagger \otimes \idi{R})} - \pi_{\textnormal{\tiny{A}}_1} \otimes \idx{\tilde{\rho}}{R}}{2}^2\\
&= d_{\textnormal{\tiny{A}}_1}\:\frac{1}{\abs{\cF}}\sum_{\idx{P}{A}\in\cF}\Trace{\textnormal{tr}_{\textnormal{\tiny{A}}_2}{(\idx{P}{A} \otimes \idi{R} \ (\idx{\tilde{\rho}}{AR}  - \pi_{\textnormal{\tiny{A}}} \otimes \idx{\tilde{\rho}}{R})\  \idx{P}{A}^\dagger \otimes \idi{R})}^2}\\
&=d_{\textnormal{\tiny{A}}_1}\:\frac{1}{\abs{\cF}}\sum_{\idx{P}{A}\in\cF}{\Trace{\left((\idx{P}{A} \otimes \idi{R})^{\otimes2} (\idx{\tilde{\rho}}{AR}  - \pi_{\textnormal{\tiny{A}}} \otimes \idx{\tilde{\rho}}{R})^{\otimes2} (\idx{P}{A}^\dagger \otimes \idi{R})^{\otimes2}\right)\:(\id_{\textnormal{\tiny{A}}_2\textnormal{\tiny{A'}}_2}\otimes\cF_{\textnormal{\tiny{A}}_1\textnormal{\tiny{A'}}_1})\otimes\idx{\cF}{R}}}\label{norep}.
\end{align}
The (indefinite) operator $\idx{\tilde{\rho}}{AR}  - \pi_{\textnormal{\tiny{A}}} \otimes \idx{\tilde{\rho}}{R}$ has CQ-structure. It can be written as
\begin{align}
\idx{\tilde{\rho}}{AR}-\pi_{\textnormal{\tiny{A}}}\otimes\idx{\tilde{\rho}}{R}
&=\sum_{i}{\proji{i}{i}{A}\otimes\idx{\tilde{\rho}}{R}^{[i]}}-\sum_{i}{\proji{i}{i}{A}\otimes\frac{1}{\idx{d}{A}}\idx{\tilde{\rho}}{R}}\\
&=\sum_{i}{\proji{i}{i}{A}\otimes\left(\idx{\tilde{\rho}}{R}^{[i]}-\frac{1}{\idx{d}{A}}\idx{\tilde{\rho}}{R}\right)}\\
&=:\sum_{i}{\proji{i}{i}{A}\otimes\idx{\tilde{\mu}}{R}^{[i]}}.
\end{align}
Then (omitting the pre-factor) equation \eqref{norep} becomes
\begin{align}
&\frac{1}{\abs{\cF}}\sum_{\idx{P}{A}\in\cF}{\Trace{\left((\idx{P}{A} \otimes \idi{R})^{\otimes2}\  (\idx{\tilde{\rho}}{AR}  - \pi_{\textnormal{\tiny{A}}} \otimes \idx{\tilde{\rho}}{R})^{\otimes2}\   (\idx{P}{A}^\dagger \otimes \idi{R})^{\otimes2}\right)\:(\id_{\textnormal{\tiny{A}}_2\textnormal{\tiny{A'}}_2}\otimes\cF_{\textnormal{\tiny{A}}_1\textnormal{\tiny{A'}}_1})\otimes\idx{\cF}{R}}}\nonumber\\
&=\frac{1}{\abs{\cF}}\sum_{i,j}{\sum_{\idx{P}{A}\in\cF}{\Trace{\left(\idx{P}{A}\proji{i}{i}{A}\idx{P}{A}^\dagger \otimes \idx{P}{A'}\proji{j}{j}{A'}\idx{P}{A'}^\dagger\right)\otimes\idx{\tilde{\mu}}{R}^{[i]}\otimes\idx{\tilde{\mu}}{R'}^{[j]}\:(\id_{\textnormal{\tiny{A}}_2\textnormal{\tiny{A'}}_2}\otimes\cF_{\textnormal{\tiny{A}}_1\textnormal{\tiny{A'}}_1})\otimes\idx{\cF}{R}}}}\\
&=\sum_{i,j}{\Trace{\left((\cP_W-\cP_H)(\proji{i}{i}{A}\otimes\proji{j}{j}{A'})\otimes\idx{\tilde{\mu}}{R}^{[i]}\otimes\idx{\tilde{\mu}}{R'}^{[j]}\right)\:
(\id_{\textnormal{\tiny{A}}_2\textnormal{\tiny{A'}}_2}\otimes\cF_{\textnormal{\tiny{A}}_1\textnormal{\tiny{A'}}_1})\otimes\idx{\cF}{R}}}\nonumber\\
&\quad+\sum_{i,j}{\Trace{\cP_H(\proji{i}{i}{A}\otimes\proji{j}{j}{A'})\otimes\idx{\tilde{\mu}}{R}^{[i]}\otimes\idx{\tilde{\mu}}{R'}^{[j]}\:(\id_{\textnormal{\tiny{A}}_2\textnormal{\tiny{A'}}_2}\otimes\cF_{\textnormal{\tiny{A}}_1\textnormal{\tiny{A'}}_1})\otimes\idx{\cF}{R}}}\label{projsumin}.
\end{align}
We added and subtracted the term $\cP_H(\proji{i}{i}{A}\otimes\proji{j}{j}{A'})$ in the last step. This allows for an application of the defining property of the $\varepsilon$-almost pairwise independent family of permutations in the first term of \eqref{projsumin}. The second term in equation \eqref{projsumin} is bounded in Section 5.3. As was already the case when we considered decoupling with $\varepsilon$-almost unitary 2-designs, this term corresponds to taking the average over the whole group. Thus in our case it is just one of the intermediate equations in the derivation of the CQ-Decoupling Theorem for Partial Trace. We will add the contribution of the second term in the end of the derivation but now we focus our attention to the first term.
A short reformulation shows that
\begin{align}
&\sum_{i,j}{\Trace{\left((\cP_W-\cP_H)(\proji{i}{i}{A}\otimes\proji{j}{j}{A'})\otimes\idx{\tilde{\mu}}{R}^{[i]}\otimes\idx{\tilde{\mu}}{R'}^{[j]}\right)\:(\id_{\textnormal{\tiny{A}}_2\textnormal{\tiny{A'}}_2}\otimes\cF_{\textnormal{\tiny{A}}_1\textnormal{\tiny{A'}}_1})\otimes\idx{\cF}{R}}}\nonumber\\
&=\sum_{i,j}{\Trace{(\cP_W-\cP_H)(\proji{i}{i}{A}\otimes\proji{j}{j}{A'})\:(\id_{\textnormal{\tiny{A}}_2\textnormal{\tiny{A'}}_2}\otimes\cF_{\textnormal{\tiny{A}}_1\textnormal{\tiny{A'}}_1})}\:\Trace{\idx{\tilde{\mu}}{R}^{[i]}\otimes\idx{\tilde{\mu}}{R'}^{[j]}\:\idx{\cF}{R}}}\\
&=\sum_{i,j}{\Trace{(\cP_W-\cP_H)(\proji{i}{i}{A}\otimes\proji{j}{j}{A'})\:(\id_{\textnormal{\tiny{A}}_2\textnormal{\tiny{A'}}_2}\otimes\cF_{\textnormal{\tiny{A}}_1\textnormal{\tiny{A'}}_1})}\:\Trace{\idx{\tilde{\mu}}{R}^{[i]}\:\idx{\tilde{\mu}}{R}^{[j]}}}\label{writeout}.
\end{align}
We bound the trace term with the Schatten 1-norm to find that
\begin{align}
&\Trace{(\cP_W-\cP_H)(\proji{i}{i}{A}\otimes\proji{j}{j}{A'})\:(\id_{\textnormal{\tiny{A}}_2\textnormal{\tiny{A'}}_2}\otimes\cF_{\textnormal{\tiny{A}}_1\textnormal{\tiny{A'}}_1})}\nonumber\\
&\leq\Norm{(\cP_W-\cP_H)(\proji{i}{i}{A}\otimes\proji{j}{j}{A'})\:(\id_{\textnormal{\tiny{A}}_2\textnormal{\tiny{A'}}_2}\otimes\cF_{\textnormal{\tiny{A}}_1\textnormal{\tiny{A'}}_1})}{1}\\
&\leq\Norm{(\cP_W-\cP_H)(\proji{i}{i}{A}\otimes\proji{j}{j}{A'})}{1}\:\Norm{(\id_{\textnormal{\tiny{A}}_2\textnormal{\tiny{A'}}_2}\otimes\cF_{\textnormal{\tiny{A}}_1\textnormal{\tiny{A'}}_1})}{\infty}\\
&=\Norm{(\cP_W-\cP_H)(\proji{i}{i}{A}\otimes\proji{j}{j}{A'})}{1}\\
&\leq\varepsilon.
\end{align}
For the last step we used the fact that $\cP_W$ constitutes an $\varepsilon$-almost pairwise independent family of permutations together with the results of the discussion of Section 6.1.1. In a similar manner one shows that
\begin{align}
(-\varepsilon)\leq\tracebig{(\cP_W-\cP_H)(\proji{i}{i}{A}\otimes\proji{j}{j}{A'})\:(\id_{\textnormal{\tiny{A}}_2\textnormal{\tiny{A'}}_2}\otimes\cF_{\textnormal{\tiny{A}}_1\textnormal{\tiny{A'}}_1})}.
\end{align}
To see where the upper bound and where the lower bound is required we write out the $\idx{\tilde{\mu}}{R}^{[i]}$ and $\idx{\tilde{\mu}}{R}^{[j]}$ in equation \eqref{writeout} and get
\begin{align}
&\sum_{i,j}{\Trace{(\cP_W-\cP_H)(\proji{i}{i}{A}\otimes\proji{j}{j}{A'})\:(\id_{\textnormal{\tiny{A}}_2\textnormal{\tiny{A'}}_2}\otimes\cF_{\textnormal{\tiny{A}}_1\textnormal{\tiny{A'}}_1})}\:\Trace{\idx{\tilde{\mu}}{R}^{[i]}\:\idx{\tilde{\mu}}{R}^{[j]}}}\nonumber\\
&=\sum_{i,j}{\Trace{(\cP_W-\cP_H)(\proji{i}{i}{A}\otimes\proji{j}{j}{A'})\:(\id_{\textnormal{\tiny{A}}_2\textnormal{\tiny{A'}}_2}\otimes\cF_{\textnormal{\tiny{A}}_1\textnormal{\tiny{A'}}_1})}\:\Trace{\idx{\tilde{\rho}}{R}^{[i]}\:\idx{\tilde{\rho}}{R}^{[j]}}}\nonumber\\
&\quad-\frac{1}{\idx{d}{A}}\sum_{i,j}{\Trace{(\cP_W-\cP_H)(\proji{i}{i}{A}\otimes\proji{j}{j}{A'})\:(\id_{\textnormal{\tiny{A}}_2\textnormal{\tiny{A'}}_2}\otimes\cF_{\textnormal{\tiny{A}}_1\textnormal{\tiny{A'}}_1})}\:\Trace{\idx{\tilde{\rho}}{R}^{[i]}\:\idx{\tilde{\rho}}{R}}}\nonumber\\
&\quad-\frac{1}{\idx{d}{A}}\sum_{i,j}{\Trace{(\cP_W-\cP_H)(\proji{i}{i}{A}\otimes\proji{j}{j}{A'})\:(\id_{\textnormal{\tiny{A}}_2\textnormal{\tiny{A'}}_2}\otimes\cF_{\textnormal{\tiny{A}}_1\textnormal{\tiny{A'}}_1})}\:\Trace{\idx{\tilde{\rho}}{R}\:\idx{\tilde{\rho}}{R}^{[i]}}}\nonumber\\
&\quad+\frac{1}{\idx{d}{A}^2}\sum_{i,j}{\Trace{(\cP_W-\cP_H)(\proji{i}{i}{A}\otimes\proji{j}{j}{A'})\:(\id_{\textnormal{\tiny{A}}_2\textnormal{\tiny{A'}}_2}\otimes\cF_{\textnormal{\tiny{A}}_1\textnormal{\tiny{A'}}_1})}\:\Trace{\idx{\tilde{\rho}}{R}\:\idx{\tilde{\rho}}{R'}}}\\
&\leq\varepsilon\cdot\sum_{i,j}{\Trace{\idx{\tilde{\rho}}{R}^{[i]}\:\idx{\tilde{\rho}}{R}^{[j]}}}+\varepsilon\cdot\frac{2}{\idx{d}{A}}\sum_{i,j}{\Trace{\idx{\tilde{\rho}}{R}^{[i]}\:\idx{\tilde{\rho}}{R}}}+\varepsilon\cdot\frac{1}{\idx{d}{A}^2}\sum_{i,j}{\Trace{\idx{\tilde{\rho}}{R}\:\idx{\tilde{\rho}}{R}}}\\
&=4\varepsilon\cdot{\Trace{\idx{\tilde{\rho}}{R}\:\idx{\tilde{\rho}}{R}}}\label{ptrCQ}.
\end{align}
Equation \eqref{ptrCQ} makes use of the CQ-structure of $\idx{\rho}{AR}$. Note that the partial trace $\idx{\tilde{\rho}}{R}$ of $\idx{\tilde{\rho}}{AR}$ is given by $\idx{\tilde{\rho}}{R}=\sum_{i}\idx{\tilde{\rho}}{R}^{[i]}$.
An application of the Cauchy-Schwarz inequality shows that
\begin{align}
\Trace{\idx{\tilde{\rho}}{R}^2}&=\Trace{\idx{\tilde{\rho}}{AR}\otimes\idi{A'}\:\idx{\tilde{\rho}}{A'R}\otimes\idi{A}}\\
&\leq\sqrt{\Trace{(\idx{\tilde{\rho}}{AR}\otimes\idi{A'})^2}\:\Trace{(\idx{\tilde{\rho}}{A'R}\otimes\idi{A})^2}}\\
&=\idx{d}{A}\:\Trace{\idx{\tilde{\rho}}{AR}^2}.
\end{align}
This bound is very bad since it does not use the given CQ-structure of $\idx{\tilde{\rho}}{AR}$ but until now, I was not able to find a significantly better bound (i.e.\:a bound that does not involve any dimension factors). With equation \eqref{ptrCQ} this yields a bound on the first term of \eqref{projsumin}
\begin{align}
&\sum_{i,j}{\Trace{\left((\cP_W-\cP_H)(\proji{i}{i}{A}\otimes\proji{j}{j}{A'})\otimes\idx{\tilde{\mu}}{R}^{[i]}\otimes\idx{\tilde{\mu}}{R'}^{[j]}\right)\:(\id_{\textnormal{\tiny{A}}_2\textnormal{\tiny{A'}}_2}\otimes\cF_{\textnormal{\tiny{A}}_1\textnormal{\tiny{A'}}_1})\otimes\idx{\cF}{R}}}\nonumber\\
&\leq4\varepsilon\idx{d}{A}\Trace{\idx{\tilde{\rho}}{AR}^2}.
\end{align}
The bound on the second term of \eqref{projsumin} is 
\begin{align}
\sum_{i,j}{\Trace{\cP_H(\proji{i}{i}{A}\otimes\proji{j}{j}{A'})\otimes\idx{\tilde{\mu}}{R}^{[i]}\otimes\idx{\tilde{\mu}}{R'}^{[j]}\:(\id_{\textnormal{\tiny{A}}_2\textnormal{\tiny{A'}}_2}\otimes\cF_{\textnormal{\tiny{A}}_1\textnormal{\tiny{A'}}_1})\otimes\idx{\cF}{R}}\leq\frac{d_{\textnormal{\tiny{A}}}-d_{\textnormal{\tiny{A}}_2}}{\idx{d}{A}-1}\:\Trace{\idx{\tilde{\rho}}{AR}^2}}.
\end{align}
We conclude taking together both results that
\begin{align}
&\frac{1}{\abs{\cF}}\sum_{\idx{P}{A}\in\cF}\Norm{\textnormal{tr}_{\textnormal{\tiny{A}}_2}{(\idx{P}{A} \otimes \idi{R} \ \rhoAR \  \idx{P}{A}^\dagger \otimes \idi{R})} - \pi_{\textnormal{\tiny{A}}_1}\otimes \idx{\rho}{R}}{1}^2\nonumber\\
&\leq d_{\textnormal{\tiny{A}}_1}\:\left(\frac{d_{\textnormal{\tiny{A}}}-d_{\textnormal{\tiny{A}}_2}}{\idx{d}{A}-1}\:\Trace{\idx{\tilde{\rho}}{AR}^2}+4\varepsilon\idx{d}{A}\Trace{\idx{\tilde{\rho}}{AR}^2} \right).
\end{align}
As always one adjusts the $\zetaR$ in $\idx{\tilde{\rho}}{AR}^2$ to get the $H_2$-entropy. Then the square root is taken on both sides followed by an application of the Jensen inequality to find that
\begin{align}
&\frac{1}{\abs{\cF}}\sum_{\idx{P}{A}\in\cF}\Norm{\textnormal{tr}_{\textnormal{\tiny{A}}_2}{(\idx{P}{A} \otimes \idi{R} \ \rhoAR \  \idx{P}{A}^\dagger \otimes \idi{R})} - \pi_{\textnormal{\tiny{A}}_1}\otimes \idx{\rho}{R}}{1}\nonumber\\
&\leq\sqrt{d_{\textnormal{\tiny{A}}_1}\left(\frac{d_{\textnormal{\tiny{A}}}-d_{\textnormal{\tiny{A}}_2}}{\idx{d}{A}-1}+4\varepsilon\idx{d}{A} \right)\:2^{-\chtwo{A}{R}{\rho}}}.
\end{align}
\chapter{Decoupling Quantum States with Permutation Operators}
The aim of this chapter is to understand whether permutations operators can be used for decoupling applications in a quantum context. We would like to generalize the discussion of Chapter 5 dropping the assumption that the state of the system has CQ-structure.\\
This chapter requires a solid understanding of the representation theory of the symmetric group $S_d$. An introduction to this topic is beyond the scope of this project but can be found in \cite{SymGroup}.
\section{The general setup}
Instead of bounding the term
\begin{align}
\int_{\mathbb{U}(A)} &{\left\| \mathcal{T}(\idx{U}{A} \otimes \idi{R} \ \rhoAR \  \idx{U}{A}^\dagger \otimes \idi{R}) - \idx{\omega}{E} \otimes \idx{\rho}{R} \right\|}_1 \d{U}
\end{align}
as is done in the Decoupling Theorem, one might be interested in an upper bound on expressions of the type
\begin{align}
\frac{1}{\abs{\mathbb{P}}}\sum_{P\in\mathbb{P}(A)}{\left\| \mathcal{T}(\idx{P}{A} \otimes \idi{R} \ \rhoAR \  \idx{P}{A}^\dagger \otimes \idi{R}) - \idx{\omega}{E} \otimes \idx{\rho}{R} \right\|}_1\label{dcplwprm}.
\end{align}
It turns out that for arbitrary $\idx{\rho}{AR}$ and $\cTAE$ the computational effort required for the solution of this problem is big. Moreover the occurring bounds are difficult to interpret in terms of entropies. Nevertheless the presented method is general in the sense that it can be used to obtain upper bounds on \eqref{dcplwprm} for any $\idx{\rho}{AR}$ and any $\cTAE$. It even allows to bound the distance from states different than $\idx{\omega}{R}\otimes\idx{\rho}{R}$.
Generally the calculations in this section will be very close to the calculations done in the proof of the Decoupling Theorem.
Conceptually we therefore can follow the discussion of Chapter 2.  We will study three variations of formula \eqref{dcplwprm} in this thesis, which we consider to be interesting. The first one shows how much classicalizing a map can be made by a pre-concatenation of a permutation operator. We prove
\begin{align}
\frac{1}{\abs{\mathbb{P}}}\sum_{P\in\mathbb{P}(A)}{\left\| \mathcal{T}(\idx{P}{A} \otimes \idi{R} \ (\idx{\Phi}{AR} - \idx{T}{AR})\ \idx{P}{A}^\dagger \otimes \idi{R}) \right\|}_1 
\leq\sqrt{d_A\frac{\idx{d}{R}-1}{\idx{d}{A}-1}\:2^{-\chmin{A'}{E}{\omega}}},\nonumber
\end{align}
where we define $\idx{\Phi}{AR}:=\frac{1}{\idx{d}{R}}\sum_{i,j}^{\idx{d}{R}}{\proji{i}{j}{A}\otimes\proji{i}{j}{R}}$ and $\idx{T}{AR}:=\frac{1}{\idx{d}{R}}\sum_{i}^{\idx{d}{R}}{\proji{i}{i}{A}\otimes\proji{i}{i}{R}}$.
The second one is an important special case of a general decoupling theorem with permutations. There we bound
\begin{align}
\frac{1}{\abs{\mathbb{P}}}\sum_{P\in\mathbb{P}(A)}{\left\| \mathcal{T}(\idx{P}{A} \otimes \idi{R} \ (\idx{\Phi}{AR} - \idx{\pi}{AR})\ \idx{P}{A}^\dagger \otimes \idi{R}) \right\|}_1\leq\sqrt{2\frac{\idx{d}{A}^2}{\idx{d}{R}}\: \frac{\idx{d}{R}-1}{\idx{d}{A}-1}\:2^{-\chmin{A'}{E}{\omega}}}\nonumber,
\end{align}
where $\idx{\Phi}{AR}$ is defined as above and $\idx{\pi}{AR}:=\frac{1}{\idx{d}{R}^2}\sum_{i,j}^{\idx{d}{R}}{\proji{i}{i}{A}\otimes\proji{j}{j}{R}}$.
Finally we generalize on the discussion of Section~5.2 dropping the assumption that the $A$ system is classical. This provides us with a geralization of the Hash Lemma (as stated for instance in \cite{TSSR10}) to the fully quantum context. We obtain that
\begin{align}
&\frac{1}{\mathbb{\abs{P}}}\sum_{P\in\mathbb{P}(A)}{\Norm{\tr_{A_2}{(\idx{P}{A}\otimes\idi{R}\:\rhoAR\:\idx{P}{A}^\dagger\otimes\idi{R})}-\pi_{A_1}\otimes\rhoR}{1}}
\leq\sqrt{2\:{d}_{A_1}\:2^{-\chmin{A}{R}{\rho}}}.\nonumber
\end{align}
The structure of the derivations is as always. We first derive ``decoupling lemmata'', where instead of the Schatten 1-norm, the above expressions contain the Schatten 2-norm. Then we use the H\"older inequality to go to the Schatten 1-norm and to obtain bounds in terms of entropic quantities. Finally the Jensen Inequality is applied and we lower bound the different entropies with the $\idx{H}{min}$-entropy.
Since we would like to study two different special cases we keep the discussion completely general for the moment. We will specialize to the above cases as soon as this becomes necessary. To keep the discussion general we introduce the hermitian operator $\idx{\lambda}{AR}$ and write
\begin{align}
&\frac{1}{\abs{\mathbb{P}}}\sum_{P\in\mathbb{P}(A)}{\left\| \mathcal{T}(\idx{P}{A} \otimes \idi{R} \ \idx{\lambda}{AR}\ \idx{P}{A}^\dagger \otimes \idi{R}) \right\|}_2^2\nonumber\\
&=\frac{1}{\abs{\mathbb{P}}}\sum_{P\in\mathbb{P}(A)}\Trace{\mathcal{T}(\idx{P}{A} \otimes \idi{R} \ \idx{\lambda}{AR}\ \idx{P}{A}^\dagger \otimes \idi{R})^2}\label{usetoclal}\\
&=\frac{1}{\abs{\mathbb{P}}}\sum_{P\in\mathbb{P}(A)}\Trace{\mathcal{T}^{\otimes2}\left((\idx{P}{A} \otimes \idi{R})^{\otimes2} \ (\idx{\lambda}{AR})^{\otimes2}\ (\idx{P}{A}^\dagger \otimes \idi{R})^{\otimes2}\right)\:\idx{\cF}{ER}}\\
&=\frac{1}{\abs{\mathbb{P}}}\sum_{P\in\mathbb{P}(A)}\Trace{(\idx{\lambda}{AR})^{\otimes2}\ \left((\idx{P}{A}^\dagger)^{\otimes2}(\cT^\dagger)^{\otimes2}[\idx{\cF}{E}](\idx{P}{A})^{\otimes2}\right)\otimes\idx{\cF}{R}}\\
&=\Trace{(\idx{\lambda}{AR})^{\otimes2}\ \frac{1}{\abs{\mathbb{P}}}\sum_{P\in\mathbb{P}(A)}\left((\idx{P}{A}^\dagger)^{\otimes2}(\cT^\dagger)^{\otimes2}[\idx{\cF}{E}](\idx{P}{A})^{\otimes2}\right)\otimes\idx{\cF}{R}}\label{followpl}.
\end{align}
The evaluation of the term  
\begin{align}
\sum_{P\in\mathbb{P}(A)}\left((\idx{P}{A}^\dagger)^{\otimes2}(\cT^\dagger)^{\otimes2}[\idx{\cF}{E}](\idx{P}{A})^{\otimes2}\right)
=\sum_{P\in\mathbb{P}(A)}\left((\idx{P}{A})^{\otimes2}(\cT^\dagger)^{\otimes2}[\idx{\cF}{E}](\idx{P}{A}^\dagger)^{\otimes2}\right)
\end{align}
constitutes the major part of this chapter and is performed in the following section.
\section{The mathematical backbone for decoupling theorems with permutations}
This section is partitioned in five subsections in which we present different ingredients, which taken together will allow us to calculate $\sum_{P\in\mathbb{P}(A)}{(\idx{P}{A})^{\otimes2}(\cT^\dagger)^{\otimes2}[\idx{\cF}{E}](\idx{P}{A}^\dagger)^{\otimes2}}$. The first subsection introduces some basic notions from representation theory and shows how they are related to our problem. The following three subsections deal with the analysis of the commutant of a given representation and specialize this to our concrete example. In the last subsection we compute $\sum_{P\in\mathbb{P}(A)}{(\idx{P}{A})^{\otimes2}(\cT^\dagger)^{\otimes2}[\idx{\cF}{E}](\idx{P}{A}^\dagger)^{\otimes2}}$. The rest of this chapter is then organized in two further sections, ``Distance from classicality'' and ``Decoupling with permutation operators'', in which we apply the result of this section.
\subsection{Basics from representation theory}
As already mentioned in the beginning an introduction to the area of representation theory cannot be given in this thesis. Nevertheless it is inevitable to introduce some notational conventions and fix the terminology of the following subsections.\\
Any element $\sigma$ of the symmetric group $S_d$ can be written in \textit{cycle notation}. There, a permutation is represented by a sequence of cycles $(...)\:(...)\:...\:(...)$ with each cycle containing a sequence of elements of $\{1,...,d\}$. Each entry of a cycle is mapped by the corresponding permutation to the following entry in the cycle. The last entry of a cycle is mapped to its first one. For instance the cycle $(i\:j\:k)$ corresponds to a permutation in $S_3$ that maps $i$ to $j$, $j$ to $k$ and $k$ back to $i$. For a given permutation one can count the different cycles in it. We call the tuple $\left((1)^{k_1},(2)^{k_2},...,(d)^{k_d}\right)$ the \textit{cycle structure} of a permutation with $k_1$ cycles of length one, $k_2$ cycles of length two etc. Note that the integers $k_i$ are constrained to satisfy $\sum_{i}ik_{i}=d$. It is easy to show that the cycle structure of some permutation in $S_d$ is invariant under the conjugation of this permutation with an arbitrary element of $S_d$. Two permutations are in the same conjugacy class if and only if their cycle structure is the same. It is therefore possible to label the conjugacy classes of $S_d$ by tuples $\left((1)^{k_1},(2)^{k_2},...,(d)^{k_d}\right)$. Furthermore, we call a non increasing sequence of natural numbers $\lambda=\left(\lambda_1\geq\lambda_2\geq...\geq\lambda_n\right)$ with $\sum_{i}^{n}\lambda_i=d$ a \textit{partition} of $d$. From the above labeling of conjugacy classes of $S_d$ we see that it is alternatively possible to label these classes via partitions of $d$. A function from a group into some field which is constant on each conjugacy class of that group is called a \textit{class function}.\\
An extremely important concept in Quantum Mechanics is the one of representing the action of some group (as for example $S_d$) on a vector space.
\begin{definition}(Representation)
Let $V$ be a $d$-dimensional $\mathbb{C}$-vector space. A (linear) representation of a group $G$ is a group homomorphism
\begin{align}
X:\:G&\rightarrow Gl(V)\nonumber\\
g&\mapsto X(g)\nonumber,
\end{align}
where with $Gl(V)$ we denote the General Linear Group of invertible linear maps from $V$ onto itself.
\end{definition}
In our context it will often be sufficient to fix a basis for $V$ and think of the representation as attaching to each group element an invertible $d\times d$\:-\:matrix. This type of representations will be called matrix representations. We have already seen the defining representation of the group $S_d$, which can be seen as a matrix representation assigning to each element $\sigma\in S_d$ the corresponding permutation operator $P(\sigma)$ (see Definition 9) via
\begin{align}
P:\:S_d&\rightarrow Gl(d\times d,\mathbb{C})\nonumber\\
\sigma&\mapsto P(\sigma)\nonumber,
\end{align}
where with $Gl(d\times d,\mathbb{C})$ we denote the group of invertible $d\times d$\:-\:matrices. Another interesting representation is the \textit{swap representation} of $S_2$. For $V$ being a $d$-dimensional vector space one defines
\begin{align}
S:\:S_2&\rightarrow Gl(V\otimes V)\nonumber\\
\sigma&\mapsto S(\sigma):=\begin{cases}
\id\  \text{if}\ \sigma=e\\
\cF\ \text{otherwise}
\end{cases}\nonumber
\end{align}
where we wrote $e$ for the neutral element of $S_2$ and $\cF$ is the swap operator on $V\otimes V$.
In our concrete setup the following representation will be of particular interest. We denote by $S_d\times S_2$ the group of pairs of elements from $S_d$ and $S_2$ and define for $d$-dimensional $V$ the representation 
\begin{align}
R:\:S_d\times S_2&\rightarrow Gl(V\otimes V)\nonumber\\
(\sigma,\pi)&\mapsto R((\sigma,\pi)):=(P(\sigma)\otimes P(\sigma),S(\pi))\nonumber,
\end{align}
where the pair $(P(\sigma)\otimes P(\sigma),S(\pi))$ acts on $V\otimes V$ by multiplication: For $\ket{i}\otimes\ket{j}\in V\otimes V$ one has
\begin{align}
(P(\sigma)\otimes P(\sigma),S(\pi))(\ket{i}\otimes\ket{j}):=(P(\sigma)\otimes P(\sigma))\cdot S(\pi)(\ket{i}\otimes\ket{j}).
\end{align}
Since the actions of $P(\sigma)\otimes P(\sigma)$ and $S(\pi)$ commute the representation is well defined.\\
Later we will decompose this representation into irreducible representations of $S_d\times S_2$. For this it is convenient to label the irreducible representations of $S_d$ using partitions of $d$. (The irreducible representations are in one to one correspondence to the conjugacy classes of $S_d$.) For example we will write $\pi_{(d-1,1)}$ for the irreducible representation $\pi$ of $S_d$ belonging to the partition $(d-1,1)$ (or the Young Frame with $d-1$ boxes in the first row and one box in the second one).\\
For a given representation we define its ``character'' and its ``commutant''. Thess objects will be of interest for our problem.
\begin{definition}(Character \cite{SymGroup})
Given a representation $X$ of $G$ on a $\mathbb{C}$-vectorspace, the character of $X$ is a map 
\begin{align}
\chi\::\:G&\rightarrow\mathbb{C}\nonumber\\
g\:&\mapsto \chi(g):=\trace{X(g)},\nonumber
\end{align}
where the trace is taken of one of the matrices representing the map $X(g)$.
\end{definition}
The characters are well defined since two different matrices $A$, $B$ representing the same linear map from $V$ into iself are related via
$B=S\:A\:S^{-1}$ for some fixed $S$. Therefore $\trace{B}=\trace{A}$.
From the definition it is also clear that characters are class functions.
\begin{definition}(Commutant \cite{SymGroup}) Let $\textnormal{Mat}(d\times d,\mathbb{C})$ be the set of all $d\times d$\:-\:matrices with complex entries and let $Gl(d\times d,\mathbb{C})\subset\textnormal{Mat}(d\times d,\mathbb{C})$ be the subset of invertible matrices. For a matrix representation ${X:\:G\rightarrow Gl(d\times d,\mathbb{C})}$ the corresponding commutant is
\begin{align}
\textnormal{Com}(X)=\{T\in\textnormal{Mat}(d\times d,\mathbb{C})\ :\ T\cdot X(g)=X(g)\cdot T\quad\forall g\in G\}\nonumber.
\end{align}
\end{definition}
From the definition it is evident that $\textnormal{Com}(X)$ has the structure of an algebra. We now consider the term 
\begin{align}
\sum_{P\in\mathbb{P}(A)}{(\idx{P}{A})^{\otimes2}(\cT^\dagger)^{\otimes2}[\idx{\cF}{E}](\idx{P}{A}^\dagger)^{\otimes2}}=\sum_{P\in\mathbb{P}(A)}{(\idx{P}{A})^{\otimes2}\left(\sum_{i,j}{\cT^\dagger(\proji{i}{j}{A})\otimes\cT^\dagger(\proji{j}{i}{A'})}\right)(\idx{P}{A}^\dagger)^{\otimes2}}\label{writeoutT}
\end{align}
and note that 
\begin{align}
\sum_{P\in\mathbb{P}(A)}{(\idx{P}{A})^{\otimes2}(\cT^\dagger)^{\otimes2}[\idx{\cF}{E}](\idx{P}{A}^\dagger)^{\otimes2}}\quad\in\quad\textnormal{Com}(R).
\end{align}
This is because for any $(\sigma,\pi)\in S_d\times S_2$ we have that
\begin{align}
&R((\sigma,\pi))\left(\sum_{P\in\mathbb{P}(A)}{(\idx{P}{A})^{\otimes2}(\cT^\dagger)^{\otimes2}[\idx{\cF}{E}](\idx{P}{A}^\dagger)^{\otimes2}}\right)R((\sigma,\pi))^{-1}\nonumber\\
&=(P(\sigma)\otimes P(\sigma))\cdot S(\pi)\left(\sum_{P\in\mathbb{P}(A)}{(\idx{P}{A})^{\otimes2}(\cT^\dagger)^{\otimes2}[\idx{\cF}{E}](\idx{P}{A}^\dagger)^{\otimes2}}\right) S(\pi)^{-1}\cdot(P(\sigma)\otimes P(\sigma))^{-1}\\
&=(P(\sigma)\otimes P(\sigma))\left(\sum_{P\in\mathbb{P}(A)}{(\idx{P}{A})^{\otimes2}(\cT^\dagger)^{\otimes2}[\idx{\cF}{E}](\idx{P}{A}^\dagger)^{\otimes2}}\right) (P(\sigma)\otimes P(\sigma))^{\dagger}\label{usswappro}\\
&=\sum_{P\in\mathbb{P}(A)}{(\idx{P}{A})^{\otimes2}(\cT^\dagger)^{\otimes2}[\idx{\cF}{E}](\idx{P}{A}^\dagger)^{\otimes2}}\label{useperminv}.
\end{align}
In equation \eqref{usswappro} we used the fact that $S(\pi)$ is either the identity or the swap operator. The invariance under the conjugation with the identity is trivial whereas one can see from equation \eqref{writeoutT} that the whole sum is symmetric in the systems $A$ and $A'$ and therefore it is also invariant under the conjugation with the swap operator. Furthermore in equation \eqref{useperminv} we used the fact that the summation takes place over all permutation operators. If we multiply a permutation operator with another we still get a permutation operator. The map that corresponds to the left multiplication with a permutation operator is bijective. So the whole sum still goes over all permutation operators.\\
Moreover the term $(\cT^\dagger)^{\otimes2}[\idx{\cF}{E}]$ is hermitian, such that 
\begin{align}
\left(\sum_{P\in\mathbb{P}(A)}{(\idx{P}{A})^{\otimes2}(\cT^\dagger)^{\otimes2}[\idx{\cF}{E}](\idx{P}{A}^\dagger)^{\otimes2}}\right)^\dagger=
\sum_{P\in\mathbb{P}(A)}{(\idx{P}{A})^{\otimes2}(\cT^\dagger)^{\otimes2}[\idx{\cF}{E}](\idx{P}{A}^\dagger)^{\otimes2}}
\end{align}
holds for the whole sum. We denote by $\textnormal{Com}(R)^\dagger$ the set of the hermitian matrices in $\textnormal{Com}(R)$. This set has the structure of a vector space and 
\begin{align}
\sum_{P\in\mathbb{P}(A)}{(\idx{P}{A})^{\otimes2}(\cT^\dagger)^{\otimes2}[\idx{\cF}{E}](\idx{P}{A}^\dagger)^{\otimes2}}\ \in\ \textnormal{Com}(R)^\dagger.
\end{align}
This property strongly restricts the possible results of the summation.
\subsection{The structure of the commutant}
We have seen that the commutant of the representation $R$, $\textnormal{Com}(R)$, is relevant in our context. In this chapter we first analyze the structure of the commutant of a general representation and then see what are the implications on  $\textnormal{Com}(R)$.\\
It is a general fact in representation theory (Maschke's Theorem) that any (finite dimensional, unitary) representation of a finite group can be decomposed into a direct sum of irreducible representations. In particular in case of a matrix representation the matrix $X(g)$ for any $g\in G$ can be written as
\begin{align}
X(g)=
\begin{pmatrix}
 X^{[1]}(g) & 0   & \cdots & 0\\
 0 & X^{[2]}(g) &\cdots & 0\\
 \vdots & \vdots & \ddots &\vdots\\
 0 & 0 &  \cdots  & X^{[i]}(g)
\end{pmatrix}
\label{firstrepdec}
\end{align}
in a basis which corresponds to the invariant subspaces. The matrices $X^{[j]}(g)$ belong to irreducible sub representations of $X(g)$ and the block diagonal structure of $X(g)$ reflects the invariance of the corresponding irreducible subspaces. Thus for any matrix representation there exists a basis in which it can be decomposed into 
\begin{align}
X=\bigoplus_{i}{m_i\:X^{(i)}},
\end{align}
where (in contrast to \eqref{firstrepdec}) the $X^{(i)}$ are pairwise inequivalent and have multiplicity $m_i$.
This fact together with several applications of the fundamental Schur Lemma can be used to show the following theorem about the commutant of some matrix representation.\\
\\
\textbf{Theorem:} (Structure of the commutant (\cite{SymGroup} page 26))\textit{
Let G be a finite group and X be a (finite dimensional) matrix representation of G. Assume X decomposes into inequivalent, irreducible representations as $X=\bigoplus_{i}^{k}{m_i\:X^{(i)}}$ and $X^{i}$ has dimension $d_i$, then 
\begin{align}
\textnormal{Com}(X)=\{\bigoplus_{i}^{k}\:\left(M_{m_i}\otimes\id_{d_i}\right)\ |\ M_{m_i}\in\textnormal{Mat}(m_i\times m_i,\mathbb{C})\}\nonumber
\end{align}
}\\
\\
The above theorem implies that the dimension of the commutant algebra $\textnormal{Com}(X)$ only depends on the multiplicities of the irreducible representations occurring in the decomposition of $X$.
For us in the first place the commutant of $R$ is interesting. The dimension of $\textnormal{Com}(R)$ determines the amount of linearly independent matrices in this algebra. We will try to write the term $\sum_{P\in\mathbb{P}(A)}{(\idx{P}{A})^{\otimes2}(\cT^\dagger)^{\otimes2}[\idx{\cF}{E}](\idx{P}{A}^\dagger)^{\otimes2}}$ as a linear combination of basis vectors of $\textnormal{Com}(R)^\dagger$. For this we first determine the dimension $\textnormal{dim}\left(\textnormal{Com}(R)\right)$ of $\textnormal{Com}(R)$. Finding then the dimension of $\textnormal{Com}(R)^\dagger$, $\textnormal{dim}\left(\textnormal{Com}(R)^\dagger\right)$ is straight forward.
\subsection{The dimension of $\textnormal{Com}(R)^\dagger$}
We have seen that the dimension of the commutant $\textnormal{Com}(R)$ is determined by the multiplicities of the irreducible representation in $R$.
To obtain these multiplicities we give an explicit decomposition of $R$ into irreducible representations in terms of characters. We write $\chi_{\lambda}$ for the character of the irreducible matrix representation $X_{\lambda}$ of $S_d$ belonging to the conjugacy class $\lambda$. Since the irreducible representations of $S_d\times S_2$ are given by all the tensor products of the irreducible representations of $S_d$ and $S_2$ it is possible to label the characters of the irreducible representations of $S_d\times S_2$ via $\chi_{\lambda,\:\mu}$ where $\mu$ denotes a partition of $2$ (\cite{SymGroup}, Theorem 1.11.3). The character $\chi_{\lambda,\:\mu}$ can explicitly be calculated to be $\chi_{\lambda,\:\mu}=\chi_{\lambda}\:\chi_{\mu}$ with $\chi_{\lambda}$ and $\chi_{\mu}$ being characters for $S_d$ and $S_2$ respectively.\\
\\
\textbf{Claim:} In the setup of the previous subsection with $d\geq4$, we have for the character $\chi_{R}$ of $R$ that
\begin{align}
\chi_{R}=2\:\chi_{(d),\:(2)}+2\:\chi_{(d-1,1),\:(2)}+\chi_{(d-1,1),\:(1,1)}+\chi_{(d-2,1,1),\:(1,1)}+\chi_{(d-2,2),\:(2)}.\nonumber
\end{align}
\\
First we give a proof of this claim and afterwards we will see how this result can be used to compute $\textnormal{dim}(\textnormal{Com}(R)^\dagger)$. The characters of the irreducible representations of a group form an orthonormal basis for the class functions of that group. The above claim corresponds to an expansion of the class function $\chi_{R}$ into that basis. Since generally the coordinates of any vector written in some basis are unique, we conclude that if we can explicitly validate the above expansion, then that decomposition is automatically unique. All characters are class functions and therefore it is sufficient to check the claim on any conjugacy class of $S_d\times S_2$. We write a conjugacy class in $S_d$ in cycle notation labeling it with $a:=\left((1)^{k_1},(2)^{k_2},...,(d)^{k_d}\right)$ and a class in $S_2$ with $b:=\left((1)^{l_1},(2)^{l_2}\right)$. One can reformulate the claim making the dependence of the functions on the conjugacy classes explicit.  Denoting with $e$ the neutral element of $S_2$ we get
\begin{align}
\chi_{R}(a,b)&=\trace{P(a)\otimes P(a)\cdot S(b)}\\
&=\Trace{P(a)\otimes P(a)}\delta_{be}+(1-\delta_{be})\:\Trace{P(a)\otimes P(a)\:\cF}\\
&=\Trace{P(a)}^2\delta_{be}+(1-\delta_{be})\:\Trace{P(a)^2}\\
&=k_1^2\delta_{be}+(1-\delta_{be})\:(k_1+2\cdot k_2).
\end{align}
$\delta_{be}$ is the function on $S_2$ which is one on $e$ and zero else. It can be easily rewritten in terms of the characters of the irreducible representations of $S_2$.
\begin{align}
\delta_{be}=\frac{1}{2}\:(\chi_{(2)}(b)+\chi_{(1,1)}(b))
\end{align}
and
\begin{align}
1-\delta_{be}=\frac{1}{2}\:(\chi_{(2)}(b)-\chi_{(1,1)}(b)),
\end{align}
which implies that
\begin{align}
\chi_{R}(a,b)=\frac{1}{2}\:k_1^2\:\left(\chi_{(2)}(b)+\chi_{(1,1)}(b)\right)+\frac{1}{2}\:\left(\chi_{(2)}(b)-\chi_{(1,1)}(b)\right)\:\left(k_1+2\cdot k_2\right).\label{irreplft}
\end{align}
The characters on the right hand side of the claim
$$\chi_{(d),\:(2)}, \chi_{(d-1,1),\:(2)}, \chi_{(d-1,1),\:(1,1)}, \chi_{(d-2,1,1),\:(1,1)}, \chi_{(d-2,2),\:(2)}$$
can all be evaluated writing them out as products of characters of $S_d$ and $S_2$ and afterwards applying the Murnaghan-Nakayama rule (\cite{SymGroup}, Theorem 4.10.2) to calculate 
$$\chi_{(d)}, \chi_{(d-1,1)}, \chi_{(d-2,1,1)}\ \textnormal{and}\ \chi_{(d-2,2)}.$$
It is a conceptually simple but partially elaborate task (see Appendix D) to obtain
\begin{align}
\chi_{(d)}(a)&=1\\
\chi_{(d-1,1)}(a)&=k_1-1\\
\chi_{(d-2,1,1)}(a)&=\frac{1}{2}\:(k_1-1)(k_1-2)-k_2\\
\chi_{(d-2,2)}(a)&=\frac{1}{2}\:k_1(k_1-3)+k_2
\end{align}
using the Murnaghan-Nakayama rule. With this the right hand side of our claim becomes
\begin{align}
&\textnormal{\small{$2\:\chi_{(d),(2)}(a,b)+2\:\chi_{(d-1,1),(2)}(a,b)+\chi_{(d-1,1),(1,1)}(a,b)+\chi_{(d-2,1,1),(1,1)}(a,b)+\chi_{(d-2,2),(2)}(a,b)$}}\nonumber\\
&\textnormal{\small{=$2\:\chi_{(d)}(a)\:\chi_{(2)}(b)+2\:\chi_{(d-1,1)}(a)\:\chi_{(2)}(b)+\chi_{(d-1,1)}(a)\:\chi_{(1,1)}(b)$}}\nonumber\\
&\quad\textnormal{\small{$+\chi_{(d-2,1,1)}(a)\:\chi_{(1,1)}(b)+\chi_{(d-2,2)}(a)\:\chi_{(2)}(b)$}}\\
&\textnormal{\small{=$2\:\chi_{(2)}(b)+2\:(k_1-1)\:\chi_{(2)}(b)+(k_1-1)\:\chi_{(1,1)}(b)+\left(\frac{1}{2}\:(k_1-1)(k_1-2)-k_2\right)\:\chi_{(1,1)}(b)$}}\nonumber\\
&\quad\textnormal{\small{$+\left(\frac{1}{2}\:k_1(k_1-3)+k_2\right)\:\chi_{(2)}(b)$}}\\
&\textnormal{\small{=$2\:\chi_{(2)}(b)+2\:(k_1-1)\:\chi_{(2)}(b)+(k_1-1)\:\chi_{(1,1)}(b)+\left(\frac{1}{2}\:(k_1-1)(k_1-2)\right)\:\chi_{(1,1)}(b)$}}\nonumber\\
&\quad\textnormal{\small{$+\left(\frac{1}{2}\:k_1(k_1-3)\right)\:\chi_{(2)}(b)+\frac{1}{2}\:\left(\chi_{(2)}(b)-\chi_{(1,1)}(b)\right)\cdot(2\:k_2)$}}\\
&\textnormal{\small{=$2\:\chi_{(2)}(b)+2\:(k_1-1)\:\chi_{(2)}(b)+(k_1-1)\:\chi_{(1,1)}(b)+\left(\frac{1}{2}k_1^2-\frac{3}{2}k_1+1\right)\:\chi_{(1,1)}(b)$}}\nonumber\\
&\quad\textnormal{\small{$+\left(\frac{1}{2}k_1^2-\frac{3}{2}k_1\right)\:\chi_{(2)}(b)+\frac{1}{2}\:\left(\chi_{(2)}(b)-\chi_{(1,1)}(b)\right)\cdot(2\:k_2)$}}\\
&\textnormal{\small{=$\frac{1}{2}\:\left(\chi_{(2)}(b)+\chi_{(1,1)}(b)\right)\:k_1^2+\frac{1}{2}\:\left(\chi_{(2)}(b)-\chi_{(1,1)}(b)\right)\cdot(2\:k_2)$}}\nonumber\\
&\quad\textnormal{\small{$+2k_1\:\chi_{(2)}(b)+k_1\:\chi_{(1,1)}(b)-\frac{3}{2}k_1\:\chi_{(1,1)}(b)-\frac{3}{2}k_1\:\chi_{(2)}(b)$}}\\
&\textnormal{\small{=$\frac{1}{2}\:\left(\chi_{(2)}(b)+\chi_{(1,1)}(b)\right)\:k_1^2+\frac{1}{2}\:\left(\chi_{(2)}(b)-\chi_{(1,1)}(b)\right)\cdot(2\:k_2)+\frac{1}{2}\:\left(\chi_{(2)}(b)-\chi_{(1,1)}(b)\right)\:k_1$}}.
\end{align}
Comparing this result with equation \eqref{irreplft} concludes the proof of our claim.\\
Given the multiplicities in the decomposition of $R$ into irreducible representations, we now can apply the theorem about the structure of the commutant form the previous subsection. There exists some basis in which any matrix $A\in\textnormal{Com}(R)$ can be written as
\begin{align}
A=
\left(
   \begin{array}{ccccc}
     M_{2\times 2}& 0& 0& 0& 0\\
     0& N_{2\times 2}\otimes\id_{d-1}& 0& 0& 0\\
     0& 0& u\:\id_{d-1}& 0& 0\\
     0& 0& 0& v\:\id_{\frac{1}{2}(d-1)(d-2)}& 0\\
     0& 0& 0& 0& w\:\id_{\frac{1}{2} d(d-3)}
   \end{array}\right)\label{commata},
\end{align}
where $M_{2\times 2}, N_{2\times 2}\in\textnormal{Mat}(2\times2,\mathbb{C})$ and $u,v,w\in\mathbb{C}$.
Counting the free parameters in $A$ we conclude that $\textnormal{Com}(R)$ is a 
\begin{align}
2\cdot2+2\cdot2+1+1+1=11
\end{align}
dimensional, complex vector space. Moreover, we see from equation \eqref{commata} that any matrix $A^\dagger\in\textnormal{Com}(R)^\dagger$ can be written as
\begin{align}
A^\dagger=
\left(
   \begin{array}{ccccc}
     M^\dagger_{2\times 2}& 0& 0& 0& 0\\
     0& N^\dagger_{2\times 2}\otimes\id_{d-1}& 0& 0& 0\\
     0& 0& u\:\id_{d-1}& 0& 0\\
     0& 0& 0& v\:\id_{\frac{1}{2}(d-1)(d-2)}& 0\\
     0& 0& 0& 0& w\:\id_{\frac{1}{2} d(d-3)}
   \end{array}\right),
\end{align}
where this time the matrices $M^\dagger$ and $N^\dagger$ are hermitian $2\times 2$ matrices and the parameters $u,v,w$ are real. Thus $\textnormal{Com}(R)^\dagger$ is an eleven dimensional, real vector space.
\newpage
\subsection{A basis for $\textnormal{Com}(R)^\dagger$}
\textbf{Theorem:} (Basis for $\textnormal{Com}(R)^\dagger$)\textit{ Let $R$ be the representation defined in Subsection 7.2.1 and let $d\geq4$. We write $\ket{e}:=\sum_{i}{\ket{i}}$.
The following list of matrices forms a basis for $\textnormal{Com}(R)^\dagger$.
\begin{align}
\cA_1&=\sum_{i}\proj{i}{i}\otimes\sum_{j}\proj{j}{j}=\id\otimes\id\nonumber\\
\cA_2&=\sum_{i,j}\proj{i}{j}\otimes\sum_{i,j}\proj{i}{j}=\proj{e}{e}\otimes\proj{e}{e}\nonumber\\
\cA_3&=\sum_{i}\proj{i}{i}\otimes\sum_{i,j}\proj{i}{j}+\sum_{i,j}\proj{i}{j}\otimes\sum_{i}\proj{i}{i}=\id\otimes\proj{e}{e}+\proj{e}{e}\otimes\id\nonumber\\
\cA_4&=\sum_{i,j}\proj{i}{j}\otimes\proj{i}{j}=d\:\Phi\nonumber\\
\cA_5&=\sum_{i,j}\proj{i}{j}\otimes\proj{j}{i}=\cF\nonumber\\
\cA_6&=\sum_{i}\proj{i}{i}\otimes\proj{i}{i}=d\:T\nonumber\\
\cA_7&=\sum_{i,j}\proj{i}{i}\otimes\proj{i}{j}+\sum_{i,j}\proj{i}{i}\otimes\proj{j}{i}+\sum_{i,j}\proj{i}{j}\otimes\proj{i}{i}+\sum_{i,j}\proj{j}{i}\otimes\proj{i}{i}\nonumber\\
\cA_8&=\sum_{i}\proj{i}{e}\otimes\proj{i}{e}+\sum_{i}\proj{e}{i}\otimes\proj{e}{i}\nonumber\\
\cA_9&=\sum_{i}\proj{e}{i}\otimes\proj{i}{e}+\sum_{i}\proj{i}{e}\otimes\proj{e}{i}\nonumber\\
\nonumber\\
\cA_{10}&=\textnormal{i}\left(\sum_{i,j}\proj{i}{i}\otimes\proj{i}{j}-\sum_{i,j}\proj{i}{i}\otimes\proj{j}{i}+\sum_{i,j}\proj{i}{j}\otimes\proj{i}{i}-\sum_{i,j}\proj{j}{i}\otimes\proj{i}{i}\right)\nonumber\\
\cA_{11}&=\textnormal{i}\left(\sum_{i}\proj{i}{e}\otimes\proj{i}{e}-\sum_{i}\proj{e}{i}\otimes\proj{e}{i}\right)\nonumber
\end{align}
}\\
\\
From the structure of the matrices it is evident, that they all are hermitian. Furthermore they are invariant under the conjugation with the swap operator. Therefore they are in the commutant of the representation $S$ (see Subsection 7.2.1). Finally it is easy to see that each of the above matrices is invariant under conjugation with $P\otimes P$ for any $P\in\mathbb{P}$. Hence, we conclude
\begin{align}
\cA_{i}\ \in\ \textnormal{Com}(R)^\dagger\quad\forall{i}.
\end{align}
In the previous subsection we have already seen that the dimension of $\textnormal{Com}(R)^\dagger$ is 11. Thus if the matrices are linearly independent, we know that the list is complete. The linear independence is the only thing left to show to conclude the proof that the $\{\cA_i\}_{i\in\{1,...,11\}}$ form a basis. For this we note that for any two hermitian matrices $A$ and $B$ we have the Schmidt scalar product $\braket{A}{B}=\trace{A\cdot B}$. We compute the Gramian matrix $G$ whose entries are $G_{ij}:=\braket{\cA_j}{\cA_i}$ for the Schmidt scalar product. This gives that
\begin{align}
&G=\nonumber\\
&\textnormal{\footnotesize{$\left(
   \begin{array}{ccccccccccc}
     d^2 & d^2 & 2d^2 & d & d & d & 4d & 2d & 2d & 0 & 0 \\
     d^2 & d^4 & 2d^3 & d^2 & d^2 & d & 4d^2 & 2d^3 & 2d^3 & 0 & 0 \\
     2d^2 & 2d^3 & 2d^2+2d^3 & 2d & 2d & 2d & 4d+4d^2 & 4d^2 & 4d^2 & 0 & 0 \\
     d & d^2 & 2d & d^2 & d & d & 4d & 2d^2 & 2d & 0 & 0 \\
     d & d^2 & 2d & d & d^2 & d & 4d & 2d & 2d^2 & 0 & 0 \\
     d & d & 2d & d & d & d & 4d & 2d & 2d & 0 & 0 \\
     4d & 4d^2 & 4d+4d^2 & 4d & 4d & 4d & 12d+4d^2 & 4d+4d^2 & 4d+4d^2 & 0 & 0 \\
     2d & 2d^3 & 4d^2 & 2d^2 & 2d & 2d & 4d+4d^2 & 2d^2+2d^3 & 4d^2 & 0 & 0 \\
     2d & 2d^3 & 4d^2 & 2d & 2d^2 & 2d & 4d+4d^2 & 4d^2 & 2d^2+2d^3 & 0 & 0 \\
     0 & 0 & 0 & 0 & 0 & 0 & 0 & 0 & 0 & 4d^2-4d & 4d^2-4d \\
     0 & 0 & 0 & 0 & 0 & 0 & 0 & 0 & 0 & 4d^2-4d & 2d^3-2d^2 
   \end{array}
\right)$}}
\label{Grammat},
\end{align}
where $d$ is the dimension of the vector space $V$ and thus $d^2$ is the dimension of the representation $R$. Later it will be useful to have $G$ in block matrix form. We introduce the matrices $G_1$ and $G_2$ such that
\begin{align}
G=
\left(
   \begin{array}{cc}
     G_1 & 0  \\
     0 & G_2 
   \end{array}
\right)\label{invgramexbl}
\end{align}
as above. (The zeros in the above matrix now correspond to matrices.) Inverting the matrices $G_1$ and $G_2$ depending on the dimension shows that
\begin{align}
&(G_1)^{-1}=\textnormal{\scriptsize{$\frac{1}{d(d-1)(d-2)(d-3)}$}}\:\cdot\nonumber\\
&\textnormal{\scriptsize{$\left(
   \begin{array}{ccccccccc}
     d^2-3d+1 & 1 & -d+2 & 1 & 1 & -d^2+d & d-1 & -1 & -1  \\
     1 & 1 & -1 & 1 & 1 & -6 & 2 & -1 & -1 \\
     -d+2 & -1 & \frac{1}{2}(d-1) & -1 & -1 & 2d & -\frac{1}{2}(d+1) & 1 & 1  \\
     1 & 1 & -1 & d^2-3d+1 & 1 & -d^2+d & d-1 & -d+2 & -1  \\
     1 & 1 & -1 & 1 & d^2-3d+1 & -d^2+d & d-1 & -1 & -d+2  \\
     -d^2+d & -6 & 2d & -d^2+d & -d^2+d & d^3+d^2 & -d^2-d & 2d & 2d  \\
     d-1 & 2 & -\frac{1}{2}(d+1) & d-1 & d-1 & -d^2-d & \frac{1}{4}d^2+\frac{1}{4}d+1 & -\frac{1}{2}(d+1) & -\frac{1}{2}(d+1) \\
     -1 & -1 & 1 & -d+2 & -1 & 2d &-\frac{1}{2}(d+1) & \frac{1}{2}(d-1) & 1  \\
     -1 & -1 & 1 & -1 & -d+2 & 2d & -\frac{1}{2}(d+1) & 1 & \frac{1}{2}(d-1)  
   \end{array}
\right)$}}\label{invgramsub}
\end{align}
and
\begin{align}
(G_2)^{-1}=\frac{1}{d(d-1)(d-2)}\:\cdot
\left(
   \begin{array}{cc}
     \frac{d}{4} & -\frac{1}{2}\\
     -\frac{1}{2} & \frac{1}{2}  
   \end{array}
\right)
\end{align}
We conclude that the matrix $G$ is invertible if and only if $d\geq4$. The fact that the Gramian matrix of the list $\{\cA_i\}_{i\in\{1,...,11\}}$ is invertible for $d\geq4$ implies that the list is linearly independent in this case. Then the list $\{\cA_i\}_{i\in\{1,...,11\}}$ constitutes a basis of $\textnormal{Com}(R)^\dagger$.
\subsection{Evaluating the term $\sum_{P\in\mathbb{P}(A)}{(\idx{P}{A})^{\otimes2}(\cT^\dagger)^{\otimes2}[\idx{\cF}{E}](\idx{P}{A}^\dagger)^{\otimes2}}$}
The previous subsections have shown that $\sum_{P\in\mathbb{P}(A)}{(\idx{P}{A})^{\otimes2}(\cT^\dagger)^{\otimes2}[\idx{\cF}{E}](\idx{P}{A}^\dagger)^{\otimes2}}$ is an element of an 11-dimensional, real vector space. In addition the last subsection gave an explicit basis for that space. We now expand $\sum_{P\in\mathbb{P}(A)}{(\idx{P}{A})^{\otimes2}(\cT^\dagger)^{\otimes2}[\idx{\cF}{E}](\idx{P}{A}^\dagger)^{\otimes2}}$ into that basis. That is, we write 
\begin{align}
\frac{1}{\abs{\mathbb{P}}}\sum_{P\in\mathbb{P}(A)}{(\idx{P}{A})^{\otimes2}(\cT^\dagger)^{\otimes2}[\idx{\cF}{E}](\idx{P}{A}^\dagger)^{\otimes2}}=\sum_{j}^{11}{\alpha_j\cA_j}\label{expandper}
\end{align}
with real coefficients $\alpha_k$ depending on $\cT$. We would like to have an explicit formula for the coefficients in terms of $(\cT^\dagger)^{\otimes2}[\idx{\cF}{E}]$. This can be achieved calculating
\begin{align}
\Trace{\sum_{P\in\mathbb{P}(A)}{(\idx{P}{A})^{\otimes2}(\cT^\dagger)^{\otimes2}[\idx{\cF}{E}](\idx{P}{A}^\dagger)^{\otimes2}}\cdot\cA_k}
&=\sum_{P\in\mathbb{P}(A)}\Trace{{(\cT^\dagger)^{\otimes2}[\idx{\cF}{E}]\cdot(\idx{P}{A}^\dagger)^{\otimes2}}\cA_k(\idx{P}{A})^{\otimes2}}\\
&=\sum_{P\in\mathbb{P}(A)}\Trace{{(\cT^\dagger)^{\otimes2}[\idx{\cF}{E}]\cdot\cA_k}}\\
&=\abs{\mathbb{P}}\:\Trace{{(\cT^\dagger)^{\otimes2}[\idx{\cF}{E}]\cdot\cA_k}}.
\end{align}
Together with equation \eqref{expandper} this implies that
\begin{align}
\Trace{{(\cT^\dagger)^{\otimes2}[\idx{\cF}{E}]\cdot\cA_k}}&=\sum_{j}^{11}\alpha_j\:\Trace{\cA_j\cA_k}\\
&=\sum_{j}^{11}G_{kj}\:\alpha_j,
\end{align}
where $G$ is the Gramian matrix of the Schmidt scalar product of the $\cA_k$ as in equation \eqref{Grammat}.
In the previous subsection we already calculated the inverse of the Gramian matrix. We can use it to invert the above equation and to obtain the coefficients $\alpha_k$.
\begin{align}
\alpha_j&=
\sum_{k}^{11}{(G^{-1})_{jk}\:\Trace{{(\cT^\dagger)^{\otimes2}[\idx{\cF}{E}]\cdot\cA_k}}}.
\end{align}
Finally we can write
\begin{align}
\frac{1}{\abs{\mathbb{P}}}\sum_{P\in\mathbb{P}(A)}{(\idx{P}{A})^{\otimes2}(\cT^\dagger)^{\otimes2}[\idx{\cF}{E}](\idx{P}{A}^\dagger)^{\otimes2}}
=\sum_{i,j}^{11}{(G^{-1})_{ij}\:\Trace{{(\cT^\dagger)^{\otimes2}[\idx{\cF}{E}]\cdot\cA_i}}\:\cA_j}\label{forhighdim}.
\end{align}
This is the core result of this section. The formula allows us to generally calculate terms of the type \eqref{usetoclal} and thus provides the possibility to obtain decoupling lemmata with permutation operators similar to the Decoupling Lemma of Section 2.1.
Strictly speaking this is valid only in dimensions higher than 3 and a similar discussion is required to find such formulas in the case that $d\in\{1,2,3\}$. Evidently this cases are easier to treat since the basis vector $\cA_i$ can be ``restricted'' to lower dimensions. In quantum information theory typically we are concerned with systems of high dimension (as for example in the discussion of the heat bath in Chapter 3) we therefore will leave it with formula \eqref{forhighdim}. Technically this formula is much more difficult to work with than its unitary counterpart 
equation \eqref{intgrlpr}. Instead of two linearly independent vectors that span the commutant algebra in the unitary case here we have eleven. Hence, formula \eqref{forhighdim} contains 121 terms (of which many are zero corresponding to the zero entries in the inverse Gramian matrix). This is the reason why we cannot just write down a decoupling theorem with an easy upper bound in terms of physical quantities. Instead, as already mentioned in the introduction to this chapter, we do the computations for three interesting special cases.
\section{Distance from classicality}
This section aims at answering the question, whether or not a map $\cTAE$ can be made classicalizing with a pre concatenation of a permutation operator. We note that for a map $\cTAE$ and its classicalized version $\cTAE^{cl}$ one can write the Choi-Jamiolkowski representation of $\cTAE^{cl}$ as (see \eqref{intrT})
\begin{align}
\idx{\omega}{A'E}^{cl}&=\cTAE^{cl}(\idx{\Phi}{AA'})\\
&=\cTAE(\idx{T}{AA'})\label{usetomeasdist},
\end{align}
For a fixed permutation operator $P$ one can compare the Choi-Jamiolkowski representations of the map $\cTAE\circ P\cdot$\ and its classicalized version. If the distance of the two Choi-Jamiolkowski representations is measured in some norm (for example the Schatten 1-norm) this naturally implies a measure of distance on the set the corresponding maps: One can define the distance between two maps in some norm as being the distance of the Choi-Jamiolkowski representations. If we use the Schatten 1-norm on the state space the induced norm on the set of maps is the $\idx{\Delta}{PRO}$
norm (see \cite{DistMeasure}, Paragraph 4). With equation \eqref{usetomeasdist} (for $\idx{d}{R}=\idx{d}{A}$) the term 
\begin{align}
{\left\| \mathcal{T}(\idx{P}{A} \otimes \idi{R} \ (\idx{\Phi}{AR}-\idx{T}{AR})\ \idx{P}{A}^\dagger \otimes \idi{R}) \right\|}_1\nonumber
\end{align}
gives the distance between the Choi-Jamiolkowski representation of $\cTAE\circ P\cdot$\ and its classicalized version and can be interpreted as giving the $\idx{\Delta}{PRO}$-distance of the corresponding maps.
Our result has applications especially in coding theory. We hope that it implies a classical one-shot coding theorem, similar to the quantum coding theorems presented in \cite{Fred:PHD}. We follow equation \eqref{followpl} and have
\begin{align}
&\frac{1}{\abs{\mathbb{P}}}\sum_{P\in\mathbb{P}(A)}{\left\| \mathcal{T}(\idx{P}{A} \otimes \idi{R} \ (\idx{\Phi}{AR}-\idx{T}{AR})\ \idx{P}{A}^\dagger \otimes \idi{R}) \right\|}_2^2\nonumber\\
&=\Trace{(\idx{\Phi}{AR}-\idx{T}{AR})^{\otimes2}\ \frac{1}{\abs{\mathbb{P}}}\sum_{P\in\mathbb{P}(A)}\left((\idx{P}{A}^\dagger)^{\otimes2}(\cT^\dagger)^{\otimes2}[\idx{\cF}{E}](\idx{P}{A})^{\otimes2}\right)\otimes\idx{\cF}{R}}\\
&=\sum_{i,j}^{11}{(G^{-1})_{ij}\:\Trace{{(\cT^\dagger)^{\otimes2}[\idx{\cF}{E}]\cdot\cA_i}}\:\Trace{(\idx{\Phi}{AR}-\idx{T}{AR})^{\otimes2}\ \cA_j\otimes\idx{\cF}{R}}}
\end{align}
Of the 121 summands in the above sum many are zero. The inverse Gramian matrix $G^{-1}$ has block diagonal structure (equation \eqref{invgramexbl}). Hence any summand with $(i,j)\in\{1,...,9\}\times\{10,11\}$ or $(i,j)\in\{10,11\}\times\{1,...,9\}$ is zero. Therefore we can write
\begin{align}
&\sum_{i,j}^{11}{(G^{-1})_{ij}\:\Trace{{(\cT^\dagger)^{\otimes2}[\idx{\cF}{E}]\cdot\cA_i}}\:\Trace{(\idx{\Phi}{AR}-\idx{T}{AR})^{\otimes2}\ \cA_j\otimes\idx{\cF}{R}}}\\
&=\sum_{i,j}^{9}{(G^{-1})_{ij}\:\Trace{{(\cT^\dagger)^{\otimes2}[\idx{\cF}{E}]\cdot\cA_i}}\:\Trace{(\idx{\Phi}{AR}-\idx{T}{AR})^{\otimes2}\ \cA_j\otimes\idx{\cF}{R}}}\nonumber\\
&\quad+\sum_{i,j\in\{10,11\}}{(G^{-1})_{ij}\:\Trace{{(\cT^\dagger)^{\otimes2}[\idx{\cF}{E}]\cdot\cA_i}}\:\Trace{(\idx{\Phi}{AR}-\idx{T}{AR})^{\otimes2}\ \cA_j\otimes\idx{\cF}{R}}}\label{expldisaplastops}.
\end{align}
The operators $\cA_{10}$ and $\cA_{11}$ consist of a product of the imaginary unit and an antisymmetric matrix. They engender the imaginary part of $$\frac{1}{\abs{\mathbb{P}}}\sum_{P\in\mathbb{P}(A)}{(\idx{P}{A})^{\otimes2}(\cT^\dagger)^{\otimes2}[\idx{\cF}{E}](\idx{P}{A}^\dagger)^{\otimes2}}.$$
Considering equation \eqref{followpl} it is evident that in our setup, where we calculate a norm \eqref{usetoclal} (and $\idx{\lambda}{AR}$ is symmetric), these operators never can contribute. More precisely, for $i,j\in\{10,11\}$ the second trace term in the summands of \eqref{expldisaplastops}, $\Trace{(\idx{\Phi}{AR}-\idx{T}{AR})^{\otimes2}\ \cA_j\otimes\idx{\cF}{R}}$ is zero, because the trace of a product of a symmetric and an antisymmetric matrix vanishes. Thus the operators $\cA_{10}$ and $\cA_{11}$ are irrelevant in our context. We stated them for completeness but, since we don't expect any contributions, after a blank line in Subsection 7.2.4. Moreover instead of working with the full Gramian matrix it suffices to work with $(G_1)^{-1}$, i.\:e.\:
\begin{align}
&\sum_{i,j}^{11}{(G^{-1})_{ij}\:\Trace{{(\cT^\dagger)^{\otimes2}[\idx{\cF}{E}]\cdot\cA_i}}\:\Trace{(\idx{\Phi}{AR}-\idx{T}{AR})^{\otimes2}\ \cA_j\otimes\idx{\cF}{R}}}\nonumber\\
&=\sum_{i,j}^{9}{({G_1}^{-1})_{ij}\:\Trace{{(\cT^\dagger)^{\otimes2}[\idx{\cF}{E}]\cdot\cA_i}}\:\Trace{(\idx{\Phi}{AR}-\idx{T}{AR})^{\otimes2}\ \cA_j\otimes\idx{\cF}{R}}}\label{calcalltrms}.
\end{align}
The terms $\Trace{(\idx{\Phi}{AR}-\idx{T}{AR})^{\otimes2}\ \cA_j\otimes\idx{\cF}{R}}$ can be calculated easily yielding
\begin{align}
\Trace{(\idx{\Phi}{AR}-\idx{T}{AR})^{\otimes2}\ \cA_1\otimes\idx{\cF}{R}}&=0\\
\Trace{(\idx{\Phi}{AR}-\idx{T}{AR})^{\otimes2}\ \cA_2\otimes\idx{\cF}{R}}&=1-\frac{1}{\idx{d}{R}}\\
\Trace{(\idx{\Phi}{AR}-\idx{T}{AR})^{\otimes2}\ \cA_3\otimes\idx{\cF}{R}}&=0\\
\Trace{(\idx{\Phi}{AR}-\idx{T}{AR})^{\otimes2}\ \cA_4\otimes\idx{\cF}{R}}&=0\\
\Trace{(\idx{\Phi}{AR}-\idx{T}{AR})^{\otimes2}\ \cA_5\otimes\idx{\cF}{R}}&=1-\frac{1}{\idx{d}{R}}\\
\Trace{(\idx{\Phi}{AR}-\idx{T}{AR})^{\otimes2}\ \cA_6\otimes\idx{\cF}{R}}&=0\\
\Trace{(\idx{\Phi}{AR}-\idx{T}{AR})^{\otimes2}\ \cA_7\otimes\idx{\cF}{R}}&=0\\
\Trace{(\idx{\Phi}{AR}-\idx{T}{AR})^{\otimes2}\ \cA_8\otimes\idx{\cF}{R}}&=0\\
\Trace{(\idx{\Phi}{AR}-\idx{T}{AR})^{\otimes2}\ \cA_9\otimes\idx{\cF}{R}}&=2\left(1-\frac{1}{\idx{d}{R}}\right).
\end{align}
Equation \eqref{calcalltrms} then becomes 
\begin{align}
&\sum_{i,j}^{9}{({G_1}^{-1})_{ij}\:\Trace{{(\cT^\dagger)^{\otimes2}[\idx{\cF}{E}]\cdot\cA_i}}\:\Trace{(\idx{\Phi}{AR}-\idx{T}{AR})^{\otimes2}\ \cA_j\otimes\idx{\cF}{R}}}\nonumber\\
&=\left(1-\frac{1}{\idx{d}{R}}\right)\sum_{i}^{9}{\Trace{{(\cT^\dagger)^{\otimes2}[\idx{\cF}{E}]\cdot\cA_i}}\:\left(({G_1}^{-1})_{i2}+({G_1}^{-1})_{i5}+2({G_1}^{-1})_{i9}\right)}\label{evalgramtrms}
\end{align}
The matrix ${G_1}^{-1}$ is known \eqref{invgramsub} and the terms $({G_1}^{-1})_{i2}+({G_1}^{-1})_{i5}+2({G_1}^{-1})_{i9}$ are easily evaluated. We get
\begin{align}
({G_1}^{-1})_{12}+({G_1}^{-1})_{15}+2({G_1}^{-1})_{19}&=0\\
({G_1}^{-1})_{22}+({G_1}^{-1})_{25}+2({G_1}^{-1})_{29}&=0\\
({G_1}^{-1})_{32}+({G_1}^{-1})_{35}+2({G_1}^{-1})_{39}&=0\\
({G_1}^{-1})_{42}+({G_1}^{-1})_{45}+2({G_1}^{-1})_{49}&=0\\
({G_1}^{-1})_{52}+({G_1}^{-1})_{55}+2({G_1}^{-1})_{59}&=\frac{1}{\idx{d}{A}(\idx{d}{A}-1)}\\
({G_1}^{-1})_{62}+({G_1}^{-1})_{65}+2({G_1}^{-1})_{69}&=\frac{-1}{\idx{d}{A}(\idx{d}{A}-1)}\\
({G_1}^{-1})_{72}+({G_1}^{-1})_{75}+2({G_1}^{-1})_{79}&=0\\
({G_1}^{-1})_{82}+({G_1}^{-1})_{85}+2({G_1}^{-1})_{89}&=0\\
({G_1}^{-1})_{92}+({G_1}^{-1})_{95}+2({G_1}^{-1})_{99}&=0
\end{align}
and equation \eqref{evalgramtrms} becomes
\begin{align}
&\left(1-\frac{1}{\idx{d}{R}}\right)\sum_{i}^{9}{\Trace{{(\cT^\dagger)^{\otimes2}[\idx{\cF}{E}]\cdot\cA_i}}\:\left(({G_1}^{-1})_{i2}+({G_1}^{-1})_{i5}+2({G_1}^{-1})_{i9}\right)}\\
&=\frac{1}{\idx{d}{R}\idx{d}{A}}\:\frac{\idx{d}{R}-1}{\idx{d}{A}-1}\left(\Trace{{(\cT^\dagger)^{\otimes2}[\idx{\cF}{E}]\cdot\cA_5}}-\Trace{{(\cT^\dagger)^{\otimes2}[\idx{\cF}{E}]\cdot\cA_6}}\right)\\
&=\frac{1}{\idx{d}{R}\idx{d}{A}}\:\frac{\idx{d}{R}-1}{\idx{d}{A}-1}\left(\Trace{{(\cT^\dagger)^{\otimes2}[\idx{\cF}{E}]\idx{\cF}{A}}}-\sum_{i}\Trace{{(\cT^\dagger)^{\otimes2}[\idx{\cF}{E}]\:\left(\proji{i}{i}{A}\otimes\proji{i}{i}{A'}\right)}}\right)\\
&=\frac{1}{\idx{d}{R}\idx{d}{A}}\:\frac{\idx{d}{R}-1}{\idx{d}{A}-1}\left(\Trace{{\idx{\cF}{E}\cT^{\otimes2}(\idx{\cF}{A})}}-\sum_{i}\Trace{\cT(\proji{i}{i}{A})\cT(\proji{i}{i}{A'})}\right)\label{applyCJinv}\\
&=\frac{1}{\idx{d}{R}\idx{d}{A}}\:\frac{\idx{d}{R}-1}{\idx{d}{A}-1}\left(\idx{d}{A}^2\:\Trace{\idx{\cF}{E}\:\Ptrace{AA'}{\idx{\omega}{AE}^{\otimes2}(\idi{EE'}\otimes\idx{\cF}{A})}}-\sum_{i}\Trace{\cT(\proji{i}{i}{A})\cT(\proji{i}{i}{A'})}\right)\\
&=\frac{1}{\idx{d}{R}\idx{d}{A}}\:\frac{\idx{d}{R}-1}{\idx{d}{A}-1}\left(\idx{d}{A}^2\:\Trace{(\idx{\cF}{E}\otimes\idi{AA'})\:\idx{\omega}{A'E}^{\otimes2}\:(\idi{EE'}\otimes\idx{\cF}{A})}-\sum_{i}\Trace{\cT(\proji{i}{i}{A})\cT(\proji{i}{i}{A'})}\right)\\
&=\frac{1}{\idx{d}{R}\idx{d}{A}}\:\frac{\idx{d}{R}-1}{\idx{d}{A}-1}\left(\idx{d}{A}^2\:\Trace{\idx{\omega}{AE}^2}-\sum_{i}\Trace{\cT(\proji{i}{i}{A})\cT(\proji{i}{i}{A'})}\right)\\
&=\frac{1}{\idx{d}{R}\idx{d}{A}}\:\frac{\idx{d}{R}-1}{\idx{d}{A}-1}\left(\idx{d}{A}^2\:\Trace{\idx{\omega}{A'E}^2}-\idx{d}{A}^2\:\Trace{(\idx{\omega}{A'E}^{cl})^2}\right).\label{lastcla}
\end{align}
Equation \eqref{applyCJinv} uses the explicit form of the inverse of the Choi-Jamiolkowski isomorphism to express $\cTAE$ via its Choi-Jamiolkowski representation (\textit{Lemma 5} in \cite{Semthesis}). The last step \eqref{lastcla} is by the definition of the Choi-Jamiolkowski representation of the classicalizing version of $\cTAE$ and is shown in detail following equation \eqref{classmpviaclassCJ}. An easy reformulation can be used to see that
\begin{align}
\Trace{\idx{\omega}{A'E}^2}-\Trace{(\idx{\omega}{A'E}^{cl})^2}=\Norm{\idx{\omega}{A'E}-\idx{\omega}{A'E}^{cl}}{2}^2.
\end{align}
We arrive at the following lemma.\\
\\
{\textbf{Lemma:} (Distance from classicality)\\ \textit{Introduce the states $\idx{\Phi}{AR}:=\frac{1}{\idx{d}{R}}\sum_{i,j}^{\idx{d}{R}}{\proji{i}{j}{A}\otimes\proji{i}{j}{R}}$ and $\idx{T}{AR}:=\frac{1}{\idx{d}{R}}\sum_{i}^{\idx{d}{R}}{\proji{i}{i}{A}\otimes\proji{i}{i}{R}}$ with $\idx{d}{A}\geq\idx{d}{R}$ and $\idx{d}{A}\geq4$.
Let $\cTAE\in\textnormal{Hom}(\linops{\HA},\linops{\HE})$ be a linear map with Choi-Jamiolkowski representation $\idx{\omega}{A'E}\in\hermops{\idx{\cH}{E}\otimes\idx{\cH}{A'}}$, then
\begin{align}
\frac{1}{\abs{\mathbb{P}}}\sum_{P\in\mathbb{P}(A)}&{\left\| \mathcal{T}(\idx{P}{A} \otimes \idi{R} \ (\idx{\Phi}{AR} - \idx{T}{AR})\ \idx{P}{A}^\dagger \otimes \idi{R}) \right\|}_2^2 \nonumber\\
&=\frac{\idx{d}{A}}{\idx{d}{R}}\:\frac{\idx{d}{R}-1}{\idx{d}{A}-1}\:\Norm{\idx{\omega}{A'E}-\idx{\omega}{A'E}^{cl}}{2}^2,\nonumber
\end{align}
where the summation goes over all permutation operators.
}}\\
\\
Using the H\"older inequality~\eqref{hldrfk} as in the previous chapters we obtain the following result.\\
\\
{\textbf{Theorem:} (Distance from classicality)\\ \textit{Introduce the states $\idx{\Phi}{AR}:=\frac{1}{\idx{d}{R}}\sum_{i,j}^{\idx{d}{R}}{\proji{i}{j}{A}\otimes\proji{i}{j}{R}}$ and $\idx{T}{AR}:=\frac{1}{\idx{d}{R}}\sum_{i}^{\idx{d}{R}}{\proji{i}{i}{A}\otimes\proji{i}{i}{R}}$ with $\idx{d}{A}\geq4$.
Let $\cTAE\in\textnormal{Hom}(\linops{\HA},\linops{\HE})$ be a completely positive map with Choi-Jamiolkowski representation $\idx{\omega}{A'E}\in\subnormstates{\idx{\cH}{E}\otimes\idx{\cH}{A'}}$, then
\begin{align}
\frac{1}{\abs{\mathbb{P}}}\sum_{P\in\mathbb{P}(A)}{\left\| \mathcal{T}(\idx{P}{A} \otimes \idi{R} \ (\idx{\Phi}{AR} - \idx{T}{AR})\ \idx{P}{A}^\dagger \otimes \idi{R}) \right\|}_1 
\leq\sqrt{\idx{d}{A}\:\frac{\idx{d}{R}-1}{\idx{d}{A}-1}}\:2^{-\frac{1}{2}\chtwo{A'}{E}{\omega}},\nonumber
\end{align}
where the summation goes over all permutation operators.
}}\\
\\
\section{Decoupling with permutations operators}
We follow equation \eqref{followpl} and have
\begin{align}
&\frac{1}{\abs{\mathbb{P}}}\sum_{P\in\mathbb{P}(A)}{\left\| \mathcal{T}(\idx{P}{A} \otimes \idi{R} \ (\idx{\Phi}{AR}-\idx{\pi}{AR})\ \idx{P}{A}^\dagger \otimes \idi{R}) \right\|}_2^2\nonumber\\
&=\Trace{(\idx{\Phi}{AR}-\idx{\pi}{AR})^{\otimes2}\ \frac{1}{\abs{\mathbb{P}}}\sum_{P\in\mathbb{P}(A)}\left((\idx{P}{A}^\dagger)^{\otimes2}(\cT^\dagger)^{\otimes2}[\idx{\cF}{E}](\idx{P}{A})^{\otimes2}\right)\otimes\idx{\cF}{R}}\\
&=\sum_{i,j}^{9}{((G_1)^{-1})_{ij}\:\Trace{{(\cT^\dagger)^{\otimes2}[\idx{\cF}{E}]\cdot\cA_i}}\:\Trace{(\idx{\Phi}{AR}-\idx{\pi}{AR})^{\otimes2}\ \cA_j\otimes\idx{\cF}{R}}}\label{plufin}
\end{align}
The last equation makes use of the same arguments as equation \eqref{calcalltrms}. Evaluating the terms $\Trace{(\idx{\Phi}{AR}-\idx{\pi}{AR})^{\otimes2}\ \cA_j\otimes\idx{\cF}{R}}$ this time yields
\begin{align}
\Trace{(\idx{\Phi}{AR}-\idx{\pi}{AR})^{\otimes2}\ \cA_1\otimes\idx{\cF}{R}}&=0\\
\Trace{(\idx{\Phi}{AR}-\idx{\pi}{AR})^{\otimes2}\ \cA_2\otimes\idx{\cF}{R}}&=1-\frac{1}{\idx{d}{R}}\\
\Trace{(\idx{\Phi}{AR}-\idx{\pi}{AR})^{\otimes2}\ \cA_3\otimes\idx{\cF}{R}}&=0\\
\Trace{(\idx{\Phi}{AR}-\idx{\pi}{AR})^{\otimes2}\ \cA_4\otimes\idx{\cF}{R}}&=\frac{1}{\idx{d}{R}}\left(1-\frac{1}{\idx{d}{R}}\right)\\
\Trace{(\idx{\Phi}{AR}-\idx{\pi}{AR})^{\otimes2}\ \cA_5\otimes\idx{\cF}{R}}&=1-\frac{1}{\idx{d}{R}^2}\\
\Trace{(\idx{\Phi}{AR}-\idx{\pi}{AR})^{\otimes2}\ \cA_6\otimes\idx{\cF}{R}}&=\frac{1}{\idx{d}{R}}\left(1-\frac{1}{\idx{d}{R}}\right)\\
\Trace{(\idx{\Phi}{AR}-\idx{\pi}{AR})^{\otimes2}\ \cA_7\otimes\idx{\cF}{R}}&=\frac{4}{\idx{d}{R}}\left(1-\frac{1}{\idx{d}{R}}\right)\\
\Trace{(\idx{\Phi}{AR}-\idx{\pi}{AR})^{\otimes2}\ \cA_8\otimes\idx{\cF}{R}}&=\frac{2}{\idx{d}{R}}\left(1-\frac{1}{\idx{d}{R}}\right)\\
\Trace{(\idx{\Phi}{AR}-\idx{\pi}{AR})^{\otimes2}\ \cA_9\otimes\idx{\cF}{R}}&=2\left(1-\frac{1}{\idx{d}{R}^2}\right).
\end{align}
Plugging all these terms into \eqref{plufin} gives:
\begin{align}
&\sum_{i,j}^{9}{((G_1)^{-1})_{ij}\:\Trace{{(\cT^\dagger)^{\otimes2}[\idx{\cF}{E}]\cdot\cA_i}}\:\Trace{(\idx{\Phi}{AR}-\idx{\pi}{AR})^{\otimes2}\ \cA_j\otimes\idx{\cF}{R}}}\nonumber\\
&\textnormal{\scriptsize{$=\frac{1}{\idx{d}{R}}\left(1-\frac{1}{\idx{d}{R}}\right)\:\sum_{i}^{9}{\Trace{{(\cT^\dagger)^{\otimes2}[\idx{\cF}{E}]\cdot\cA_i}}}$}}\cdot\nonumber\\
&\quad\textnormal{\scriptsize{$\left(\idx{d}{R}(G_1^{-1})_{i2}+(G_1^{-1})_{i4}+(\idx{d}{R}+1)(G_1^{-1})_{i5}+(G_1^{-1})_{i6}+4(G_1^{-1})_{i7}+2(G_1^{-1})_{i8}+2(\idx{d}{R}+1)(G_1^{-1})_{i9}\right)$}}\label{nodieadlkdf}
\end{align}
The terms
\begin{align}
\textnormal{\scriptsize{$\left(\idx{d}{R}(G_1^{-1})_{i2}+(G_1^{-1})_{i4}+(\idx{d}{R}+1)(G_1^{-1})_{i5}+(G_1^{-1})_{i6}+4(G_1^{-1})_{i7}+2(G_1^{-1})_{i8}+2(\idx{d}{R}+1)(G_1^{-1})_{i9}\right)$}}\nonumber
\end{align}
are easily computed using \eqref{invgramsub}.
\begin{align}
&\textnormal{\scriptsize{$\left(\idx{d}{R}(G_1^{-1})_{12}+(G_1^{-1})_{14}+(\idx{d}{R}+1)(G_1^{-1})_{15}+(G_1^{-1})_{16}+4(G_1^{-1})_{17}+2(G_1^{-1})_{18}+2(\idx{d}{R}+1)(G_1^{-1})_{19}\right)$}}\nonumber\\
&=\textnormal{\scriptsize{$\frac{-1}{\idx{d}{A}(\idx{d}{A}-1)}$}}\\
&\textnormal{\scriptsize{$\left(\idx{d}{R}(G_1^{-1})_{22}+(G_1^{-1})_{24}+(\idx{d}{R}+1)(G_1^{-1})_{25}+(G_1^{-1})_{26}+4(G_1^{-1})_{27}+2(G_1^{-1})_{28}+2(\idx{d}{R}+1)(G_1^{-1})_{29}\right)$}}\nonumber\\
&=0\\
&\textnormal{\scriptsize{$\left(\idx{d}{R}(G_1^{-1})_{32}+(G_1^{-1})_{34}+(\idx{d}{R}+1)(G_1^{-1})_{35}+(G_1^{-1})_{36}+4(G_1^{-1})_{37}+2(G_1^{-1})_{38}+2(\idx{d}{R}+1)(G_1^{-1})_{39}\right)$}}\nonumber\\
&=0\\
&\textnormal{\scriptsize{$\left(\idx{d}{R}(G_1^{-1})_{42}+(G_1^{-1})_{44}+(\idx{d}{R}+1)(G_1^{-1})_{45}+(G_1^{-1})_{46}+4(G_1^{-1})_{47}+2(G_1^{-1})_{48}+2(\idx{d}{R}+1)(G_1^{-1})_{49}\right)$}}\nonumber\\
&=0\\
&\textnormal{\scriptsize{$\left(\idx{d}{R}(G_1^{-1})_{52}+(G_1^{-1})_{54}+(\idx{d}{R}+1)(G_1^{-1})_{55}+(G_1^{-1})_{56}+4(G_1^{-1})_{57}+2(G_1^{-1})_{58}+2(\idx{d}{R}+1)(G_1^{-1})_{59}\right)$}}\nonumber\\
&=\textnormal{\scriptsize{$\frac{\idx{d}{R}}{\idx{d}{A}(\idx{d}{A}-1)}$}}\\
&\textnormal{\scriptsize{$\left(\idx{d}{R}(G_1^{-1})_{62}+(G_1^{-1})_{64}+(\idx{d}{R}+1)(G_1^{-1})_{65}+(G_1^{-1})_{66}+4(G_1^{-1})_{67}+2(G_1^{-1})_{68}+2(\idx{d}{R}+1)(G_1^{-1})_{69}\right)$}}\nonumber\\
&=\textnormal{\scriptsize{$\left(1-\frac{\idx{d}{R}}{\idx{d}{A}}\right)\frac{1}{\idx{d}{A}-1}$}}\\
&\textnormal{\scriptsize{$\left(\idx{d}{R}(G_1^{-1})_{72}+(G_1^{-1})_{74}+(\idx{d}{R}+1)(G_1^{-1})_{75}+(G_1^{-1})_{76}+4(G_1^{-1})_{77}+2(G_1^{-1})_{78}+2(\idx{d}{R}+1)(G_1^{-1})_{79}\right)$}}\nonumber\\
&=0\\
&\textnormal{\scriptsize{$\left(\idx{d}{R}(G_1^{-1})_{82}+(G_1^{-1})_{84}+(\idx{d}{R}+1)(G_1^{-1})_{85}+(G_1^{-1})_{86}+4(G_1^{-1})_{87}+2(G_1^{-1})_{88}+2(\idx{d}{R}+1)(G_1^{-1})_{89}\right)$}}\nonumber\\
&=0\\
&\textnormal{\scriptsize{$\left(\idx{d}{R}(G_1^{-1})_{92}+(G_1^{-1})_{94}+(\idx{d}{R}+1)(G_1^{-1})_{95}+(G_1^{-1})_{96}+4(G_1^{-1})_{97}+2(G_1^{-1})_{98}+2(\idx{d}{R}+1)(G_1^{-1})_{99}\right)$}}\nonumber\\
&=0
\end{align}
Together with equation \eqref{nodieadlkdf} the above implies that
\begin{align}
&\frac{1}{\abs{\mathbb{P}}}\sum_{P\in\mathbb{P}(A)}{\left\| \mathcal{T}(\idx{P}{A} \otimes \idi{R} \ (\idx{\Phi}{AR}-\idx{\pi}{AR})\ \idx{P}{A}^\dagger \otimes \idi{R}) \right\|}_2^2\nonumber\\
&=\textnormal{\scriptsize{$\frac{1}{\idx{d}{R}^2}\frac{\idx{d}{R}-1}{\idx{d}{A}-1}\:
\left(-\frac{1}{\idx{d}{A}}\:\Trace{{(\cT^\dagger)^{\otimes2}[\idx{\cF}{E}]\cdot\cA_1}}+\frac{\idx{d}{R}}{\idx{d}{A}}\:\Trace{{(\cT^\dagger)^{\otimes2}[\idx{\cF}{E}]\cdot\cA_5}}+\left(1-\frac{\idx{d}{A}}{\idx{d}{R}}\right)\Trace{{(\cT^\dagger)^{\otimes2}[\idx{\cF}{E}]\cdot\cA_6}}\right)$}}.
\end{align}
The occurring trace terms are evaluated as always (see \eqref{lastcla} for an example) and we get
\begin{align}
&\textnormal{\scriptsize{$\frac{1}{\idx{d}{R}^2}\frac{\idx{d}{R}-1}{\idx{d}{A}-1}\:
\left(-\frac{1}{\idx{d}{A}}\:\Trace{{(\cT^\dagger)^{\otimes2}[\idx{\cF}{E}]\cdot\cA_1}}+\frac{\idx{d}{R}}{\idx{d}{A}}\:\Trace{{(\cT^\dagger)^{\otimes2}[\idx{\cF}{E}]\cdot\cA_5}}+\left(1-\frac{\idx{d}{R}}{\idx{d}{A}}\right)\Trace{{(\cT^\dagger)^{\otimes2}[\idx{\cF}{E}]\cdot\cA_6}}\right)$}}\nonumber\\
&=\frac{\idx{d}{A}^2}{\idx{d}{R}^2}\frac{\idx{d}{R}-1}{\idx{d}{A}-1}\:
\left(-\frac{1}{\idx{d}{A}}\:\Trace{\idx{\omega}{E}^2}+\frac{\idx{d}{R}}{\idx{d}{A}}\:\Trace{\idx{\omega}{A'E}^2}+\left(1-\frac{\idx{d}{R}}{\idx{d}{A}}\right)\Trace{(\idx{\omega}{A'E}^{cl})^2}\right).
\end{align}
We formulate this result in a separate lemma:\\
\\
{\textbf{Lemma:} (Decoupling lemma for permutations)\\ \textit{Introduce the states $\idx{\Phi}{AR}:=\frac{1}{\idx{d}{R}}\sum_{i,j}^{\idx{d}{R}}{\proji{i}{j}{A}\otimes\proji{i}{j}{R}}$ and $\idx{\pi}{AR}:=\frac{1}{\idx{d}{R}^2}\sum_{i,j}^{\idx{d}{R}}{\proji{i}{i}{A}\otimes\proji{j}{j}{R}}$ with $\idx{d}{A}\geq4$.
Let $\cTAE\in\textnormal{Hom}(\linops{\HA},\linops{\HE})$ be a linear map with Choi-Jamiolkowski representation $\idx{\omega}{A'E}\in\hermops{\idx{\cH}{E}\otimes\idx{\cH}{A'}}$, then
\begin{align}
\frac{1}{\abs{\mathbb{P}}}\sum_{P\in\mathbb{P}(A)}&{\left\| \mathcal{T}(\idx{P}{A} \otimes \idi{R} \ \idx{\Phi}{AR} \ \idx{P}{A}^\dagger \otimes \idi{R})- \idx{\omega}{E}\otimes\idx{\pi}{R} \right\|}_2^2 \nonumber\\
&=\frac{\idx{d}{A}^2}{\idx{d}{R}^2}\frac{\idx{d}{R}-1}{\idx{d}{A}-1}\:
\left(\frac{\idx{d}{R}}{\idx{d}{A}}\:\Trace{\idx{\omega}{A'E}^2}-\frac{1}{\idx{d}{A}}\:\Trace{\idx{\omega}{E}^2}+\left(1-\frac{\idx{d}{R}}{\idx{d}{A}}\right)\Trace{(\idx{\omega}{A'E}^{cl})^2}\right),\nonumber
\end{align}
where the summation goes over all permutation operators.
}}\\
\\
It is interesting to note that in the case $\idx{d}{R}=\idx{d}{A}$ the comparison of this lemma with the Decoupling Lemma of Section 2.1 shows that averaging over all permutations gives the same result as averaging over all unitaries. In the case that the system $AA'$ is in the fully entangled state for any fixed operation $\cTAE$ applied to the system $A$ there is a permutation operator that decouples well.\\
Leaving out the negative terms on the right hand side and using our standard procedure, the above lemma can trivially be transformed into a Decoupling Theorem with Permutation Operators.

\section[Decoupling Quantum States with a Permutations followed by the Partial Trace]{Decoupling Quantum States with a Permutations followed by the Partial Trace\footnote{The results of this section were obtained in the week after the deadline of this Master project.}}
In quantum cryptography the pivotal Hash-Lemma is used to extract randomness from a classical source that might be correlated with a quantum system hold by an adversary \cite{rennerkoenig05,TSSR10}. The following Lemma generalizes the Hash-Lemma to the case where one wants to decouple a quantum system from a quantum adversary using classical operations only. It can be seen as an intermediate step between the crucial Fully Quantum Slepian Wolf Theorem \cite{mother} and the Hash Lemma. We write $A=(A_1A_2)$ and we consider the Schatten 2-distance. As in Equation~\ref{followpl} we have that
\begin{align}
&\frac{1}{\mathbb{\abs{P}}}\sum_{P\in\mathbb{P}(A)}{\Norm{\tr_{A_2}{(\idx{P}{A}\otimes\idi{R}\:\rhoAR\:\idx{P}{A}^\dagger\otimes\idi{R})}-\pi_{A_1}\otimes\rhoR}{2}^2}\nonumber\\
&=\Trace{\left(\rhoAR-\idx{\pi}{A}\otimes\rhoR\right)^{\otimes2}\ \frac{1}{\mathbb{\abs{P}}} \sum_{P\in\mathbb{P}(A)}{(\idx{P}{A}^{\otimes2}\:({\cF}_{A_1}\otimes\id_{A_2A_2'})\:\idx{P}{A}^{\otimes2\dagger})\otimes\idx{\cF}{R}}}\\
&=\sum_{i,j}^9{(G^{-1})_{ij}}\ \Trace{{\cF}_{A_1}\otimes\id_{A_2A_2'}\cdot{\cA}_{i}}\Trace{\left(\rhoAR-\idx{\pi}{A}\otimes\rhoR\right)^{\otimes2}\ {\cA}_{j}\otimes\idx{\cF}{R}}.
\end{align}
We compute the coefficients  
\begin{align}
&\trace{{\cF}_{A_1}\otimes\id_{A_2A_2'}\cdot{{\cA}_{1}}}=d_A d_{A_2}\\
&\trace{{\cF}_{A_1}\otimes\id_{A_2A_2'}\cdot{{\cA}_{2}}}=d_A^2\\
&\trace{{\cF}_{A_1}\otimes\id_{A_2A_2'}\cdot{{\cA}_{3}}}=2d_A d_{A_2}\\
&\trace{{\cF}_{A_1}\otimes\id_{A_2A_2'}\cdot{{\cA}_{4}}}=d_A\\
&\trace{{\cF}_{A_1}\otimes\id_{A_2A_2'}\cdot{{\cA}_{5}}}=\frac{1}{d_{A_2}}d_A^2\\
&\trace{{\cF}_{A_1}\otimes\id_{A_2A_2'}\cdot{{\cA}_{6}}}=d_A\\
&\trace{{\cF}_{A_1}\otimes\id_{A_2A_2'}\cdot{{\cA}_{7}}}=4 d_A\\
&\trace{{\cF}_{A_1}\otimes\id_{A_2A_2'}\cdot{{\cA}_{8}}}=2 d_A\\
&\trace{{\cF}_{A_1}\otimes\id_{A_2A_2'}\cdot{{\cA}_{9}}}=\frac{2}{d_{A_2}}d_A^2.
\end{align}
For convenience we shortly write $x:=d_A(d_A-1)(d_A-2)(d_A-3)$ and $c=d_A(d_{A_1}-1)(d_{A_2}-1)$ and compute the numbers $c_j:=\sum_{i}^9(G_1^{-1})_{ij}\ \Trace{{\cF}_{A_1}\otimes\id_{A_2A_2'}\cdot{{\cA}_{i}}}$ using the given form of $G^{-1}$. We get
\begin{align}
&c_2=\frac{c}{x}\\
&c_3=-\frac{c}{x}\\
&c_4=\frac{c}{x}\\
&c_5=\frac{d_A(d_{A_1}-1)(d_A-2)(d_A-3)+c}{x}\\
&c_6=d_A(d_A-5)\frac{c}{x}\\
&c_7=\frac{2c}{x}\\
&c_8=-\frac{c}{x}\\
&c_9=-\frac{c}{x}
\end{align}
and obtain that
\begin{align}
&\frac{1}{\mathbb{\abs{P}}}\sum_{P\in\mathbb{P}(A)}{\Norm{\tr_{A_2}{(\idx{P}{A}\otimes\idi{R}\:\rhoAR\:\idx{P}{A}^\dagger\otimes\idi{R})}-\pi_{A_1}\otimes\rhoR}{2}^2}\nonumber\\
&=\sum_{j}^9{c_j}\Trace{\left(\rhoAR-\idx{\pi}{A}\otimes\rhoR\right)^{\otimes2}\ \idx{\cA}{j}\otimes\idx{\cF}{R}}\label{tricky}\\
&=\frac{d_{A_1}-1}{d_{A}-1}\:\Trace{\left(\rhoAR-\idx{\pi}{A}\otimes\rhoR\right)^{\otimes2}\ \idx{\cA}{5}\otimes\idx{\cF}{R}}\\
&\quad+\frac{c}{d_A(d_A-1)}\:\Trace{\left(\rhoAR-\idx{\pi}{A}\otimes\rhoR\right)^{\otimes2}\ \idx{\cA}{6}\otimes\idx{\cF}{R}}\nonumber\\
&\quad+\frac{c}{x}\Trace{\left(\rhoAR-\idx{\pi}{A}\otimes\rhoR\right)^{\otimes2}\ \cY\otimes\idx{\cF}{R}}\nonumber,
\end{align}
where $\cY=\idx{\cA}{1}+\idx{\cA}{2}-\idx{\cA}{3}+\idx{\cA}{4}+\idx{\cA}{5}-6\idx{\cA}{6}+2\idx{\cA}{7}-\idx{\cA}{8}-\idx{\cA}{9}$. The first two terms can be evaluated directly with an application of the swap trick noting that the classicalized version of the swap operator is $\cA_6$. This gives
\begin{align}
&\frac{1}{\mathbb{\abs{P}}}\sum_{P\in\mathbb{P}(A)}{\Norm{\tr_{A_2}{(\idx{P}{A}\otimes\idi{R}\:\rhoAR\:\idx{P}{A}^\dagger\otimes\idi{R})}-\pi_{A_1}\otimes\rhoR}{2}^2}\nonumber\\
&=\frac{d_{A_1}-1}{d_{A}-1}\:\Norm{\rhoAR-\idx{\pi}{A}\otimes\rhoR}{2}^2\label{bndme}\\
&\quad+\frac{c}{d_A(d_A-1)}\:\Norm{\rhoAR^{cl}-\idx{\pi}{A}\otimes\rhoR}{2}^2\nonumber\\
&\quad+\frac{c}{x}\Trace{\left(\rhoAR-\idx{\pi}{A}\otimes\rhoR\right)^{\otimes2}\ \cY\otimes\idx{\cF}{R}}\nonumber.
\end{align}
To bound the third term we note that the vector $\cY$ was chosen in a way such that it corresponds to a multiple of the second column of $G_1^{-1}$. We have that
\begin{align}
\cY=\sum_{i}\:x\:((G_1)_{i2})^{-1}\cA_i\label{comp2}.
\end{align}
On the other hand any element of the commutant can be written in the $\cA_i$ basis via
\begin{align}
\cX=\sum_{i,j}\:((G_1)_{ij})^{-1}\Trace{\cX\cA_j}\:\cA_i\label{comp1}.
\end{align}
Comparing Equations~\eqref{comp2} and~\eqref{comp1} we conclude that
\begin{align}
\Trace{\cY\cA_j}=x\:\delta_{2j},
\end{align}
which implies $\trace{\cY^2}=x$. The third term of Equation~\eqref{bndme} can be bounded as follows. We have that  
\begin{align}
&\frac{c}{x}\Trace{\left(\rhoAR-\idx{\pi}{A}\otimes\rhoR\right)^{\otimes2}\ \cY\otimes\idx{\cF}{R}}\nonumber\\
&=\frac{c}{x}\Trace{\left(\rhoAR-\idx{\pi}{A}\otimes\rhoR\right)\otimes\idi{A'}\ \left(\idx{\rho}{A'R}-\idx{\pi}{A'}\otimes\rhoR\right)\otimes\idi{A}\ \cY\otimes\idx{\id}{R}}\label{smstr}\\
&\leq\frac{c}{x}\:\Trace{\left(\rhoAR-\idx{\pi}{A}\otimes\rhoR\right)^2}\sqrt{\Trace{\cY^2}}\label{finine}\\
&=\frac{c}{\sqrt{x}}\:\Norm{\rhoAR-\idx{\pi}{A}\otimes\rhoR}{2}^2
\end{align}
To obtain the inequality we note that Equation~\eqref{smstr} has the same structure as the right hand side of Equation~\eqref{ttexpd} and can be bounded identically. Thus, Inequality~\eqref{finine} follows from the derivations of Section 3.4. We note that for $\idx{d}{A}\geq4$  
\begin{align*}
\frac{d_A(d_{A_1}-1)(d_{A_2}-1)}{\sqrt{d_A(d_A-1)(d_A-2)(d_A-3)}}\leq1
\end{align*}
holds. Using this and plugging in $c$ and $x$ and we find
\begin{align}
&\frac{1}{\mathbb{\abs{P}}}\sum_{P\in\mathbb{P}(A)}{\Norm{\tr_{A_2}{(\idx{P}{A}\otimes\idi{R}\:\rhoAR\:\idx{P}{A}^\dagger\otimes\idi{R})}-\pi_{A_1}\otimes\rhoR}{2}^2}\nonumber\\
&\leq\frac{d_{A_1}-1}{d_{A}-1}\:\Norm{\rhoAR-\idx{\pi}{A}\otimes\rhoR}{2}^2\label{analysis}\\
&\quad+\frac{(d_{A_1}-1)(d_{A_2}-1)}{d_A-1}\:\Norm{\rhoAR^{cl}-\idx{\pi}{A}\otimes\rhoR}{2}^2\nonumber\\
&\quad+\frac{d_A(d_{A_1}-1)(d_{A_2}-1)}{\sqrt{d_A(d_A-1)(d_A-2)(d_A-3)}}\Norm{\rhoAR-\idx{\pi}{A}\otimes\rhoR}{2}^2\nonumber\\
&\leq\frac{d_{A_1}-1}{d_{A}-1}\:\tr{[\rhoAR^2]}+\frac{(d_{A_1}-1)(d_{A_2}-1)}{d_A-1}\:\tr{[(\rhoAR^{cl})^2]}+\tr{[\rhoAR^2]}.\label{usafthoeld}
\end{align}
This statement can be transformed into a decoupling theorem with the Schatten 1-norm with an application of the H\"{o}lder Inequality \eqref{hldrfk}. 
We proceed as in Section 5.2, Equations~\eqref{applyCQdec}-\eqref{usemeintheend} and introduce the positive definite and normalized operator $\zetaR$ writing
\begin{align}
&\Norm{\textnormal{tr}_{\textnormal{\tiny{A}}_2}{(\idx{P}{A} \otimes \idi{R} \ \rhoAR \  \idx{P}{A}^\dagger \otimes \idi{R})} - \pi_{\textnormal{\tiny{A}}_1} \otimes \idx{\rho}{R}}{1}\nonumber\\
&\leq\Norm{(\pi_{\textnormal{\tiny{A}}_1}\otimes\zetaR)^{-\frac{1}{4}}\left(\textnormal{tr}_{\textnormal{\tiny{A}}_2}{(\idx{P}{A} \otimes \idi{R} \ \rhoAR \  \idx{P}{A}^\dagger \otimes \idi{R})} - \pi_{\textnormal{\tiny{A}}_1} \otimes \idx{\rho}{R}\right)(\pi_{\textnormal{\tiny{A}}_1}\otimes\zetaR)^{-\frac{1}{4}}}{2}\\
&=\sqrt{d_{\textnormal{\tiny{A}}_1}}\:\Norm{\textnormal{tr}_{\textnormal{\tiny{A}}_2}{(\idx{P}{A} \otimes \idi{R} \ \idx{\tilde{\rho}}{AR} \  \idx{P}{A}^\dagger \otimes \idi{R})} - \pi_{\textnormal{\tiny{A}}_1} \otimes \idx{\tilde{\rho}}{R}}{2}\label{applyCQdeclater}
\end{align}
To keep the notation simple we introduced an operator $\idx{\tilde{\rho}}{AR}:=(\id_{A}\otimes\zetaR)^{-\frac{1}{4}}\rhoAR(\id_A\otimes\zetaR)^{-\frac{1}{4}}$. Thus, we can bound 
\begin{align}
&\frac{1}{\mathbb{\abs{P}}}\sum_{P\in\mathbb{P}(A)}{\Norm{\tr_{A_2}{(\idx{P}{A}\otimes\idi{R}\:\rhoAR\:\idx{P}{A}^\dagger\otimes\idi{R})}-\pi_{A_1}\otimes\rhoR}{1}^2}\nonumber\\
&\leq\frac{1}{\mathbb{\abs{P}}}\sum_{P\in\mathbb{P}(A)}{d_{\textnormal{\tiny{A}}_1}}\:\Norm{\textnormal{tr}_{\textnormal{\tiny{A}}_2}{(\idx{P}{A} \otimes \idi{R} \ \idx{\tilde{\rho}}{AR} \  \idx{P}{A}^\dagger \otimes \idi{R})} - \pi_{\textnormal{\tiny{A}}_1} \otimes \idx{\tilde{\rho}}{R}}{2}^2\\
&\leq d_{\textnormal{\tiny{A}}_1}\left(\frac{d_{A_1}-1}{d_{A}-1}\:\tr{[\tilde{\rho}_{AR}^2]}+\frac{(d_{A_1}-1)(d_{A_2}-1)}{d_A-1}\:\tr{[(\tilde{\rho}_{AR}^{cl})^2]}+\tr{[\tilde{\rho}_{AR}^2]}\right)\label{explwthab},
\end{align}
where Inequality~\eqref{explwthab} uses~\eqref{usafthoeld}. We now choose $\zetaR$ such that $\tr{[\tilde{\rho}_{AR}^2]}$ is minimal and we have $\tr{[\tilde{\rho}_{AR}^2]}\leq2^{-\chmin{A}{R}{\rho}}$.
Furthermore note that by the definition of the min-entropy we have that 
\begin{align*}
\rhoAR\leq2^{-\chmin{A}{R}{\rho}}\:\idi{A}\otimes\zetaR
\end{align*}
which implies that
\begin{align*}
\rho_{AB}^{cl}\leq2^{-\chmin{A}{R}{\rho}}\:\idi{A}\otimes\zetaR
\end{align*}
and therefore
\begin{align*}
\tr{[(\tilde{\rho}_{AR}^{cl})^2]}\leq\trace{\rho_{AR}}\:2^{-\chmin{A}{R}{\rho}}\leq2^{-\chmin{A}{R}{\rho}}.
\end{align*}
With this we bound the right hand side of Inequality~\eqref{explwthab}
\begin{align}
&d_{\textnormal{\tiny{A}}_1}\left(\frac{d_{A_1}-1}{d_{A}-1}\:\tr{[\tilde{\rho}_{AR}^2]}+\frac{(d_{A_1}-1)(d_{A_2}-1)}{d_A-1}\:\tr{[(\tilde{\rho}_{AR}^{cl})^2]}+\tr{[\tilde{\rho}_{AR}^2]}\right)\nonumber\\
&\leq d_{\textnormal{\tiny{A}}_1}\left(\frac{d_{A_1}-1}{d_{A}-1}\:2^{-\chmin{A}{R}{\rho}}+\frac{(d_{A_1}-1)(d_{A_2}-1)}{d_A-1}\:2^{-\chmin{A}{R}{\rho}}+2^{-\chmin{A}{R}{\rho}}\right)\label{chosemin}\\
&\leq 2\cdot d_{\textnormal{\tiny{A}}_1}\cdot2^{-\chmin{A}{R}{\rho}}.
\end{align}
\\
{\textbf{Theorem:} (Decoupling Quantum States with Classical Operations)\\ \textit{Let $\rho_{AR}\in\subnormstates{\cH_{AR}}$ be a sub normalized density operator and let $\idx{d}{A}\geq4$, then
\begin{align}
\frac{1}{\mathbb{\abs{P}}}\sum_{P\in\mathbb{P}(A)}{\Norm{\tr_{A_2}{(\idx{P}{A}\otimes\idi{R}\:\rhoAR\:\idx{P}{A}^\dagger\otimes\idi{R})}-\pi_{A_1}\otimes\rhoR}{1}}\leq\sqrt{2\:{d}_{A_1}\:2^{-\chmin{A}{R}{\rho}}},\nonumber
\end{align}
where the summation goes over all permutation operators.
}}\\
\\
We note that the above formula generalizes the Hash Lemma obtained in Section 5.2. Furthermore it implies that there is a classical operation that decouples well in the sense of the lemma. The fully classical Hash Lemma contains essentially the same upper bound as the above and is known to be tight \cite{rennerkoenig05,TSSR10}. Therefore one cannot expect significantly better bounds in the above formula. An extension of the above using approximate 2-wise independent families of permutations in the spirit of \cite{TSSR10} can be obtained using the techniques developed in the previous chapters.

\appendix
\renewcommand\appendix{\par 
   \setcounter{section}{0}%
   \setcounter{subsection}{0}%
   \setcounter{figure}{0}%
   \renewcommand\thesection{\Alph{section}}%
   \renewcommand\thefigure{\Alph{section}.\arabic{figure}}}
\chapter{H\"older Inequality}
We state H\"older inequality as given in \cite{Bhatia}:\\
\\
\textbf{Theorem:} (H\"older Inequality for Unitarily Invariant Norms)\textit{
For every unitarily invariant norm and for all square matrices A, B
$$\left\|{AB}\right\|\leq\left\|\abs{A}^p\right\|^{\frac{1}{p}}\left\|\abs{B}^q\right\|^{\frac{1}{q}}$$
for all $p>1$ and $\frac{1}{p}+\frac{1}{q}=1$.}\\
\\
If one applies the inequality twice, one arrives at the following corollary.\\
\\
\textbf{Corollary:} (H\"older Inequality for Three Matrices)\textit{
For every unitarily invariant norm and for all square matrices A, B, C
$$\left\|{ABC}\right\|\leq\left\|\abs{A}^r\right\|^{\frac{1}{r}}\left\|\abs{B}^s\right\|^{\frac{1}{s}}\left\|\abs{C}^t\right\|^{\frac{1}{t}}$$
for all $r>1$ and $\frac{1}{r}+\frac{1}{s}+\frac{1}{t}=1$.}
\\

\chapter{Jensen Inequality}
For completeness we shortly state the widely used Jensen Inequality:\\
\\
\textbf{Jensen Inequality:}
\textit{Let $(\Omega, A,\mu)$ be a measure space, with $\mu(\Omega)=1$. If $g$ is a real- valued function that is $\mu$-integrable, and if $\varphi$ is a convex function on the real numbers, then:}

$$\varphi\left(\int_\Omega g\: d\mu\right) \leq \int_\Omega \varphi \circ g\: d\mu.$$\\
\\
Note that if $\varphi$ is concave then $-\varphi$ must be convex. Therefore the Jensen Inequality is also valid for concave functions $\varphi$, but with the inequality sign reversed. We will often use the Jensen Inequality for the concave square root.
\chapter{Swap Trick}
The Swap Trick is of crucial importance throughout the derivations in this thesis. For a fixed basis $\idx{\ket{i}}{A}$ of some Hilbert space $\HA$ we introduce the swap operator $\idx{\cF}{A}$ acting on the bipartite Hilbert space $\idx{\cH}{AA'}$
\begin{align}
\idx{\cF}{A}:=\sum_{i,j}^{\idx{d}{A}}{\proji{i}{j}{A}\otimes\proji{j}{i}{A'}}.
\end{align}
We generally leave out the index of the second subsystem writing $\idx{\cF}{A}$ since it is determined by the first one already.\\
\\
\textbf{Lemma:} (Swap Trick)\textit{ Let $M$, $N\in \linops{\HA}$ and let $\mathcal{F}$ be the swap operator, then}
$$\tr(MN) \ = \ \tr((M \otimes N)\mathcal{F})$$
This result can be shown writing down $M$ and $N$ in the standard basis directly: $M = \sum\limits_{i,j} m_{ij} \proj{i}{j}$ and 
$N = \sum\limits_{k,l} n_{kl} \proj{k}{l} $. Then,

\begin{align} 
\tr((M \otimes N)\mathcal{F}) &= \tr( \sum\limits_{i,j,k,l} m_{ij} n_{kl} \proj{i}{j} \otimes \proj{k}{l} \mathcal{F}) \\
&= \tr(\sum\limits_{i,j,k,l} m_{ij} n_{kl} \proj{i}{l} \otimes \proj{k}{j})\\
&=\sum\limits_{i,j} m_{ij} n_{ji} \\
&= \tr(MN).
\end{align}
\chapter{The Murnaghan-Nakayama Rule}
The Murnaghan-Nakayama rule provides a possibility to graphically construct the values of the characters of the irreducible representations of the symmetric group $S_d$. It is a recursive rule that gives $\chi_\lambda\left((1)^{k_1},(2)^{k_2},...,(d)^{k_d}\right)$ for a given conjugacy class labeled with $\left((1)^{k_1},(2)^{k_2},...,(d)^{k_d}\right)$ and an irreducible representation labeled with a partition $\lambda$ of $d$. A \textit{skew hook} is a connected part of a Young Diagram which does not contain any $2\times2$ subset of cells 
$$\yng(2,2)$$
and that can be removed in a way such that the remaining boxes still form a smaller valid Young Diagram. For example the marked boxes in the following diagram form 4-hooks.
$$\young(\hfill\hfill\bullet,\hfill\bullet\bullet,\hfill\bullet)\qquad\qquad\young(\hfill\hfill\hfill,\hfill\bullet\bullet,\bullet\bullet)$$
For a skew hook starting in the $i$-th row of a Young Diagram and ending in the $j$-th row, we call the number $k=j-i$ the \textit{length} of the hook and the number $s=(-1)^k$ its sign.
In our example the length of the first skew hook is two, while the second one has length one.\\
The Murnaghan-Nakayama rule states that to calculate $\chi_\lambda\left((1)^{k_1},(2)^{k_2},...,(d)^{k_d}\right)$ for a partition $\lambda=(\lambda_1,...,\lambda_n)$ one can proceed with the following recursion: Choose the cycle with greatest length in $\left((1)^{k_1},(2)^{k_2},...,(d)^{k_d}\right)$. In a first step draw all possible ways of removing a skew hook of the length of that cycle from the diagram corresponding to $\lambda$ and write down $s=(-1)^k$ for each possibility (if there is no such way the contribution is zero).
Then for each obtained sub diagram draw all ways of removing a hook of length of the second from the right cycle in $\left((1)^{k_1},(2)^{k_2},...,(d)^{k_d}\right)$ again writing down the sign of each resulting diagram but this time multiply it with the sing of the parent diagram. Do this over all the cycles in $\left((1)^{k_1},(2)^{k_2},...,(d)^{k_d}\right)$, such that in the end there are no boxes left anymore. The sum of all the numbers obtained in the last step is $\chi_\lambda\left((1)^{k_1},(2)^{k_2},...,(d)^{k_d}\right)$. A more detailed discussion of the Murnaghan-Nakayama rule can be found in \cite{SymGroup}.\\
\section{The Irreps of $R$}
This section aims at giving a sketch of a proof of the relations
\begin{align}
\chi_{(d)}(a)&=1\label{appendd1}\\
\chi_{(d-1,1)}(a)&=k_1-1\label{appendd2}\\
\chi_{(d-2,1,1)}(a)&=\frac{1}{2}\:(k_1-1)(k_1-2)-k_2\label{appendd3}\\
\chi_{(d-2,2)}(a)&=\frac{1}{2}\:k_1(k_1-3)+k_2\label{appendd4}
\end{align}
which were used in Chapter 7. The first relation \eqref{appendd1} is evident since the character of the trivial representation is one on any conjugacy class.
Nevertheless we can use the Murnaghan-Nakayama rule to recover that value. The Young Diagram of the trivial representation of $S_d$ has $d$-boxes arranged as
$$\yng(5)\:...\:\yng(1).$$
For a conjugacy class $\left((1)^{k_1},(2)^{k_2},...,(d)^{k_d}\right)$ we pick the greatest cycle. There is only one possibility to erase a skew-hook from the above diagram in any case. For example if this cycle has length four this gives
$$\yng(5)\:...\:\young(\bullet\bullet\bullet\bullet).$$
We note that the length of the skew hook is zero and its sign is $+1$. We then proceed to the next from the right cycle in the class $\left((1)^{k_1},(2)^{k_2},...,(d)^{k_d}\right)$. But again there is only one possibility to erase it and in any case the resulting sing is $+1$.
This procedure can be done until no boxes are left anymore and the result is always
\begin{align}
1\cdot1\cdot1\cdot...\cdot1=1,
\end{align}
which proves \eqref{appendd1}.\\
We now consider \eqref{appendd2} and do the calculation exemplary for $S_{11}$. The generalization to $S_d$ is straight forward. We fix some conjugacy class $\left((1)^{k_1},(2)^{k_2},...,(11)^{k_{11}}\right)$. The Young Diagram corresponding to the irreducible representation is: 
$$\yng(10,1)$$
(In the general case there are $d-1$ boxes in the upper row instead of ten.) Assume for now that $k_1\geq2$. For any cycle of length greater than one there is only one possibility to erase a skew hook from the diagram. For example for a two cycle we would get:
$$\young(\hfill\hfill\hfill\hfill\hfill\hfill\hfill\hfill\hfill\bullet\bullet,\hfill)$$
Since we assumed that $k_1\geq2$ the sign will be $+1$. In this way it is possible to eliminate all cycles with length greater than one and the numerical value corresponding to the diagram will not change. The problem reduces to calculating the value of the resulting diagram with $k_1>2$ boxes on a conjugacy class which contains $k_1\geq2$ ones only. This time there are two possibilities to erase a box from the diagram and still to obtain a valid diagram. For example:
$$\young(\hfill\hfill\hfill\hfill\hfill,\bullet)\qquad\qquad\young(\hfill\hfill\hfill\hfill\bullet,\hfill)$$
While the first diagram gives one immediately the second diagram again can be decomposed. The result is a summation: Each decomposition of a diagram into sub diagrams yields an additional one. Since one can decompose the diagram $k_1-1$ times the sum is $k_1-1$. In total we have
\begin{align}
\chi_{(10,1)}\left(\left((1)^{k_1},(2)^{k_2},...,(11)^{k_{11}}\right)\right)&=k_1-1
\end{align}
for $k_1\geq2$. The cases $k_1=1$ and $k_1=0$ we check explicitly. If $k_1=1$ before we subtract the last box the situation is given by:
$$\young(\bullet\bullet\bullet\bullet\bullet\bullet\bullet\bullet\bullet\bullet,\hfill)$$
But the resulting diagram with the one box in the second row is not valid in any case so that the character is $0=1-1$. In the case $k_1=0$ there is at least a two cycle at the end yielding to a skew hook of length one. The character in this case is
\begin{align}
1\cdot1\cdot...\cdot1\cdot(-1)&=-1\\
&=0-1,
\end{align}
which shows that the formula
\begin{align}
\chi_{(10,1)}\left(\left((1)^{k_1},(2)^{k_2},...,(11)^{k_{11}}\right)\right)&=k_1-1
\end{align}
is valid in this cases, too.
The reader should have no difficulties to see that the whole argumentation did not depend on $d$. Therefore
\begin{align}
\chi_{(d-1,1)}\left(\left((1)^{k_1},(2)^{k_2},...,(d)^{k_{d}}\right)\right)&=k_1-1
\end{align}
and equation \eqref{appendd2} is proved.\\
We proceed with the validation of \eqref{appendd4}:action of \eqref{appendd4}:
\begin{align}
\chi_{(d-2,2)}(a)&=\frac{1}{2}\:k_1(k_1-3)+k_2
\end{align}
We start with considering the special case $k_2=0$, i.\:e.\:we evaluate the character on a conjugacy class that does not contain cycles of length two. As before, we do this exemplary for $S_{12}$ (for notational convenience), but the generalization to arbitrary $d$ is apparent. Assume for the moment $k_1\geq4$. As there are no two cycles in the conjugacy class by assumption the next shortest cycles after $1$-cycles are three cycles. But there is only one possibility of subtracting skew hooks with three or more boxes from the diagram:
$$\young(\hfill\hfill\hfill\hfill\hfill\hfill\hfill\bullet\bullet\bullet,\hfill\hfill)$$
In any case the skew hook has sign one, such that the numerical value of the sub diagram always gets a factor of one from the parent diagram. After subtracting all skew hooks with more than one box we end up in a situation, where we have to evaluate a diagram of the type
$$\young(\hfill\hfill\hfill\hfill\hfill,\hfill\hfill)$$
on a conjugacy class that contains $k_1$ ones only. Since this is the conjugacy class of the identity the problem can equivalently be seen as the one of calculating the dimension of the irreducible representation 
$$\young(\hfill\hfill\hfill\hfill\hfill,\hfill\hfill)$$
of $S_{k_1}$. This is done the easiest with an application of the Hook Formula \cite{SymGroup}. The result is
$$\textnormal{dim}\left(\quad\young(\hfill\hfill\hfill\hfill\hfill,\hfill\hfill)\quad\right)=\frac{1}{2}\:k_1\:(k_1-3).$$
Of course this result can also be obtained with an application of the Murnaghan-Nakayama rule. We check the cases $k_1\in\{1,2,3\}$ with explicit calculations. If $k_1=3$, then at the end there must be three boxes left but this is only possible if subtracts a skew hook in the following way:
$$\young(\hfill\bullet\bullet\bullet\bullet,\hfill\hfill)$$
But the resulting diagram after the subtraction is not a valid Young Diagram and the result is therefore zero. If $k_1=2$ at the end two boxes are left
as for example in:
$$\young(\hfill\bullet\bullet\bullet\bullet,\hfill\bullet)$$
Note that in any case the drawn hook has sign $-1$, which adds an additional factor of $-1$. Thus in this case the result is $-1$.
The case $k_1=1$ yields the result $-1$. Finally in the case of $k_1=0$ there is no possible valid skew hook that can fill all boxes at once and the result is zero. The formula
\begin{align}
\chi_{(10,2)}\left(\left((1)^{k_1},(2)^{0},...,(12)^{k_{12}}\right)\right)&=\frac{1}{2}\:k_1(k_1-3)
\end{align}
is valid in any case. And again the argumentation shown is valid for general $d$ which proofs \eqref{appendd4} for all conjugacy classes that do not contain two cycles. We also need to consider that case to complete the proof. But even if there are 2-cycles, as before, there is only way of subtracting skew hooks with three or more boxes from the diagram:
$$\young(\hfill\hfill\hfill\hfill\hfill\hfill\hfill\bullet\bullet\bullet,\hfill\hfill)$$
The prefactor is always one and we are left with a situation, where we have to evaluate a character of the type
$$\young(\hfill\hfill\hfill\hfill\hfill\hfill,\hfill\hfill)$$
on a conjugacy class that contains 2- and 1-cycles only. For the moment again assume that $k_1\geq4$. There are generally two possibilities to remove a 2-hook from such a diagram:
$$\young(\hfill\hfill\hfill\hfill\hfill\hfill,\bullet\bullet)\qquad\qquad\young(\hfill\hfill\hfill\hfill\bullet\bullet,\hfill\hfill)$$
The first one yields $1$ in any case, while the second one can again be decomposed in the same manner until there are only $k_1$ 1-cycles left.
Each decomposition adds a $1$ to the total sum, while the remaining diagram is of the type discussed above i.\:e.\:it can be treated with the formula
\begin{align}
\chi_{(k_1-2,2)}\left(\left((1)^{k_1},(2)^{0},...,(12)^{0}\right)\right)&=\frac{1}{2}\:k_1(k_1-3).
\end{align}
Since there are $k_2$ possible decompositions we obtain generalizing the above discussion to $S_d$ that for $k_1\geq4$
\begin{align}
\chi_{(d-2,2)}(a)&=\frac{1}{2}\:k_1(k_1-3)+k_2.
\end{align}
The remaining cases can be checked by explicit calculations. Analogously one verifies formula \eqref{appendd3}. The only real difference is that in this case erasing a skew hook of the type
$$\young(\hfill\hfill\hfill\hfill\hfill\hfill\hfill\hfill\hfill\hfill,\bullet,\bullet)$$
gives a $-1$ instead of a $+1$.

\addcontentsline{toc}{chapter}{Bibliography}
\bibliography{Masterthesis}
\bibliographystyle{abbrv}

\end{document}